# Requirements-Driven Automated Software Testing: A Systematic Review

FANYU WANG, CHETAN ARORA, CHAKKRIT TANTITHAMTHAVORN, KAICHENG HUANG, and ALDEIDA ALETI, Faculty of Information Technology, Monash University, Australia

Automated software testing has significant potential to enhance efficiency and reliability within software development processes. However, its broader adoption faces considerable challenges, particularly concerning alignment between test generation methodologies and software requirements. **RE**quirements-**D**riven **A**utomated **S**oftware **T**esting (REDAST) addresses this gap by systematically leveraging requirements as the foundation for automated test artifact generation. This systematic literature review (SLR) critically examines the REDAST landscape, analyzing the current state of requirements input formats, transformation techniques, generated test artifacts, evaluation methods, and prevailing limitations. We conducted a thorough analysis of 156 relevant studies selected through a rigorous multi-stage filtering process from an initial collection of 27,333 papers sourced from six major research databases. Our findings highlight the predominance of functional requirements, model-based specifications, and natural language formats. Rule-based techniques are extensively utilized, while machine learning-based approaches remain relatively underexplored. Furthermore, most existing frameworks are sequential and dependent on singular intermediate representations, and while test cases, structured textual formats, and requirements coverage are common, full automation remains rare. We identify significant gaps related to automation completeness, and dependency on input quality. This comprehensive synthesis provides a detailed overview of REDAST research and limitations, offering clear, evidence-based recommendations to guide future advancements in automated software testing.

**ACM Reference Format:**
Fanyu Wang, Chetan Arora, Chakkrit Tantithamthavorn, Kaicheng Huang, and Aldeida Aleti. 2018. Requirements-Driven Automated Software Testing: A Systematic Review. In *Proceedings of Make sure to enter the correct conference title from your rights confirmation email (Conference acronym 'XX).* ACM, New York, NY, USA, 66 pages. https://doi.org/XXXXXXX.XXXXXXX

## 1 Introduction

Software testing plays a critical role in assuring the quality, reliability, and performance of software systems [12, 16]. At its core, software testing seeks to identify and address defects, ensure that a system functions as intended, meets stakeholder requirements, and mitigates risks before deployment [13]. Central to achieving these objectives is aligning testing activities with the software's requirements, as requirements define the system's intended behaviour and scope. Effective testing demands comprehensive test artifacts that trace back to these requirements, including test cases, plans, and scenarios. As software systems grow more complex, there is an increasing need for automated approaches to bridge the gap between requirements engineering (RE) and software testing [86]. This increasing complexity of software

Authors' Contact Information: Fanyu Wang, fanyu.wang@monash.edu; Chetan Arora, chetan.arora@monash.edu; Chakkrit Tantithamthavorn, Chakkrit@monash.edu; Kaicheng Huang, khua0042@student.monash.edu; Aldeida Aleti, aldeida.aleti@monash.edu, Faculty of Information Technology, Monash University, Clayton, Victoria, Australia.

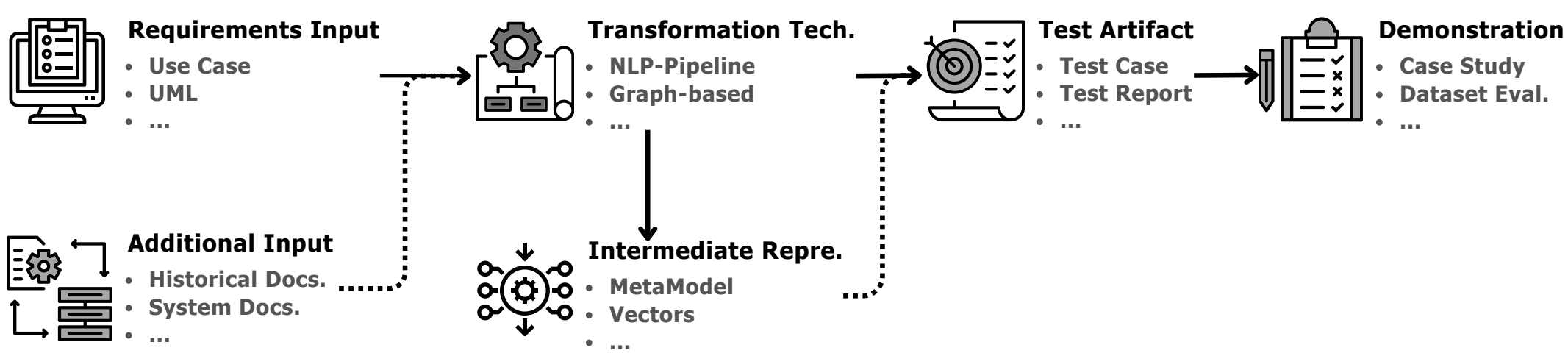

Fig. 1. The General Framework of REDAST Methodology (The dotted route show optional paths.)

systems has amplified the importance of *requirements-driven testing*—a paradigm that directly links testing activities to the software requirements.

In the requirements-driven automated software testing paradigm, the test generation process is mainly motivated by requirements, where the requirements are further interpreted, transformed, and implemented to test artifacts with the help of some additional supporting documents or tools [62]. Traditionally, this process has been predominantly manual, relying on human effort to interpret requirements and design corresponding tests [15, 30]. However, manual approaches are time-consuming, prone to human error, and struggle to scale with modern software systems' increasing size and complexity. To address these limitations, several studies have explored automated methods for generating test artifacts directly from requirements. These methods leverage model representations, heuristics, and natural language processing (NLP) techniques to translate requirements into actionable testing artifacts. However, a systematic understanding of the approaches, their challenges, and potential opportunities in requirements-driven testing remains limited.

In this paper, we study the **RE**quirements-**D**riven **A**utomated **S**oftware **T**esting (REDAST) landscape. Fig. 1 shows an overview of the REDAST process. The generation process consists of several components, including requirements as the necessary input, additional documents as the potential input, intermediate expressions as the potential step, transformation techniques as the main applied methodologies, and generated test artifacts as the final outcome. We cover the REDAST research landscape by (1) summarizing and reporting the statistics of valuable research works of automated software test generation, (2) comparing the evaluation methods for the current studies, and (3) identifying and analyzing current limitations and future opportunities of automated test generation technology in the context of the current era. We not only focus on the previous research under the traditional software engineering (SE) perspective but also introduce a new view from advance technologies to discuss the prospects and possibilities for REDAST studies. After screening 27,333 articles published over the past 30 years, we conducted a systematic review of 156 studies [1–156] in accordance with the empirical systematic literature review guidelines of Kitchenham et al. [45].

***Structure.*** **Section 2** discusses the background concepts and related work for REDAST. **Section 3** presents our methodology and process of conducting our systematic review. **Section 4** describes the taxonomy behind REDAST process. **Section 5** discusses results from our five RQs. **Section 6** examines threats to validity of our study. **Section 7** discusses the insights from our results and the REDAST research roadmap. **Section 8** concludes the paper. In the Appendix, we also show an overview of papers that cover non-functional requirements.

## 2 Background and Related Work

### 2.1 Automated Software Testing and Requirements Engineering

Requirement engineering is the initial phase in software development, guiding all subsequent stages [8, 88]. The RE process requires gathering user needs and implementing the non-structured requirements into modeling language or other formal statements [31, 70]. It encompasses various activities tailored to the specific demands of software systems, with requirements elicitation, analysis, specification, and validation being the most necessary stages [47, 109].

Software testing aims to provide objective, independent information about the quality of software and the risk of its failure to users or sponsors [3, 78]. Automated software testing is using automation techniques to use specialized tools and scripts to execute test cases on a software application without manual intervention, which can improve time efficiency and human resource efficiency [24, 71]. While the satisfaction of stakeholders is one of the priorities in software testing, the relationship between software requirements and testing becomes a critical focus in Software Development Life Cycle SDLC [11, 58]. The alignment between different stages of verification and validation, e.g., system analysis and system testing, is key for effective software quality assurance. Here, we primarily focus on the requirements specification and testing, while testing verifies that the software meets its specified requirements. This relationship is fundamental to ensuring the final product aligns with stakeholder expectations and functions correctly.

### 2.2 REDAST Secondary Studies

REDAST studies have been long investigated in past research. However, only limited studies systematically discussed REDAST-related topics. Atoum et al. [10] conducted a systematic study that examines the requirements of quality assurance and validation, where they reported a test-oriented approach. Unterkalmsteiner et al. [87] built a taxonomy for aligning requirements engineering and software testing to enhance coordination between these activities. They pointed out the importance of integrating requirements into the testing process, which includes some REDAST studies. Mustafa et al. [62] reviewed 30 selected papers between 2000-2018, offering a constrained analysis of the requirements-driven testing process from the perspectives of input artifacts and applied techniques. They solely focus on the generation of test cases based on requirements, and also investigate the transformation techniques. While there is an understandable overlap in terminology in some aspects between our study and Mustafa et al. [62] study, we undertake a much broader and comprehensive survey of all REDAST-related research, since 1992 (i.e., the concept's inception), and do not restrict the testing artefats to only test cases. Our study differs from previous REDAST research in three main ways. First, whereas Mustafa et al. [62] limit their scope to test-case generation and Unterkalmsteiner et al. [87] focus exclusively on the alignment between requirements and testing, we adopt a truly end-to-end perspective on the entire requirements-driven automated software testing process. Second, Mustafa et al.'s review considers only papers published before 2018, but our survey spans the full timeline of REDAST research—including landmark NLP advances such as BERT [25] and ChatGPT [41]—to provide a cutting-edge viewpoint. Finally, instead of offering a brief, fragmented overview, we introduce a comprehensive taxonomy that classifies every identified REDAST study and analyze each work within this framework, thereby giving readers a structured, holistic understanding of the field.

## 3 Research Methodology

In this section, we discuss the process of conducting our systematic review, e.g., our search strategy for data extraction of relevant studies, based on the guidelines of Kitchenham et al. [45] to conduct systematic literature reviews and

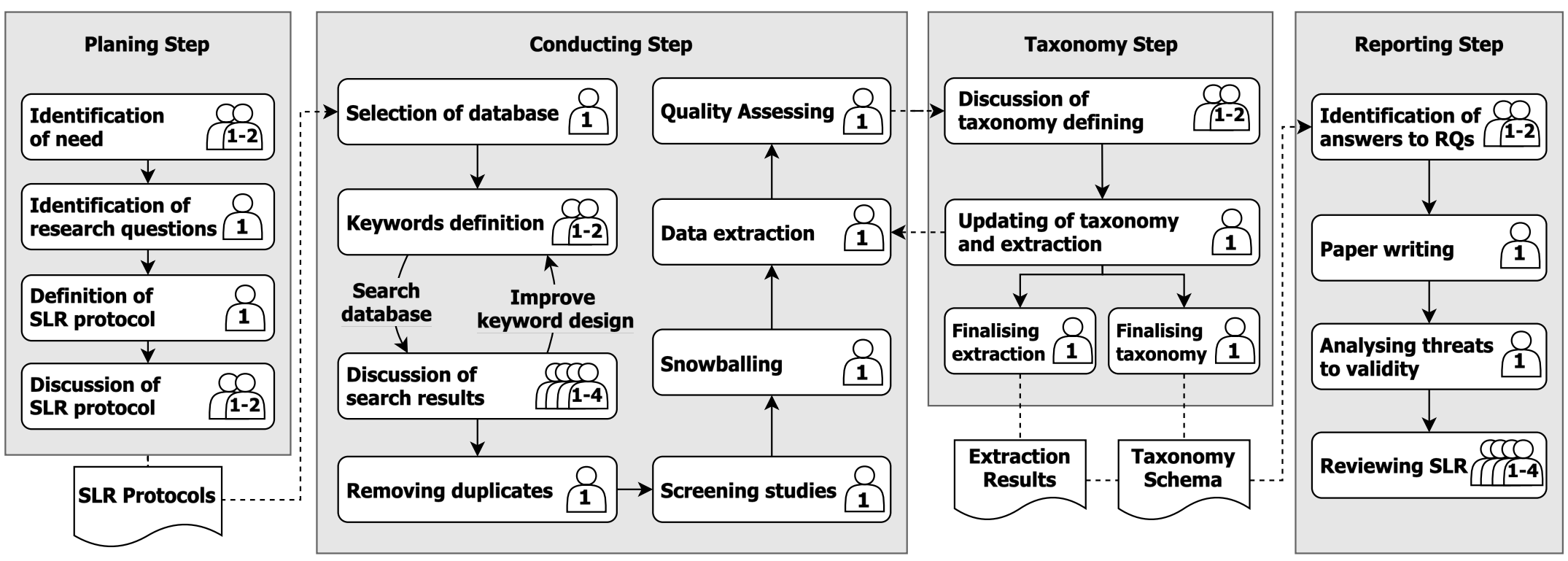

Fig. 2. Systematic Literature Review Process

Petersen et al. [68] to conduct systematic mapping studies in Software Engineering. In this systematic review, we divide our work into a four-stage procedure, including planning, conducting, building a taxonomy, and reporting the review, illustrated in Fig. 2. The four stages: (1) the *planning* stage involved identifying RQs and specifying the detailed research plan for the study; (2) the *conducting* stage involved analyzing and synthesizing the existing primary studies to answer the RQs; (3) the *taxonomy* stage consolidated a taxonomy schema for REDAST methodology and optimized the data extraction results; (4) the *reporting* stage involved reviewing, concluding, and reporting the final results.

### 3.1 Research Question Descriptions

In this study, we developed five research questions to identify the input and output, analyze technologies, evaluate metrics, identify challenges, and identify potential opportunities.

**RQ1. What are the input configurations, formats, and notations used in the requirements in requirements-driven automated software testing?** In requirements-driven testing, the input is some form of requirements specification – which can vary significantly. RQ1 maps the input for REDAST and reports on the comparison among different formats for requirements specification.

**RQ2. What are the frameworks, tools, processing methods, and transformation techniques used in requirements-driven automated software testing studies?** RQ2 explores the technical solutions from requirements to generated artifacts, e.g., rule-based transformation applying NLP pipelines and deep learning (DL) techniques, where we additionally discuss the potential intermediate representation and additional input for the transformation process.

**RQ3. What are the test formats and coverage criteria used in the requirements-driven automated software testing process?** RQ3 focuses on identifying the formulation of generated artifacts (i.e., the final output). We map the adopted test formats and analyze their characteristics in the REDAST process.

**RQ4. How do existing studies evaluate the generated test artifacts in the requirements-driven automated software testing process?** RQ4 identifies the evaluation datasets, metrics, and case study methodologies in the selected papers. This aims to understand how researchers assess the effectiveness, accuracy, and practical applicability of the generated test artifacts.

**RQ5. What are the limitations and challenges of existing requirements-driven automated software testing methods in the current era?** RQ5 addresses the limitations and challenges of existing studies while exploring future directions in the current era of technology development.

Table 1. List of Search Terms

| Terms Group | Terms |
|---|---|
| Test Group | test* |
| Requirement Group | requirement* OR use case* OR user stor* OR specification* |
| Software Group | software* OR system* |
| Method Group | generat* OR deriv* OR map* OR creat* OR extract* OR design* OR priorit* OR construct* OR transform* |

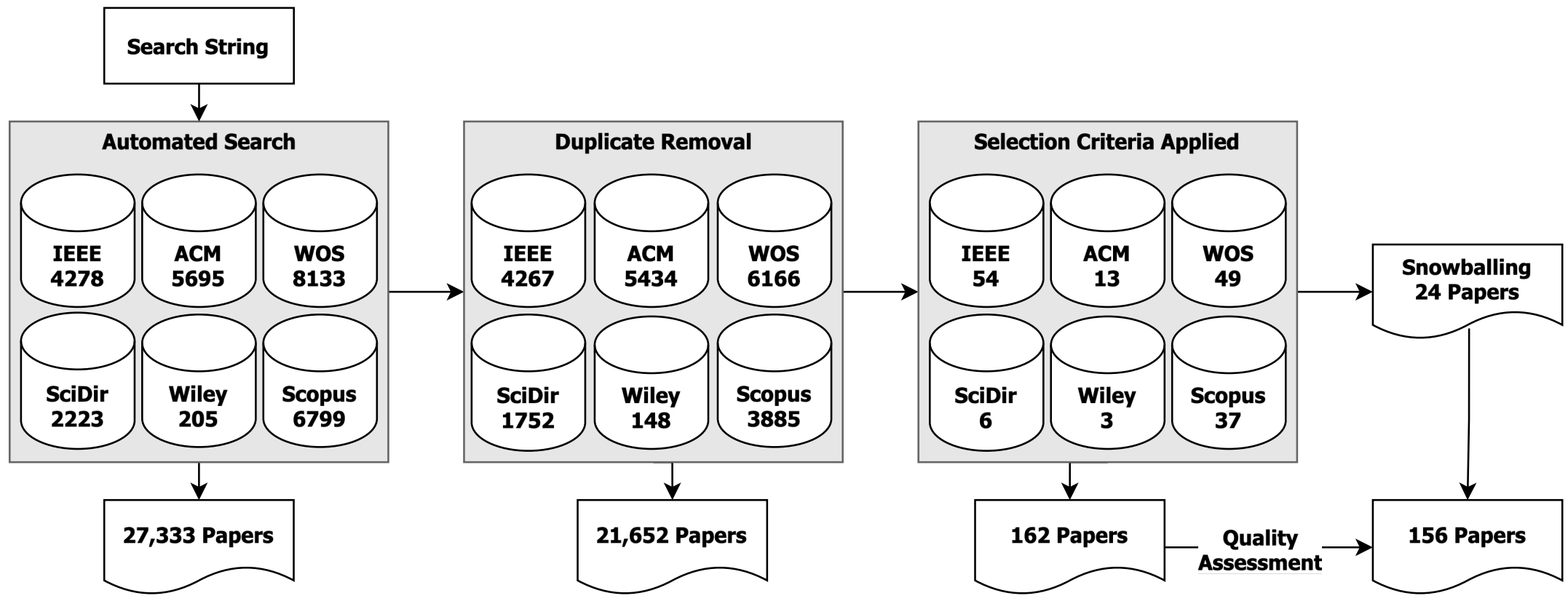

Fig. 3. Study Search Process

### 3.2 Searching Strategy

The overview of the search process is exhibited in Fig. 3, which includes all the details of our search steps.

*3.2.1 Search String Formulation.* Our RQs guided the identification of the main search terms. We designed our search string with generic keywords to avoid missing out on any related papers, where four groups of search terms are included, namely "test group", "requirement group", "software group", and "method group". In order to capture all the expressions of the search terms, we use wildcards to match the appendix of the word, e.g., "test*" can capture "testing", "tests" and so on. The search terms are listed in Table 1, decided after iterative discussion and refinement among all the authors. As a result, we finally formed the search string as follows:

**ON ABSTRACT** (("test*") **AND** ("requirement*" **OR** "use case*" **OR** "user stor*" **OR** "specifications") **AND** ("software*" **OR** "system*") **AND** ("generat*" **OR** "deriv*" **OR** "map*" **OR** "creat*" **OR** "extract*" **OR** "design*" **OR** "priorit*" **OR** "construct*" **OR** "transform*"))

The search process was conducted in September 2024, and therefore, the search results reflect studies available up to that date. We conducted the search process on six online databases: IEEE Xplore, ACM Digital Library, Wiley, Scopus, Web of Science, and Science Direct. However, some databases were incompatible with our default search string in the following situations: (1) unsupported for searching within abstract, such as Scopus, and (2) limited search terms, such as ScienceDirect. Here, for (1) situation, we searched within the title, keyword, and abstract, and for (2) situation, we separately executed the search and removed the duplicate papers in the merging process.

Table 2. Selection Criteria

| Criterion ID | Criterion Description |
|---|---|
| S01 | Papers written in English. |
| S02-1 | Papers in the subjects of "Computer Science" or "Software Engineering". |
| S02-2 | Papers published on software testing-related issues. |
| S03 | Papers published from 1991 to the present. |
| S04 | Papers with accessible full text. |

Table 3. Inclusion and Exclusion Criteria

| ID | Description |
|---|---|
| **Inclusion Criteria** | |
| I01 | Papers about requirements-driven automated system testing or acceptance testing generation, or studies that generate system-testing-related artifacts. |
| I02 | Peer-reviewed studies that have been used in academia with references from literature. |
| **Exclusion Criteria** | |
| E01 | Studies that only support automated code generation, but not test-artifact generation. |
| E02 | Studies that do not use requirements-related information as an input. |
| E03 | Papers with fewer than 5 pages (1-4 pages). |
| E04 | Non-primary studies (secondary or tertiary studies). |
| E05 | Vision papers and grey literature (unpublished work), books (chapters), posters, discussions, opinions, keynotes, magazine articles, experience, and comparison papers. |

*3.2.2 Automated Searching and Duplicate Removal.* We used advanced search to execute our search string within our selected databases, following our designed selection criteria in Table 2. The first search returned 27,333 papers. Specifically for the duplicate removal, we used a Python script to remove (1) overlapped search results among multiple databases and (2) conference or workshop papers, also found with the same title and authors in the other journals. After duplicate removal, we obtained 21,652 papers for further filtering.

*3.2.3 Filtering Process.* In this step, we filtered a total of 21,652 papers using the inclusion and exclusion criteria outlined in Table 3. This process was primarily carried out by the first and second authors. Our criteria are structured at different levels, facilitating a multi-step filtering process. This approach involves applying various criteria in three distinct phases. We employed a cross-verification method involving (1) the first and second authors and (2) the other authors. Initially, the filtering was conducted separately by the first and second authors. After cross-verifying their results, the results were then reviewed and discussed further by the other authors for final decision-making. We widely adopted this verification strategy within the filtering stages. During the filtering process, we managed our paper list using a BibTeX file and categorized the papers with color-coding through BibTeX management software[1], i.e., “red” for irrelevant papers, “yellow” for potentially relevant papers, and “blue” for relevant papers. This color-coding system facilitated the organization and review of papers according to their relevance.

The screening process is shown below,

- **1st-round Filtering** was based on the title and abstract, using the criteria I01 and E01. At this stage, the number of papers was reduced from 21,652 to 9,071.
- **2nd-round Filtering**. We attempted to include requirements-related papers based on E02 on the title and abstract level, which resulted from 9,071 to 4,071 papers. We excluded all the papers that did not focus on requirements-related information as an input or only mentioned the term "requirements" but did not refer to the requirements specification.

[1] https://bibdesk.sourceforge.io/

- **3rd-round Filtering**. We selectively reviewed the content of papers identified as potentially relevant to requirements-driven automated test generation. This process resulted in 162 papers for further analysis.

Note that, especially for third-round filtering, we aimed to include as many relevant papers as possible, even borderline cases, according to our criteria. The results were then discussed iteratively among all the authors to reach a consensus.

*3.2.4 Snowballing.* Snowballing is necessary for identifying papers that may have been missed during the search. Following Wohlin [96]'s guidelines, we conducted forward and backward snowballing with 24 additional papers.

*3.2.5 Data Extraction.* Based on the formulated RQs, we designed 38 data extraction questions[2] and created a Google Form to collect the required information from the relevant papers. The questions included 30 short-answer questions, six checkbox questions, and two selection questions. The data extraction was organized into five sections: (1) basic information: fundamental details such as title, author, venue, etc.; (2) open information: insights on motivation, limitations, challenges, etc.; (3) requirements: requirements format, notation, and related aspects; (4) methodology: details, including immediate representation and technique support; (5) test-related information: test format(s), coverage, and related elements. Similar to the filtering process, the first and second authors conducted the data extraction and then forwarded the results to the other authors to initiate the review meeting.

*3.2.6 Quality Assessment.* During the data extraction process, we encountered papers with insufficient information. To address this, we conducted a quality assessment in parallel to ensure the relevance of the papers to our objectives. This approach, also adopted in previous secondary studies [63, 76], involved designing a set of assessment questions based on guidelines by Kitchenham et al. [45]. The quality assessment questions in our study are shown below:

- **QA1**. Does this study clearly state *how* requirements drive automated test generation?
- **QA2**. Does this study clearly state the *aim* of REDAST?
- **QA3**. Does this study enable *automation* in test generation?
- **QA4**. Does this study demonstrate the usability of the method from the perspective of methodology explanation, discussion, case examples, and experiments?

QA4 originates from an open perspective in the review process, where we focused on evaluation, discussion, and explanation. Our review also examined the study's overall structure, including the methodology description, case studies, experiments, and analyses. Following this assessment, the final data extraction was based on 156 papers.

## 4 Taxonomy

In literature review studies, the taxonomy schema plays a crucial role in shaping the quality of statistical analysis and addressing RQs. To this end, we define a four-stage schema for our REDAST process based on our literature analysis and informed from Fig. 4. In each of the schemas, recognizing that the entire SE life cycle is a practical process, we incorporated multiple categorizations to enhance the structure and clarity of the schemas.

[2] https://drive.google.com/file/d/1yjy-59Juu9L3WHaOPu-XQo-j-HHGTbx_/view?usp=sharing

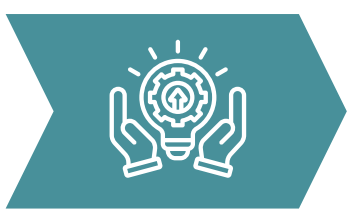

Fig. 4. Overview of Taxonomy Schema in REDAST Studies

### 4.1 Designing of Taxonomy Schema

Several studies have discussed the taxonomy categories in the RE and software testing domains in previous surveys. However, these existing schemas do not provide a comprehensive view of the REDAST process, especially for the detailed transformation process from requirements to test artifacts. In order to address the gap, we propose a hierarchical structure that encompasses (1) requirements input, (2) technical methodology, (3) test outcomes, and (4) results evaluation by referring to some related studies [62, 86]. The overview of the taxonomy schema in our literature review is illustrated in Fig. 4, where we not only exhibit the taxonomy schema in this figure but also indicate the REDAST procedure. We note that the presented taxonomy categories reflect the current state of the art in REDAST as identified in the selected literature. We acknowledge that new approaches and perspectives may emerge as research in this area advances, potentially requiring updates or extensions to these categories. Thus, our taxonomy should be viewed as a snapshot of current practices rather than a definitive or permanent classification.

### 4.2 Requirements Input Category

Requirements are the necessary input of the REDAST process. The formulation of requirements input will decisively affect the choices of the following steps, including processing technology, framework designing, and so on. In the Requirements input section (taxonomy in Table 4), we focus on the formulation of requirements based on the schema outlined in [46, 91], which corresponds to RQ1 about requirements. However, as we mentioned, there is no universal requirements categorization.

Thus, we adopted multiple categorizations in RQ1, including (1) **Requirements Type** (*Functional Requirements* and *Non-Functional Requirements*), (2) **Requirements Formats** (*Textual Specification*, *Model-based Specification*, *Formal Specification*, and *Tabular/Matrix Specification*), (3) **Requirements Notations** (*Natural Language Requirements*, *Unified Modeling Language*, *Controlled Natural Language*, *Requirements-Related Specification*, *Software Cost Reduction*, and *Domain-Specific Language*, (4) **Requirements Abstraction Model** (*Product Level*, *Feature Level*, *Function Level*, *Component Level*). (1) **Requirements Types** refer to the two main categories of widely adopted requirements in practice, i.e., functional and non-functional requirements [69]. There is a primary distinction between functional and non-functional requirements. Non-functional requirements typically necessitate additional transformation procedures to accommodate their abstract descriptions, which sets them apart from functional requirements. This categorization is crucial for discussing the differences between functional and non-functional requirements within the REDAST framework. Next, we introduce (2) **Requirements Formats**. The structural preferences of each format, which range from free-form prose to mathematically rigorous notations, directly influence the nature of parsing, intermediate representations, and oracle derivation needed during REDAST pipelines. Bruel et al. [18] present a widely recognized taxonomy for requirements representations: natural language, semi-formal, automata/graph-based, mathematical, and seamless (programming language-based) specifications. Our classification closely aligns with this framework: Textual Specification corresponds to NL representations, Model-based Specification captures automata/graph and semi-formal approaches (including graphical and state-based models), and Formal Specification maps to mathematical specifications that employ rigorous, logic-based notations, including constraint-based specification. Additionally, we introduce Tabular/Matrix Specification to account for the prevalence of tabular and matrix-based requirements artifacts (such as SCR or decision tables) in REDAST studies—a category that, while discussed in Bruel et al. as semi-formal or mathematical, is treated separately in our work to provide finer granularity. We excluded the seamless (programming-language-based) category, as observed in our primary study set and corroborated by Bruel et al. [18], such approaches are comparatively rare, particularly in REDAST process.

Table 4. RQ1 Taxonomy – Requirements Input in REDAST Studies

| RQ1 Taxonomy | Categorizations | Descriptions |
|---|---|---|
| Requirements Types | Functional Requirements | Define the functions or behaviors of a system or its components. |
| | Non-Functional Requirements | Specify criteria for evaluating system operation (e.g., performance, safety). |
| Requirements Formats | Textual Specification | Utilizes structured prose—such as “shall” statements, numbered sections, and templates—to articulate requirements [81, 106]. |
| | Model-based Specification | Embeds requirements within system models (e.g., UML, activity diagrams, or SysML diagrams) [61]. |
| | Formal Specification | Employs mathematically rigorous notations with well-defined or semi-defined syntax and semantics (e.g., Z Language, OCL) to describe system properties. |
| | Tabular/Matrix Specification | Organizes conditions and their corresponding actions in decision tables or matrices [23]. |
| Requirements Notations | Natural Language (NL) Requirements | Unrestricted prose statements describing system behavior or functionality, including use cases and user stories. |
| | Controlled Natural Language (CNL) | Restricted grammar and vocabulary to eliminate ambiguity and enable automated processing (e.g., “If `account_balance < 0`, then `freeze_account`.”). |
| | Unified Modeling Language (UML) | OMG-standard graphical language offering diagrams (class, state-machine, activity, sequence) for modeling system structure and behavior. |
| | Software Cost Reduction (SCR) | Tabular, state-based formal method relating inputs, state variables, and monitored/controlled variables by logical rules. |
| | Domain-Specific Language (DSL) | Maps requirements directly onto domain concepts, which may contain formal/mathematical equations, cause-effect graph, etc. |
| Requirements Abstraction [32] | Product Level | The most abstract level, which is unable to fit the normal definition of a requirement (e.g., testable and unambiguous) |
| | Feature Level | Explicitly but abstractly represent the features that the product supports, i.e., the requirements are an abstraction description of the software feature. |
| | Function Level | Describes what a user should be able to do, i.e., actions that are possible to perform. Note that non-functional requirements are also included in this level. |
| | Component Level | Details nature depicting information that is paraphrased or similar to how something should be solved. |

Another categorization presented in RQ1 is (3) **Requirements Notations**. Unlike requirements formats, which are broader and describe the structural form in which requirements are expressed (such as text, models, or tables), requirements notations refer to the specific syntactic or symbolic systems used to articulate requirements within those formats. Notations provide more concrete and detailed information by prescribing a particular language, set of symbols, or graphical elements for representing requirements. For example, UML is commonly used for model-based specifications. Besides, we introduce (4) **Requirements Abstraction** to investigate the requirements levels from abstract to concrete expression, based on the classification by Gorschek and Wohlin [32]. This schema is critical in REDAST because overly abstract (product/feature) requirements may cover a broad swath of behavior but lack the precision needed for direct test-artifact derivation, whereas overly concrete (component-level) requirements yield high testability but may fail to capture all relevant use cases. By tagging each requirement with its abstraction level, this study can explore transformation heuristics in REDAST framework, e.g., expanding product-level statements into finer-grained functional test templates or directly encoding component-level rules into oracles.

These schemas are expected to cover a broad range of requirements and provide a comprehensive view of the requirements input. Note that our study focuses specifically on requirements-related aspects, deliberately excluding papers focused on design-level information. For instance, Yang et al. [101] proposed an automated test scenario generation method using SysML for modeling system behavior in the system design phase. Although SysML is a commonly used specification format for requirements and system architecture, this paper was excluded from our literature review due to its emphasis on system design rather than system requirements. We also recognize that, in practice, requirements are sometimes expressed using hybrid formats that combine elements from multiple specification formats (e.g., a mix of textual and model-based representations, or formal constraints embedded within tabular formats), we did not introduce a separate "hybrid" category in our taxonomy. This decision was made to ensure clarity and comparability in our review, as hybridization is often context-specific, which is often not reported at the required level of detail in papers. Instead, for each work, we classified the requirements input based on the dominant or primary format explicitly described or utilized for automated test artifact generation. This approach maintains consistency in categorization and facilitates a clearer analysis of trends across the REDAST literature.

### 4.3 Transformation Techniques Category

Automated test generation from requirements fundamentally relies on *transformation techniques* that bridge the gap between high-level requirements and readable, understandable, and generation-friendly test artifacts. These techniques are not merely auxiliary steps; they constitute the backbone of REDAST, shaping how requirements—regardless of their type or representation—are interpreted, structured, and ultimately rendered into test cases. Depending on the usage scenario, various techniques are adopted in the transformation framework. Even though there are only limited studies explicitly discussing the transformation techniques for REDAST, we referred to the survey in related fields [6, 22, 39, 92], such as automated software testing and software generation, to finalize our schema for transformation techniques.

In RQ2 taxonomy, (1) **Transformation Procedure** refers to the stepwise process by which requirements are converted into test artifacts. (2) **Transformation Techniques** are the specific methodological approaches applied during the transformation. We additionally introduced (3) **Intermediate Representation** and (4) **Additional Input**, as several frameworks introduce intermediate artifacts to facilitate step-wise translation from requirements to test cases, which might require supplementary artifacts beyond the primary requirements to enable or enrich the transformation.

Table 5. RQ2 Taxonomy – Transformation Stages and Inputs in REDAST Studies

| RQ2 Taxonomy | Categorizations | Descriptions |
|---|---|---|
| Transformation Procedure | Parsing / Decomposition | Converts diverse requirement formats and notations into a uniform, machine-readable expression for subsequent augmentation and test generation. |
| | Augmentation / Transformation | Enhances or enriches the parsed expression—when parsing alone is insufficient—using techniques that improve semantic coverage and descriptiveness. |
| | Generation / Reconstruction | Reconstructs the transformed expressions into concrete test artifacts (e.g. via graph traversal or LLM generation), determining the final artifact form. |
| Transformation Techniques | Rule-based Method | Applies predefined templates, rules or meta-models to parse requirements, transform intermediate representations, and structure test artifacts. |
| | Tool-Supported Approach | Leverages external tools (e.g. NLP pipelines, Eclipse plugins) for parsing, processing or segmentation, enabling plugin-style adaptation of techniques. |
| | Graph-based Method | Uses graph structures (e.g. knowledge graphs, Petri nets) and traversal algorithms (e.g. BFS, DFS) to model complex system behaviors or conditions. |
| | ML-based Method | Employs machine-learning (including deep learning) for pattern/feature learning from large annotated corpora within the REDAST pipeline. |
| | Manual Intervention | Involves human expertise at points where full automation is not feasible, ensuring correctness or handling edge cases. |
| Intermediate Representation | Number of Representation | Denotes whether a single or multiple intermediate representations are used, reflecting architectural complexity. |
| | Structural/Formal Repre. | Template-based, rule-based or mathematical notations capturing requirement semantics. |
| | Model-based Representation | Meta-model driven depictions (e.g. UML, SysML) of requirements. |
| | Graph-based Representation | Node–edge structures such as diagrams or graphs representing conditions and flows. |
| | Test-Related Representation | Intermediate artifacts closely aligned to testing (e.g. test schemas) but not final test cases. |
| Additional Input | Historical Documents | Version-control logs, change logs, release notes, defect reports and issue-tracker data capturing system evolution. |
| | Requirements Documents | Raw stakeholder statements of needs and constraints. |
| | Source Code | Implementation artifacts from various SDLC stages. |
| | Supporting Documents | Supplementary design specs, API manuals, user guides and test plans. |
| | System Implementation | The deployed software (binaries, configs, runtime libraries and environment). |
| Framework Structure | Sequential | Follows a strict, ordered sequence of transformation steps, maintaining logical continuity without deviations. Each step builds upon the previous one. |
| | Conditional | Introduces decision points that enable alternative transformation paths based on specific conditions, increasing adaptability to varying requirements. |
| | Parallel | Allows simultaneous processing of different representations across multiple transformation units, significantly improving efficiency. |
| | Loop | Incorporates iterative cycles for continuous refinement, ensuring enhanced quality through repeated validation and adjustment. |

Finally, the (5) **Framework Structure** provides the overall logical arrangement of the transformation process and helps illustrate the logical connection between each step in the transformation process. Table 5 shows the taxonomy schema of RQ2, which includes (1) **Transformation Procedure** (*Parsing / Decomposition Stage*, *Augmentation / Transformation Stage*, and *Generation / Reconstruction Stage*), (2) **Transformation Techniques** (*Rule-based Method*, *Tool-Supported Approach*, **Graph-based Method**, and *Test-Related Representation*), (3) **Intermediate Representation** (*Number of Representation*, *Structural / Formal Representation*, *Model-based Representation*, *Graph-based Representation*, and *Test-Related Representation*), (4) **Additional Input** (*Historical Documents*, *Requirements Documents*, *Source Code*, *Supporting Documents*, *System Implementation*), and (5) **Framework Structure** (*Sequential*, *Conditional*, *Parallel*, and *Loop*). We note that the **Intermediate Representation** and **Additional Input** are optional components.

### 4.4 Test Artifacts Category

In the *Test Artifacts* section, we aim to focus on the formulation of generated testing artifacts. The illustration of RQ3 is shown in Table 6. The branches within the test artifacts category, such as test format, artifacts, stages, and coverage, have been explored in previous surveys [9, 36, 85, 92]. Comparing with RQ1 Taxonomy, RQ4 Taxonomy introduces similar categorization-(1) **Test Formats** to describe the preference of different test artifacts at the format level in selected studies. Besides, we introduce four testing stages for test artifacts, where (2) **Testing stages** categorization describes testing at different abstraction levels. (3) **Test Artifacts** further standardizes the terminology for the diverse outputs generated throughout the testing process, providing a unified framework for analyzing items such as test plans, test cases, test scenarios, test models, and test reports across the literature. In the original studies, the terminology used to describe test artifacts may have overlaps, which can potentially lead to confusion. To clarify, we have standardized these terms in our taxonomy. For example, while the term "test suite" is frequently employed in the literature to refer to a collection of test cases, we have opted to use "test case" to maintain consistency and avoid ambiguity in our categorization. Lastly, (4) **Coverage Methods** categorizes the different strategies or criteria used to assess the thoroughness of the generated test artifacts, e.g., requirements coverage, and model-based coverage, reflecting the evaluation dimension of test quality in the REDAST literature. (1) **Test Format** (*Structural Textual Format*, *Tabular Format*, *Model-based Format*, *Code-based Format*, *Logic-based Format*, and *Standardized/Specialized Format*, (2) **Test Stages**, (*Planning Stage*, *Designing Stage*, *Execution Stage*, and *Reporting Stage*), we aim to use this categorization to reflects the abstraction levels of the test artifacts, where the planning stage is at the most abstracted level, followed by designing stage, execution stage, and reporting stage, (3) **Test Artifacts** (*Test Plan*, *Test Case*, *Test Scenario*, *Test Vector*, *Test Model*, *Test Oracle*, *Test Script*, *Test Data/Bench*, *Test Report*, *Test Analysis*, and (4) **Coverage Methods** (*Requirements Coverage*, *Behavior/Scenario Coverage*, *Path Coverage*, *Function Coverage*, *Decision/Branch Coverage*, *Model-based Coverage*, *Statement Coverage*, *Rule-based Coverage*, and *Code Coverage*), which categorize the criteria and strategies used to measure the thoroughness and adequacy of test artifacts.

### 4.5 Results Demonstration Category

REDAST always introduces a case demonstration or a dataset evaluation to assess the quality of the generated test artifacts. The taxonomy of RQ4 is exhibited in Table 7. Here, we include evaluation as a separate category in the taxonomy schema to introduce the quality assessment and efficiency demonstration methods for the proposed pipelines and generated test artifacts in the selected study. Specifically, the taxonomy consists of (1) **Evaluation Types** (*Conceptual Case Demonstration*, *Real Case Demonstration*, *Dataset Evaluation*, *NA*), (2) **Usability** (*Methodology Explanation*, *Case Example*, *Experiment*, *Discussion*), and (3) **Demonstration Systems** (*Automotive System, Control System, Order System,*

Table 6. RQ3 Taxonomy – Test Artifacts in REDAST Studies

| RQ3 Taxonomy | Categorizations | Descriptions |
|---|---|---|
| Test Formats | Structural Textual Format | Free-form or template-based prose descriptions of test artifacts (e.g. natural language, structured templates). |
| | Tabular Format | Decision tables or matrices that organize test conditions and actions. |
| | Model-based Format | Meta-Model representations (e.g. state-transition models, test path diagram) of testing procedures. |
| | Code-based Format | Program or script code implementing test logic. |
| | Logic-based Format | Formal or mathematical expressions encoding test constraints and conditions. |
| | Standardized/Specialized Format | Tool-specific formats for dedicated testing frameworks or platforms. |
| Test Stages [92] | Planning Stage | Defines test objectives, scope, schedule and resources (e.g. test plan document). |
| | Designing Stage | Produces test cases, scenarios, vectors or models aligned to the plan. |
| | Execution Stage | Runs test scripts or cases against the system under test, generating execution data. |
| | Reporting Stage | Summarizes outcomes (pass/fail, defects, coverage) in reports and analyses. |
| Test Artifacts | Test Plan [20] | Document detailing objectives, scope, schedule and resources for a testing effort. |
| | Test Case [4] | Defined inputs, execution preconditions and expected outcomes for a test. |
| | Test Scenario [57] | High-level narrative of a functionality or workflow to be tested. |
| | Test Vector [2] | Specific set of input values or parameters used during test execution. |
| | Test Model [14] | Abstract blueprint (e.g. state machine) used to derive families of test cases. |
| | Test Oracle [75] | Mechanism that determines pass/fail criteria for each test. |
| | Test Script [103] | Executable sequence of steps and data inputs, usually in code. |
| | Test Case Prioritization | Ranking or the sequence of execution of test cases. |
| | Test Data/Bench | Data sets or fixtures used to drive test executions. |
| | Test Report | Document summarizing execution results, coverage metrics and defects. |
| | Test Analysis | Qualitative and quantitative evaluation of test outcomes and artifacts. |
| Coverage Methods [5, 77, 82] | Requirements Coverage | Percentage of requirements with at least one executed test case. |
| | Behavior/Scenario Coverage | Percentage of distinct behavioral scenarios exercised by tests. |
| | Path Coverage | Percentage of unique control-flow paths executed. |
| | Function Coverage | Percentage of functions invoked at least once. |
| | Decision/Branch Coverage | Percentage of decision points (if/loop branches) visited. |
| | Model-based Coverage | Coverage of elements in formal models (states, transitions). |
| | Statement Coverage | Percentage of executable statements executed. |
| | Rule-based Coverage | Percentage of business or domain rules exercised. |
| | Code Coverage | General umbrella for any metric measured on source code. |

*Safety System, etc.*). RQ4 aims to evaluate the methods used in the selected studies. Additionally, we have introduced a new category called **Usability** to address how various sections of the selected papers perform in terms of usability. We recognize that, due to the research standards of the time, some papers may not formally present their analyses and methodologies. Therefore, this categorization helps us identify more user-friendly studies for readers.

### 4.6 Future and Limitation Category

Note that even RQ5 is not implicitly shown in our framework in Fig. 1. However, to maintain consistency with other taxonomy sections, this section provides a high-level categorization of the limitations and future directions explicitly mentioned in the primary studies, serving as a precursor to the deeper analysis presented in later sections.. We define the key facets of RQ5 as (1) **Limitation** (*Limitation of framework design*, *Limitation of evaluation or demonstration*,

Table 7. RQ4 Taxonomy – Results Demonstration in REDAST Studies

| RQ4 Taxonomy | Categorizations | Descriptions |
| --- | --- | --- |
| Evaluation Types | Conceptual Case Demonstration | Conceptually constructs a software system for reflecting the framework procedure and execution logic. |
| | Real Case Demonstration | Leverages a real project or system from industry to demonstrate the efficiency from a practical perspective. |
| | Dataset Evaluation | Instead of the project- or system-level demonstration, dataset evaluation adopts an existing dataset or benchmark to show quantitative results of the method. |
| | NA | No Demonstration methods included. |
| Usability | Methodology Explanation | The methodology section is expected to describe the framework and pipeline in a detailed and narrative manner. |
| | Case Example | With the demonstration of a case example, readers can better understand the working logic and the exact representation in each step. |
| | Experiment | Experiment is the key part in demonstrating usability. We acknowledge that, owing to the research standards of the time, some papers were obliged to rely on experimental evidence that is not sufficiently persuasive for evaluating their methods. |
| | Discussion | The Discussion section is the post-evaluation extension part, where the results from the experiment will be further explored. The interpretation helps readers to understand the key issues in the studies better. |
| Demonstration Systems | Automotive Systems | Vehicle control and ADAS software. |
| | Control System | Generic industrial feedback loops. |
| | Order System | End-to-end retail order processing. |
| | Safety System | Hazard detection and emergency shutdown. |
| | ATM System | ATM authentication and cash dispensing. |
| | Business System | Enterprise CRM/ERP functions. |
| | Resource Management System | Tracks and allocates shared resources. |
| | Workforce System | Employee records and shift scheduling. |
| | Healthcare System | E-health records and clinical workflows. |
| | Aerospace System | Avionics or satellite control software. |
| | Authentication System | Centralized identity and SSO. |
| | Library System | Catalog and lending operations. |
| | Mobile System | Typical smartphone-app architecture. |
| | Banking System | Core banking transactions and compliance. |
| | Education System | Learning management and course delivery. |
| | Database System | Internal DBMS engine components. |
| | Examination System | Online testing and auto-grading. |

*Scalabillity to larger systems or complex usage scenarios*, *Over-relying on input quality*, *Over complexity of the methodology*, *Lack of automation*, *Requirements ambiguities*, *Limitation of implementation*, *Highly time-costing*, and *Additional cost of requirements specification*), (2) **Automation Level** (*Fully Automated*, *Highly Automated*, *Semi-Automated*, and *Low-Automated*), and (3) **Future Direction** (*Extension to other coverage criteria or requirements*, *Future validation*, *Completeness improvement*, *Extension of other techniques*, *Extension to other domains or systems*, *Extension to other phases or test patterns*, *Performance improvement*, *Real tool development*, *Benchmark construction*, *Traceability improvement*, *Robustness Improvement*, and *NA*).

Table 8. RQ5 – Future and Limitations in REDAST Studies

| RQ5 Taxonomy | Categorizations | Descriptions |
|---|---|---|
| Limitation | Framework-Design Limitation | Shortcomings rooted in the framework design (e.g., rigid pipelines, limited component modularity). |
| | Evaluation/Demonstration Limitation | Insufficient empirical evidence (e.g., toy example, small dataset). |
| | Scalability Limitation | Inability of the framework to maintain effectiveness in large-scale systems or complex usage scenarios. |
| | Incomplete Requirements Coverage | Inability to cover some types or formats of requirements. |
| | Input-Quality Over-Dependence | performance degrades significantly with ambiguous, incomplete, or inconsistent inputs. |
| | Methodology Over-Complexity | Excessive algorithmic or tooling complexity (e.g., intricate preprocessing chains, heavyweight models). |
| | Automation Gaps | Residual manual effort required at one or more stages, e.g., heuristics in methodology, scripting, environment setup in test environment, and re-specifying requirements. |
| | Requirements Ambiguity | Ambiguities in requirements specification. |
| | Implementation Limitation | The proposed framework is difficult to implement. |
| | Time Cost | High processing, training, or annotation time. |
| | Additional Specification Cost | Extra effort or tooling needed for requirements reformatting. |
| Automation | Fully Automated | (End-to-End) Manual operation is not needed in the framework. |
| | Highly Automated | (Automation-Dominant) Manual operation is not necessary. |
| | Semi-Automated | (Automation-Supported) Manual operations are necessary. |
| | Low Automated | (Minimal Automation) Manual operations are primarily relied on. |
| Future Direction | Coverage/Requirement Extension | Plans to broaden the technique to additional requirement types or coverage methods. |
| | Further Validation | Intent to extend the evaluation and experiment to more contexts. |
| | Completeness Improvement | Enhancing the solution to make it span more REDAST stages. |
| | Extension of Other Techniques | Incorporating alternative or advanced methods. |
| | Extension to Other Domains or Systems | Applying the approach to different industrial sectors. |
| | Extension to Other Phases or Test Patterns | Adapting the method for additional testing phases. |
| | Performance Improvement | Optimising runtime, memory, scalability, or accuracy. |
| | Real Tool Development | Packaging research prototypes into production-ready tooling. |
| | Benchmark Construction | Creating/publishing standard datasets, or evaluation frameworks. |
| | Traceability Improvement | Strengthening bidirectional links among requirements and generated test artifacts. |
| | Robustness Improvement | Increasing resilience to noisy, ambiguous, or evolving requirements and to environmental changes. |

## 5 Results

In this section, we present the results of the SLR on requirements-driven automated software testing, where the results are structured based on our RQs.

### 5.1 Trend: General Results of the Publications

Before addressing the research questions, we present the publication trends of the 156 primary studies on REDAST, including study distribution, venue names, and features of the selected studies.

The study distribution is analyzed in two parts: (a) study distribution by the publication year and (b) study distribution by publication type, which are illustrated in Figures 5a and 5b, respectively.

Table 9. Publication Venues with Two or More Studies in Selected Papers (Trend)

| Venue Names | Type | Num. |
|---|---|---|
| IEEE International Requirements Engineering Conference (RE) | Conference | 6 |
| IEEE International Conference on Software Quality, Reliability, and Security (QRS) | Conference | 6 |
| IEEE International Conference on Software Testing, Verification, and Validation (ICST) Workshops | Workshop | 5 |
| IEEE Transactions on Software Engineering (TSE) | Journal | 4 |
| Software Quality Journal (SQJ) | Journal | 3 |
| Science of Computer Programming | Journal | 3 |
| IEEE International Conference on Software Testing, Verification and Validation (ICST) | Conference | 3 |
| International Conference on Quality Software (QSIC) | Conference | 3 |
| International Conference on Evaluation of Novel Approaches to Software Engineering (ENASE) | Conference | 3 |
| Innovations in Systems and Software Engineering | Journal | 3 |
| IEEE International Workshop on Requirements Engineering and Testing (RET) | Workshop | 3 |
| ACM SIGSOFT Software Engineering Notes | Journal | 3 |
| Journal of Systems and Software (JSS) | Journal | 2 |
| IEEE International Symposium on Software Reliability Engineering (ISSRE) | Conference | 2 |
| IEEE International Requirements Engineering Conference Workshops (REW) | Workshop | 2 |
| International Journal of System Assurance Engineering and Management | Journal | 2 |
| International Conference on Enterprise Information Systems (ICEIS) | Conference | 2 |
| International Conference on Emerging Trends in Engineering and Technology (ICETET) | Conference | 2 |
| Electronic Notes in Theoretical Computer Science | Journal | 2 |
| Australian Software Engineering Conference (ASWEC) | Conference | 2 |
| International Journal of Advanced Computer Science and Applications | Journal | 2 |
| Electronics | Journal | 2 |

In analyzing the distribution of REDAST studies by year, we observed that the first study was published in 1993, followed by a steady annual increase. A notable surge occurred around 2008, prompting further investigation into the technical differences between studies conducted before and after this period. Two key conclusions emerged: (1) An increased adoption of NLP-pipeline-based methods from 2008 onward. Prior to 2008, only 20% of studies employed these methods, whereas from 2008 to 2013, the proportion rose to 31.9%. (2) A decline in the use of graph-based methods after 2008. The prevalence of graph-based approaches decreased from 35% before 2008 to 19.14% in the subsequent period from 2008 to 2013. By comparing with the landmark studies published between 2006 and 2008, including works by Hinton et al., [38] and Van der Maaten and Hinton[89], we believe that deep learning technologies, including LLMs and Convolutional Neural Networks (CNNs), can promote the adoption of NLP-pipeline-based methods in SE domain.

Regarding the publication type distribution, most studies were published in conferences (56.41%) and journals (32.69%). Our selected studies were published across 115 different venues. We present the venues with two or more published studies in Table 9. The IEEE International Requirements Engineering Conference, the IEEE International Conference on Software Quality, Reliability, and Security, and the IEEE Transactions on Software Engineering have the highest percentages in their respective categories, which is unsurprising given their top-level reputation in SE.

Besides the trending information, we investigated the target software in the selected studies. The results of the target software are exhibited in Table 10. Most of the selected studies are designed for general software (100/156), followed by embedded systems (14/156), web services systems (7/156), safety-critical systems (6/156), and so on.

Table 10. Overview of Target Software in Selected Papers (Trend)

| Target Software Systems | Paper ID | Num. |
|---|---|---|
| General Software | [2, 3, 5–10, 13–19, 23–25, 30, 31, 34–37, 39–43, 46, 51–55, 57, 62, 63, 65, 68–71, 73–76, 78, 79, 81–86, 89, 90, 93, 95, 96, 99–102, 105, 110, 112–117, 119–125, 127, 130–134, 136–141, 143, 145–148, 152–155] | 100 |
| Embedded System | [29, 33, 38, 60, 61, 91, 94, 107, 109, 135, 142, 144, 150, 156] | 14 |
| Web Services System | [1, 4, 11, 21, 56, 67, 108] | 7 |
| Safety-Critical System | [27, 50, 72, 97, 118, 126] | 6 |
| Timed Data-flow Reactive System | [80, 104, 128, 129] | 4 |
| Reactive System | [45, 103, 151] | 3 |
| Real-time Embedded System | [26, 64, 111] | 3 |
| Product Line System | [32, 48] | 2 |
| Object-Oriented System | [28, 66] | 2 |
| Telecommunication Application | [22, 92] | 2 |
| Automotive System | [59, 98] | 2 |
| Space Application | [44] | 1 |
| SOA-based System | [87] | 1 |
| Labeled Transition System | [49] | 1 |
| Healthcare Application | [106] | 1 |
| Event-driven System | [149] | 1 |
| Cyber-Physical System | [58] | 1 |
| Core Business System | [20] | 1 |
| Concurrent System | [12] | 1 |
| Complex Dynamic System | [88] | 1 |
| Aspect-Oriented Software | [47] | 1 |
| Agricultural Software | [77] | 1 |

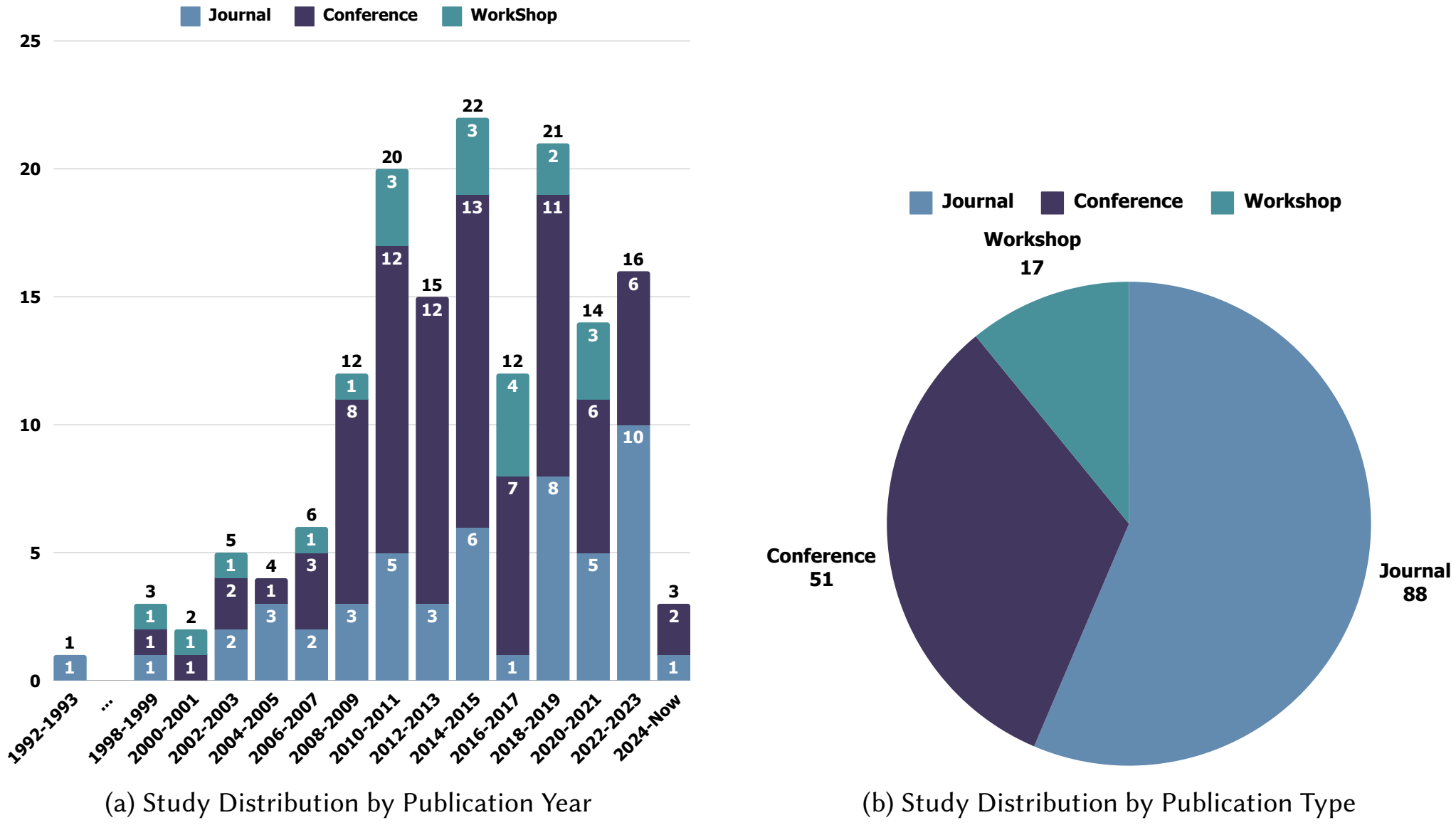

(a) Study Distribution by Publication Year

(b) Study Distribution by Publication Type

Fig. 5. Study Distribution (Trend)

Table 11. Requirements Abstraction Levels in Selected Studies (RQ1)

| Abstraction Level | Paper IDs | Num. |
|---|---|---|
| Function (3rd Abstract) | [1, 2, 4–13, 15–33, 35–39, 41, 43–47, 49–54, 56–58, 60–66, 68–70, 72, 78–81, 88, 91, 92, 94, 96, 97, 100, 102–104, 106, 109, 112, 114, 117, 119, 120, 122, 123, 125–131, 133, 134, 136–139, 142, 145–148, 150–152, 155, 156] | 108 |
| Feature (2nd Abstract) | [11, 14, 20, 30, 34, 39, 41, 46, 48, 55–57, 59, 62, 67, 68, 70, 71, 73–77, 82–85, 87, 89, 90, 93–95, 98–101, 105, 106, 108, 110, 113, 115, 116, 118, 121, 132, 135, 140, 141, 143, 153–155] | 53 |
| Component (4th Abstract) | [3, 14, 40, 42, 49, 50, 61, 65, 66, 69, 86, 98, 107, 111, 124, 144, 149–151] | 19 |
| Product (1st Abstract) | [48, 55, 59, 68] | 4 |

### 5.2 RQ1: Requirements Specification Formulation

Based on our taxonomy schema in Section 4, we explore the techniques for requirements type, format, and notation in the selected studies. Moreover, by comparing the characteristics of the studies, we analyze the features in different adoptions of specification techniques.

*5.2.1 Requirements Types.* Almost all of the REDAST studies cover functional requirements. There are also studies that cover non-functional requirements (NFRs) in addition to functional requirements. The full list of the studies supporting NFRs is shown in Table 44 (Appendix). The table summarizes how selected studies address NFRs in REDAST process. It highlights the modeling, transformation, and testing strategies for key NFRs such as safety, reliability, security, performance, and usability across different approaches. We cover the cross-analysis of requirements types and notations, and target systems in later sub-sections.

*5.2.2 Requirements Abstraction Level.* Gorschek and Wohlin [32] classify software requirements into product, feature, function, and component levels, which describe the requirements abstraction levels from abstract to concrete. The abstraction level results of the selected studies are exhibited in Table 11, where function-level (108 studies) is the major abstraction level adopted in REDAST studies, followed by feature-level (53 studies), component-level (19 studies), and product-level (4 studies). Note that some studies crossly cover two abstraction levels. For example, P56 [56] respectively cover feature and function levels in tabular requirements specification and UML state-machine, where the tabular requirements specification and scenario models are derived from the spec and attached to UML state-machine and sequence diagrams. Specifically, we discuss each abstraction level in detail as follows:

- *Function-Level* requirements correspond to the third abstraction layer in the defined abstraction levels and encompass what are commonly called system requirements—both functional and non-functional. From a REDAST perspective, function-level requirements hit the sweet spot between abstraction and detail: they are concrete enough to drive software design and validation, yet sufficiently general to ensure broad test coverage. For instance, in P72 [72], system constraints are encoded in OCL (UML), with OCL expressions capturing both functional and safety predicates, resulting in abstract sequences, and transformed into test cases giving high coverage. Similarly, P81 [81] demonstrates automated recovery of functional requirements by mining trace links among requirements, source code, and then transforming to multi-objective optimization orders test cases via latent semantic indexing.
- *Feature-Level* requirements occupy the second abstraction layer in our abstraction schema. They describe the high-level capabilities or features the system must support, without prescribing the specific functions needed to realize them. While this abstraction helps define system scope and supports better test coverage, feature-level descriptions lack the concrete detail required for directly generating test artifacts. Consequently, REDAST studies at this level typically apply synthesis techniques to refine abstract feature specifications into executable tests. For example, P105 [105] formalizes requirements as scenarios and dependency diagrams, then synthesizes test cases by traversing all scenario paths and their dependencies. Similarly, P118 [118] models security misuse and use cases, expands their mitigation flows into explicit pre- and post-conditions, and finally converts these conditions into concrete security test steps.

- *Component-Level* requirements are a detailed narrative of how the system should realize the functions. The Component-Level can also act as the atomic breakdown of the Function-Level requirements, which provides more details and sets limits to the functional description. These requirements can be easily converted into testable objectives, but as for generating test artefacts, they should be organized and integrated to give a comprehensive description of the system. The typical requirements specification at this level is a formal assertion or equations. In P124 [124], functional requirements are rewritten as low-level assume/assert scenarios embedded in C monitor code for bounded-model-checking workflows, which act as the test scripts proving each requirement holds. P86 [86] employs a multi-stage transformation, including acquiring, decomposition, combining, and verification, to reconstruct UML to a Petri-net for testing generation.
- *Product-Level* requirements represent the highest, most abstract layer in our abstraction schema. Only four studies in our SLR address requirements at this level, and each of those also incorporates feature- or function-level detail, highlighting that purely product-level statements rarely suffice for REDAST purposes. By design, product-level requirements articulate broad goals or strategic directions rather than concrete behaviors, making them ill-suited for directly generating test artifacts. This might also be true for other forms of high abstraction requirements, e.g., user requirements. Despite being central to the development and testing process, higher abstraction level requirements are not addressed in the existing literature.

*5.2.3 Requirements Specification Format.* Requirements specification is categorized into six different types: textual, model-based, constraint-based, formal (mathematical), tabular (matrix-based), and other specifications, which the detailed definitions of these categories are specified in Section 4.2. The specification formats adopted in the selected studies are presented in Table 12. Note that we found some studies that use the transformation method further to convert raw requirement input into other requirement formats. Here, we only consider the first raw requirement input in this section. For example, P76 [76] adopts model-based specification as the raw input for requirements. However, the requirements are further converted to model-based requirements. To clarify our objective, we only consider the requirements of the first input in RQ1. The intermediate representations are discussed in RQ2.

- *Textual Specification.* 105 studies adopted textual specification methods. Our analysis shows that textual specification is the most commonly used format in REDAST studies. Textual specification, written in natural language (NL), is the predominant and common choice in most of the requirements specifications. NL-based requirements are generally favored in REDAST studies due to their accessibility and ease of understanding, which supports both requirements description and parsing NL is commonly used in these studies due to its advanced explainability and flexibility. Due to its expressiveness, the textual specification acts as a medium of communication between users and developers during the requirement elicitation process [26]. Besides the textual specification itself, NL is widely integrated into other specification formats, including formal and tabular requirements. We separately discuss this category by distinguishing textual specifications from others and whether the requirements follow natural language logic [55, 79]. Other formats especially involve specific specification rules or templates compared to textual specifications.

Table 12. Requirements Format in Selected Studies (RQ1)

| Requirements Formats | Paper IDs | Num. |
|---|---|---|
| Textual | [2, 5–19, 22–26, 28, 31, 32, 34, 36–39, 41–48, 51, 52, 54, 55, 57, 58, 60–65, 67, 68, 70, 71, 75–77, 80–85, 89, 90, 93–95, 98, 99, 102, 104–106, 108, 110, 112, 113, 115–118, 120–125, 127–135, 140, 142, 144, 146–148, 151, 152, 155] | 105 |
| Model-Based | [1, 4, 7, 8, 10, 11, 14–17, 20, 21, 25, 27, 28, 30, 33, 35, 46, 49, 50, 53, 56, 57, 65, 66, 69, 73–75, 78, 84, 86, 87, 92, 93, 95, 100, 105, 106, 111, 115, 118, 126, 132, 138–140, 143, 145, 147–149, 156] | 54 |
| (Semi-) Formal | [3, 21, 27, 33, 45, 58, 59, 72, 78, 79, 88, 96, 97, 101, 106, 107, 109, 119, 149, 150] | 20 |
| Tabular (Matrix-Based) | [29, 33, 78, 89, 103, 149, 154] | 7 |
| Other | [40, 91, 114, 136, 137, 141, 153] | 7 |

- *Model-based Specification.* 54 studies specify requirements using model-based specifications. Comparing with textual requirements, model-based approaches, constructed in meta-models to represent and analyze requirements, have better abstraction capabilities for illustrating the behaviors (e.g., P1 [1], P27 [27], P35 [35], etc.), activities (e.g., P25 [25], P139 [139], P146 [146], etc.), etc., of a software system [53]. Model-based requirements emerged as the second most prevalent specification format in our review, largely because they facilitate professional communication between domain experts and software engineers by translating high-level needs into technical artifacts[67]. Furthermore, model-based specifications inherently accommodate other requirement formats—such as textual statements or constraint rules—by incorporating them as attributes within the overarching model, thereby enhancing the inclusivity and integration of diverse specification styles.
- *Formal Specification.* In the selected studies, we identified 20 papers that used formal or semi-formal specifications. The formal specification aims to translate system properties as limits, conditions, or relationships that must hold within the system into a precise and unambiguous specification [64, 74]. Typical formal requirements specification are assertions (e.g., P109 [109]), controlled language (e.g., P79 [79], P108 [108]), and so on. As for semi-formal specifications, DSL is a typical constraint-based specification, e.g., P88 [88], P95 [95], and P96 [96]. Besides the advantage of the precise expression, the mathematical equation-based requirements also bring the advantage of error or bug detection in the early requirements elicitation process, the translated formal expression will help one to identify ambiguities, omissions, etc. [21, 90].
- *Tabular (Matrix-based) Specification.* We identified 7 studies that adopted tabular specifications. Tabular specification methods can formalize requirements in a structured and organized manner, where each row represents a requirement, and each column represents a specific attribute or aspect of the requirement [74]. Tabular specification can greatly improve traceability and make it more friendly for verification and validation.

In Fig. 6, we analyze the trend of requirements format preferences over time, as illustrated in Fig. 6. Specifically, before 2008, although textual requirements were already widely employed in the REDAST methodology, their proportion did not significantly dominate among the six requirements formats. This observation aligns with the study distribution discussed in Section 5.1. After 2008, textual requirements gradually became the preferred choice for requirements specification. Furthermore, as indicated by the increasing trend in the "other" requirements format, we observed a growing adoption of diverse requirements formats in recent years. This trend suggests an increasing diversification in requirements selection over time.

*5.2.4 Requirements Specification Notation.* The requirements notation provides an extended detail of the requirements format. We identified over 60 different requirement notations across the selected studies. Some variations of standard requirement notations were categorized into root forms or similar types. A summary of these notations is presented in Table 13. Our findings indicate that UML and natural language (NL) requirements specifications are the most frequently used notations, appearing in 50 and 49 studies, respectively. They are followed by Domain-Specific Languages (DSLs) in 32 studies, Controlled Natural Languages (CNLs) in 23 studies, and Software Change Request (SCR) in 3 studies. This distribution aligns with the trends observed in the requirements format. However, while textual requirements specifications dominate the overall requirements format distribution, NL requirements do not represent the sole majority in the notation distribution. This suggests that, although NL specifications are widely adopted in requirements notations, pure NL requirements without constraints or structure are not the only prevalent form.

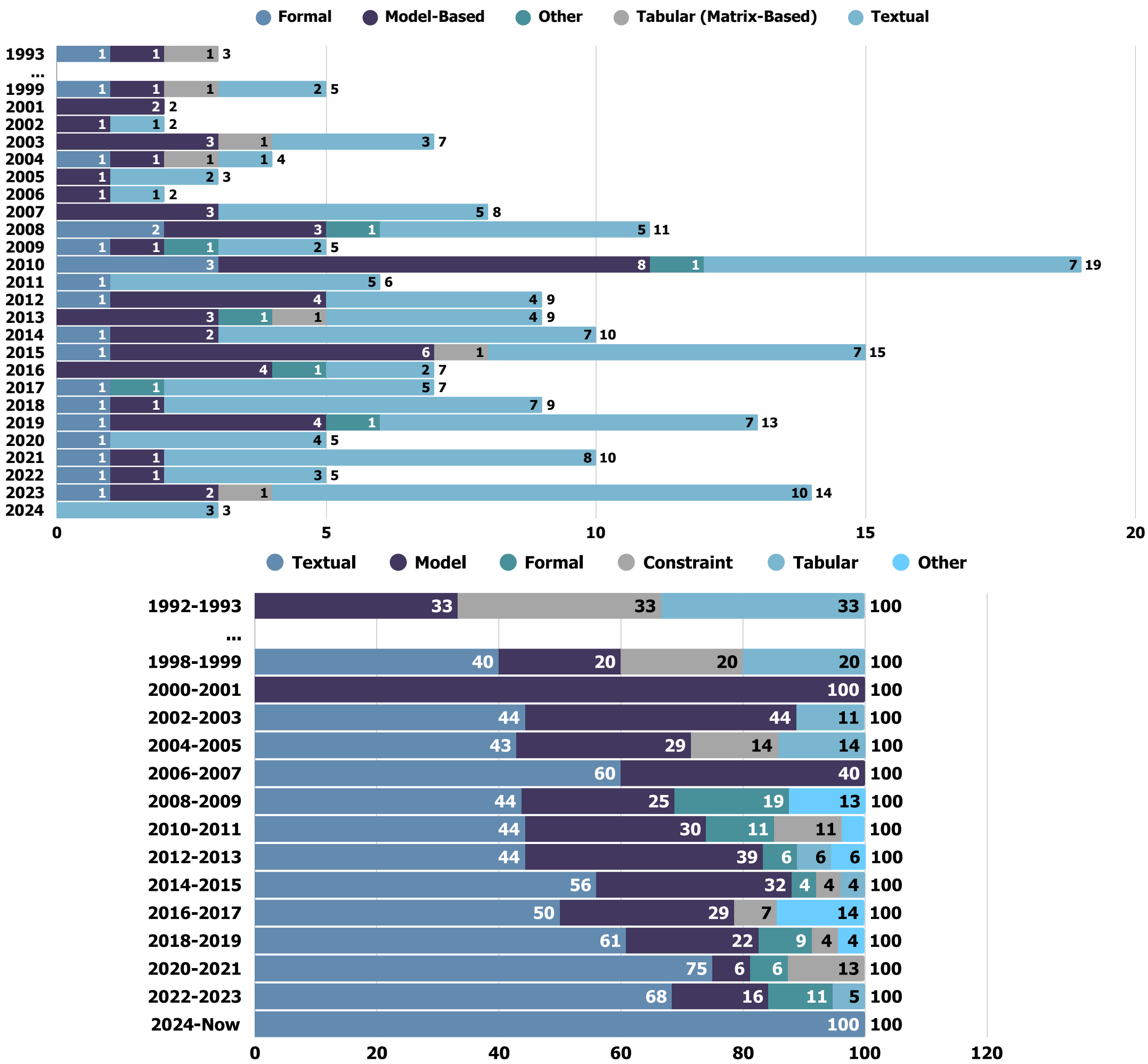

Fig. 6. Trend of Requirements Input by Years (Up: Quantities; Down:Percentage)

Table 13. Requirements Notation Results of Selected Studies (RQ1)

| Requirements Notations | Paper IDs | Num. |
|---|---|---|
| UML | [4, 7, 8, 10, 11, 15–17, 35, 46, 49, 53, 56, 57, 66, 73–75, 86, 87, 93, 95, 106, 111, 115, 118] (UML), [13, 19, 36, 43, 48, 70, 76] (Use Case Description Model), [14] (Interaction Overview Diagram), [25, 140, 147] (Activity Diagram), [28, 105, 148] (Sequence Diagram), [30] (Extended SysML), [50] (Safety SysML State Machine), [59] (OCL-Combined AD), [62] (State-based Use Case) [65] (Class Diagram), [72, 101] (OCL), [126] (Functional Diagram), [145] (Statechart Diagram), [156] (Modeling and Analysis of Real-Time and Embedded Systems) | 50 |
| NL Requirements | [23] (PURE dataset), [26, 34] (Textual User Story), [37, 39, 77, 89] (Usage Scenario), [38, 42, 44, 54, 55, 81, 83, 90, 94, 98–100, 102, 109, 112, 113, 117, 121–125, 130, 131, 133–135, 144, 146, 155] (NL Requirements), [40] (Test requirements), [41, 52, 60, 85, 120] (Textual Use Case), [63, 67, 71, 116] (Template-based), [142] (Technical Requirements Specification), [152] (Positive and negative pair) | 49 |
| DSLs | [1] (Custom Metamodel), [20, 92] (Cause-Effect Graph), [21] (OWL-S Model), [27] (SCADE Specification), [29] (State Transition Table), [31, 51, 107, 138] (Requirements-Specification Language), [47] (Aspect-Oriented PetriNet), [64] (Structured Requirements Specification), [69, 110, 132] (Requirements Modeling Language), [79, 119, 150] (Formal Equation), [84] (Contract Language for Functional PF Requirements), [88, 96, 97] (DSL), [91] (Safety Requirements), [103] (Expressive Decision Table), [114] (Customer-assigned priorities, Developer-assigned priorities), [136] (Customer Assigned Priority), [137] (Requirements Dependency Mapping), [139] (Behavior Tree), [141] (Stakeholder Priority), [143] (Domain-Specific Modeling Language) [153] (Risk Factor), [154] (Requirement Traceability Matrix) | 32 |
| CNLs | [2, 9, 45, 80, 104, 108, 127–129, 151] (CNL), [3] (Constraint-based Requirements Specification), [5, 22, 24, 32, 61, 82] (Restricted Use Case Modeling Language), [6] (Use Case Specification Language), [12, 68] (Semi-Formal Requirements Description), [18] (Semi-Structured NL), [58] (Formal NL Specification), [65] (Restricted-form of NL) | 23 |
| SCR | [33, 78, 149] | 3 |

*5.2.5 RQ1 Cross-Analysis between Requirements Formats and Notations.* In Table 12 and 13, we noticed that the trends in these two tables are almost aligned, where the textual & model-based and NL requirements & UML are the top-2 welcomed requirements formats and notations. NL specifications support a variety of processing techniques—such as condition detection (e.g., P23 [23]) and semantic analysis (e.g., P63 [63]). However, inherent inconsistency and incompleteness in NL requirements [80] hinders automated parsing and transformation tools. UML is a broad family of (graphical) notations that offers strong expressiveness, flexibility, and unambiguity,though it ranks relatively low in understandability and greater manual effort, as compared to NL requirements. We identified 43 studies employing UML artifacts (e.g., use-case and activity diagrams), which facilitate traceability, mapping, and decomposition of functional requirements [37, 59]. For instance, P56 [56] introduces a tabular-based UML extension for enhanced traceability, while P17 [17] uses use-case diagrams to depict requirement scenarios. As for CNL, OCL, etc., these notations appear in REDAST approaches at lower proportions. They combine semantic expressions with predefined rules or templates, enabling strong expressiveness, flexibility, and understandability while improving unambiguity and verification/traceability (e.g., P2 [2]). Finally, formal notations, RSL, and DSL are renowned for strict logical rigor, yielding the highest unambiguity and verification/traceability but scoring lowest on understandability, particularly suited to safety-critical systems.

*5.2.6 RQ1 Cross-Analysis between Requirements Format and Target Software Type.* We present the cross results of the target software and the requirements input in Table 14. The needs in different target software reflect the corresponding features of *textual requirements dominate across all categories.* This trend suggests that, due to the flexibility and simplicity of textual requirements, textual requirements can handle most usage scenarios in REDAST. *Model-based requirements are preferred for structured systems. Formal requirements are crucial for high-reliability domains*, wherein formal requirements can additionally satisfy the needs of strict verification and validations, e.g., for formal requirements, Safety-Critical Systems (P72 [72]), and Automotive Systems (P59 [59]), for semi-formal requirements, Safety-Critical Systems (P27 [27], P72 [72], P97 [97]), and Event-driven Systems (P149 [149]).

Table 14. Joint Distribution of Requirements Format and Target Software

| Target Software | Textual | Model-based | Formal | Other | Tabular (Matrix) |
|---|---|---|---|---|---|
| **General Software** | 75 | 34 | 6 | 3 | 6 |
| **Embedded System** | 7 | 2 | 4 | 2 | 1 |
| **Web Services** | 3 | 5 | 1 | 0 | 0 |
| **Safety-Critical** | 1 | 4 | 3 | 0 | 0 |
| **Timed Data-flow** | 4 | 0 | 0 | 0 | 0 |
| **Reactive** | 2 | 0 | 1 | 1 | 0 |
| **Real-time Embedded** | 2 | 1 | 0 | 0 | 0 |
| **Product Line** | 2 | 0 | 0 | 0 | 0 |
| **Object-Oriented** | 1 | 2 | 0 | 0 | 0 |
| **Telecom** | 1 | 1 | 0 | 0 | 0 |
| **Automotive** | 1 | 0 | 1 | 0 | 0 |

Table 15. Joint Distribution of Requirements Types and Target Systems and Notations

| Req. Type | Embedded | General | Product Line | Real-time Embedded | SOA-based | Safety-Critical | Web Services |
|---|---|---|---|---|---|---|---|
| **NFR** | 2 | 12 | 1 | 1 | 1 | 2 | 2 |

| Requirements Type | CNLs | DSLs | NL Req. | UML |
|---|---|---|---|---|
| **NFR** | 3 | 8 | 3 | 7 |

*5.2.7 RQ1 Cross-Analysis between Requirements Type and Target Systems & Requirements Notations.* In this section, we analyze the requirements types with target systems and requirements notations. Note that we only list the categories in target systems that cover NFRs, as FRs are covered across all selected papers. The distribution of NFRs across target systems and notations (Tab. 15) shows that NFRs remain a minority overall but are relatively more prominent in safety-critical and real-time embedded contexts. This pattern underscores that when software quality attributes such as performance, reliability, or security are mission-critical, practitioners are more inclined to model and automate NFRs, even though the absolute numbers remain small. When we examine how these NFRs are notated, DSLs and UML seem most popular. This suggests that NFRs still demand structured, semi-formal representations to be machine-actionable, limiting their adoption. To change this, future work should focus on lowering the barrier to NFR specification: for example, by defining reusable DSL fragments for common quality attributes, embedding constraint annotations directly in UML models, or devising lightweight CNL templates that guide authors toward parameterized, testable NFR statements.

*5.2.8 RQ1 Cross-Analysis between Requirements Abstraction Level and Requirements Notations.* This section examines the relationship between the abstraction levels of requirements and their corresponding notations. The abstraction level indicates the granularity of the requirements, and we aim to evaluate the representational capabilities of various specification notations. Regardless of the number of studies considered, all noted requirements exhibit strong general representational abilities across different abstraction levels. DSLs, CNLs, and SCRs, despite representing a smaller proportion, account for $\sim 40\%$ of function-level requirements. However, at the feature level, denoting a more abstract specification, this proportion significantly decreases to $\sim 20\%$. This suggests that DSLs, CNLs, and SCRs are less effective in representing abstract requirements. In conclusion, UML and NL requirements demonstrate robust generalized representation capabilities across the four different abstraction levels, while DSLs and CNLs are more suited for concrete requirements.

Table 16. Joint Distribution of Requirements Abstraction Level and Requirements Notation

| Abstraction Level | UML | NL Requirements | DSLs | CNLs | SCR |
|---|---|---|---|---|---|
| **Function (3rd Abstract)** | 33 | 30 | 24 | 20 | 2 |
| **Feature (2nd Abstract)** | 23 | 20 | 8 | 3 | 0 |
| **Component (4th Abstract)** | 7 | 5 | 3 | 4 | 1 |
| **Product (1st Abstract)** | 2 | 1 | 0 | 1 | 0 |

**RQ1 Key Takeaways**

• NL specifications remain lingua franca, but mainly in general-purpose business software. Model-based formats (e.g., UML diagrams) are also prevalent in specification for REDAST process. Formal specifications are also often used in safety-critical and automotive projects.
• Most REDAST studies focus on FRs, but NFRs—such as safety, reliability, security, and performance—receive attention only in safety-critical and real-time embedded systems. This indicates that while FRs are the norm, NFRs are important where software quality is mission-critical.
• Over the years, REDAST research has seen increasing diversification in requirements specification methods. While textual requirements have always been prevalent, there is a growing adoption of alternative formats (e.g., tabular, formal, and hybrid specifications), suggesting that the RE community is seeking more expressive and flexible ways to specify requirements.

### 5.3 RQ2: Transformation Technology in REDAST

Transformation technology, as discussed in Section 4, is a key component of the REDAST process. After reviewing the selected papers, we present the results of transformation technology in four categories: technique type, framework structure, intermediate representation, and additional input. Finally, to better explain our design of these categories, we provide several examples to interpret our analysis process of RQ2.

*5.3.1 Transformation Techniques.* This section outlines the three stages of the selected studies: parsing/decomposition, augmentation/transformation, and generation/reconstruction. We categorize these stages into high-level categories based on the transformation techniques used. While we recognize that this categorization cannot encompass all the intricate details of each study, we will discuss the advantages and disadvantages of each type of technique in the context of the different stages. It is important to note that several studies, due to their complex transformation strategies, employed multiple techniques within a single stage. For example, P53 [53] utilized both machine-learning-based and tool-supported techniques during its parsing stage.

***Parsing / Decomposition Stage.*** Requirements specifications—often written in diverse notations—cannot be directly consumed by a REDAST pipeline. Hence, the first step is to decompose each specification into its atomic elements. For example, P4 [4] implements an XML parser that extracts attributes from XML-formatted requirements. Table 17 summarizes the parsing techniques used across the literature: tool-supported methods dominate (83 studies), followed by rule-based approaches (44 studies), manual decomposition (41 studies), machine-learning solutions (10 studies), and graph/search algorithms (6 studies). Within tool-supported parsing, we also highlight the NLP-based approaches, which use traditional NLP components, such as, tokenization, splitting and POS tagging [73].

***Augmentation / Transformation Stage.*** This optional, intermediate stage enriches or reorganizes the parsed representations—applying domain rules or restructuring information—before generation. Although some pipelines feed parsed output directly into test-artifact generators, many augment the intermediate model to improve coverage or traceability. Table 18 shows that rule-based transformations are most prevalent (84 studies), then tool-supported techniques (50 studies), graph/search methods (29 studies), ML approaches (6 studies), and manual interventions (2 studies).

***Generation / Reconstruction Stage*** In the final stage, all intermediate representations are aggregated to synthesize the actual test artifacts. (Note that a few works focus solely on artifact analysis or traceability and do not include this stage.) As Table 19 illustrates, external tools are most commonly employed for artifact generation (74 studies), followed by rule-based generators (55 studies), graph/search-based methods (35 studies), and ML-driven techniques (5 studies).

Table 17. Parsing / Decomposition Stage in Transformation Process (RQ2)

| Techniques | Paper IDs | Num. |
|---|---|---|
| Tool-Supported | [2, 4, 5, 7, 14–17, 20, 21, 25, 27, 29, 33, 40, 42, 49–51, 53, 55, 56, 59–62, 65, 66, 68, 69, 72–75, 77–82, 85, 86, 88, 90–92, 97, 99, 103, 104, 106, 108, 109, 116, 125, 127–129, 131, 132, 135, 140, 142, 147–151, 154], [19, 22, 24, 32, 41, 43, 52, 70, 76, 94, 117, 121, 122, 130, 134, 152, 154, 155] (NLP-based Tools) | 84 |
| Rule-based | [3, 8, 9, 12, 13, 22, 26, 28, 30, 31, 36–38, 41, 44–46, 51, 55, 57, 58, 60, 71, 83, 84, 87, 99, 102, 111, 113–119, 122, 130, 133, 136, 144, 145, 147, 151] | 44 |
| Manual | [1, 6, 9–11, 34, 35, 39, 48, 54, 58, 64, 67, 88, 89, 93, 95, 96, 98, 100, 101, 105, 107, 110, 118–120, 124, 126, 135, 137–139, 141, 143, 145, 146, 148, 150, 153, 156] | 41 |
| Machine-Learning-based | [18, 23, 52, 63, 112, 117, 122, 123, 154, 155] | 10 |
| Graph / Search-based | [47, 59, 86, 112, 113, 156] | 6 |

Table 18. Augmentation / Transformation Stage in Transformation Process (RQ2)

| Techniques | Paper IDs | Num. |
|---|---|---|
| Rule-based | [1, 4, 5, 8–19, 21, 24, 29, 30, 34, 36–43, 45–47, 49, 51, 52, 54, 55, 57, 59, 62, 64, 65, 67, 69, 70, 73, 74, 82, 83, 85–87, 93, 97, 99, 100, 102, 103, 111, 113–116, 119, 121, 125–127, 130, 132–135, 139, 140, 144, 145, 149, 151–156] | 83 |
| Tool-Supported | [2, 20, 22, 24, 25, 27, 31, 54, 56, 60–62, 66, 74–80, 84, 93, 95–97, 101, 105, 106, 108–110, 119, 122, 124, 126, 129, 142, 143, 145, 151], [32, 34, 55, 68, 82, 89, 104, 107, 127, 128] (NLP-based Tools) | 50 |
| Graph / Search-based | [3, 6–8, 23, 26, 28, 33, 35, 42, 44, 48, 50, 53, 58, 65, 72, 86, 91, 92, 94, 104, 111, 116, 122, 128, 131, 141, 147] | 29 |
| Machine-Learning-based | [71, 81, 116, 117, 123, 155] | 6 |
| Manual | [130, 148] | 2 |
| Other | [103] (Fuzzing), [153] (Mathematical Computation) | 2 |

Table 19. Generation / Reconstruction Stage in Transformation Process (RQ2)

| Techniques | Paper IDs | Num. |
|---|---|---|
| Tool-Supported | [1, 4, 15–17, 21, 22, 24, 25, 27–32, 34, 37, 40, 43, 44, 46, 48–50, 56, 59–62, 68–70, 72, 73, 75–80, 82, 92, 93, 95, 96, 98, 100, 101, 103–111, 116, 119–121, 124, 126–132, 135, 137, 142, 143, 151] | 74 |
| Rule-based | [2, 3, 6–9, 11, 19, 20, 27, 34, 38, 41, 45, 47, 51–54, 57, 58, 62, 64, 67, 71, 83, 84, 87, 90, 91, 97, 102, 112–114, 116, 118–120, 123–125, 131, 133, 136, 138–140, 145, 147, 150, 152–154, 156] | 55 |
| Graph / Search-based | [5, 8, 10, 12, 14, 18, 23, 26, 33, 35, 36, 42, 55, 59, 61, 65, 66, 74, 81, 85, 86, 89, 94, 99, 105, 111, 112, 115, 116, 122, 134, 137, 141, 148, 156] | 35 |
| Machine-Learning-based | [13, 39, 91, 117, 123, 146, 150, 154, 155] | 5 |
| Other | [88] (Mathematical Computation) | 1 |

Table 20. Typical Transformation Techniques in Different Categories (RQ2)

| Stages | Technique Categories | Typical Techniques |
|---|---|---|
| Parsing/Decomposition | Tool-Supported | Model Checker (TaRGeT Tool [2, 62], SPSS Modeler [79], Model Parser [14, 56, 59, 147]), NL Parser (POS Tagging [22, 41, 42, 70, 112, 129, 152], Dependency Parser [22, 41, 117], Tokenization [24, 43, 121, 130]), Structural Language Parser (XMI/XML Parser [7, 15, 131], CNL Parser [80, 127, 128, 151]), DSL Parser [97, 116] |
| | Rule-based | External Dictionary [9, 44, 117], Model-based Rule [28, 111, 145, 147], Structural Language Rule [84, 136, 151], NL Template [13, 99, 115, 118] |
| | Graph/Search-based | PetriNet [47, 86], Knowledge Graph [112, 122], Graph Traversal (Breadth-First Searching) [156] |
| | ML-based | Trained Classifier [18, 23, 122], BERT [117, 123, 155], Knowledge Representation [112], Clustering [154], NER [52] |
| Augmentation/Transformation | Tool-Supported | MDE/model-to-model frameworks (QVT [25, 74], ATL [76]), Dedicated test-generation tools (TaRGeT [2, 31], [22]), Model-checker & formal-tool integration [27, 56, 80, 142], Domain-specific test harnesses [60, 101], Dynamic & runtime analysis tools [66, 105], NLP transformation tools [32, 68, 82], Ontology/graph-DB tools [116, 122] |
| | Rule-based | Metamodel-driven mappings [1, 30, 43, 97], UML-based rule mappings [4, 10, 12, 13, 17, 49], Cause-Effect Graph (CEG) mappings [18, 20, 53], Control-flow & constraint extraction [11, 19, 29, 36, 102], Pattern/change-detection rules [15, 21, 30], Natural-language/grammar-based rules [34, 62, 152], Template-driven transformation sequences [125] |
| | Graph/Search-based | State-space exploration & path-finding [26, 28, 33, 92, 145], Decision-table & sequence generation [7, 44, 52], Dependency graph extraction [3, 23, 42, 141], Tree- or post-order traversal [53, 58, 72] |
| | ML-based | Clustering & similarity grouping [71, 81], LM-based Augmentation [123], Information-Retrieval [155] |
| Generation/Reconstruction | Tool-Supported | CI/Test framework integration [1, 15, 108, 110], Model-based testing tools [22, 56, 59, 61], Constraint-solver engines [24, 27, 40, 50], Docs/traceability outputs [4, 25, 30, 32] |
| | Rule-based | Classification-tree & decision-table [9, 15, 38], Sequence/template mapping [11, 20, 47], Cause-effect combinatorics [18], Grammar/CNL/regex formulation [2, 34, 62] |
| | Graph/Search-based | DFS/BFS traversal [5, 10, 74], Path-finding & spanning sets [12, 23, 26], State-space/Petri-net [36, 65], Dependency graphs [42] |
| | ML-based | Genetic/meta-heuristic search [13, 137], Clustering & prioritization [39, 154], LLM-driven scenario drafting [117, 155], Markov/statistical models [88, 146] |

***Discussion of Transformation Techniques.*** Among the three stages, we adopt a similar categorization. However, we observe that the distribution of transformation techniques varies across the REDAST process, which suggests that: (1) Rule-based and tool-supported techniques are predominantly adopted across all stages: tool-supported approaches play a key role in the parsing and generation stages, whereas rule-based methods are more prevalent in the augmentation stage. (2) Manual operations persist in the transformation process, particularly during parsing, where over 25% of studies require manual intervention. (3) Machine-learning–based approaches remain in their infancy within REDAST, with only a few studies employing such methods in any of the three stages. (4) Graph- and search-based techniques are mainly utilized in the augmentation and generation stages.

We present the typical techniques and corresponding papers in Table 20. Combining the distributions of different types of transformation techniques, we further discuss the pros and cons for each transformation technique:

- *Rule-based techniques* are among the most commonly used methods within the REDAST process. By incorporating rules/heuristics—such as dictionaries and model-based rules during the parsing stage; pattern/change-detection and control-flow rules in the augmentation stage; and regex/grammar formulations and decision rules in the generation stage—these techniques allow for the integration of expert knowledge throughout the REDAST workflow. Given the relatively constrained structure of requirements specifications, rule-based methods offer an effective and efficient means of parsing and decomposing these specifications for test generation or further analysis. While rule-based techniques provide structure and ease of implementation, they face challenges in handling complex scenarios, particularly those involving NL requirements. To address this, some studies combine rule-based methods with other approaches to manage out-of-dictionary cases—an ongoing issue in the NLP domain [7, 54]. For instance, P22 [22] uses NLP pipelines to parse NL requirements and applies rules to map textual expressions to intermediate representations.
- *Tool-Supported Approaches* do not represent a single technique, but rather refer to studies that utilize existing tools instead of developing custom components. A common example is the NLP pipeline, which typically incorporates open-source toolkits such as NLTK or Stanford CoreNLP. However, we found that the NLP-based tools are primarily used in the parsing and augmentation stages, which suggests that (1) traditional NLP-based tools cannot solely support the REDAST studies and (2) NLP-based tools are more advanced in parsing and augmentation rather than generation. The other techniques, such as rule-based or graph-based techniques, should be cooperatively employed. Other tools, like TaRGeT or certain meta-model engines, support not only individual functions but the entire augmentation and generation stages of the REDAST process. Like rule-based methods, tool-supported approaches also offer convenience and efficiency. However, their reliance on external tools limits their adaptability, as these tools often lack customization or localization for specific end tasks, ultimately constraining their flexibility in practice.
- *Graph/Search-based Techniques*, unlike the previous two categories, are not general-purpose solutions for all stages of the REDAST process. The limitations of graph/search-based techniques stem from two main factors: (1) They require input in the form of graph- or diagram-based representations. If the input is not already in this format, a graph construction step must be introduced; (2) These techniques are generally more effective for transformation or generation tasks than for parsing, as raw requirements often lack the structured information needed for meaningful graph analysis. A representative approach in this category is the use of knowledge graphs (KGs). For instance, P122 [122] employs a knowledge graph to represent the pre- and post-conditions of requirements for test generation. In their method, NLP pipelines and named entity recognition (NER) are first used to decompose the input natural language requirements to facilitate KG construction. By leveraging graph/search-based techniques, logical relationships among requirements and test artifacts can be visually and structurally captured through graph dependencies, paths, or node connections—offering both flexibility in test generation and enhanced control over the process.
- *Machine-Learning-based Techniques* have gained prominence over the past decade, particularly following the rise of deep learning. Compared to other transformation techniques, ML-based methods significantly enhance the parsing and understanding of various types of input representations—especially natural language (NL) requirements. For example, P154 [154] employs SentenceBERT [72] to index NL requirements, then uses semantic similarity to retrieve relevant context, followed by a large language model (LLM) for test scenario generation. The powerful representational capacity of ML-based techniques enables a new paradigm for generating test artifacts, offering both flexibility and effectiveness. Despite their promise, current ML-based techniques face several limitations in the REDAST process: (1) Traditional ML models typically require large amounts of task-specific training or fine-tuning data, which is often unavailable in the software engineering (SE) domain; (2) Only a limited number of ML-based approaches are capable of effectively interpreting non-NL requirements and generating corresponding test artifacts, which further restricts their applicability.

- *Manual Intervention* is not a technical method, but it warrants separate mention due to its impact on the automation of the REDAST process. Human involvement typically occurs during the parsing stage to resolve ambiguities, highlighting that accurate parsing remains the most challenging and critical step in the REDAST workflow. While manual input can improve precision in specific cases, it inherently limits scalability and undermines the goal of full automation. A detailed discussion on automation will be presented in RQ4.

*5.3.2 Framework Details.* In this subsection, we cover the remaining parts of the transformation process, i.e., additional input, intermediate representation types and the framework structure.

***Additional Input.*** While our SLR studies the REDAST process, the inputs are not constrained to the requirements specification only. We found that some studies introduce additional documents as input to improve the functional coverage of their methodologies. This section illustrates the details of additional input in REDAST studies, including:

- *System Implementation.* Implementation details provide concrete evidence from real scenarios and thus critically inform test-artifact generation. Four studies treat system implementation as an auxiliary input. P69 [69] introduces an *Implementation Execution Environment* as the evaluation for the final test artifacts to generate the comparison results. P108 [108] use the configuration files, which refers to the implementation details of the developed tools, as the additional input to ensure the tool is working properly and configure stylization properties. P109 [109] introduces a parallel route from implementation design, where the implementation details will be applied on the abstract guarded assertions (a type of representation supporting test generation) to implement the abstract assertions to concrete representations. P124 [123] introduces the implementation for checking and filtering requirements on if the requirements are violated by the implementation design, which can exclude the non-functional requirements in the parsing stage.
- *Historical Documents.* Archived requirements and past test reports inject valuable context into both test generation and prioritization. P32 [32] leverages historical test execution logs—outcomes of previous test cases—to prioritize system tests for new product releases. P122 [122] uses past test reports, which contain necessary information such as initialization conditions, to generate KGs and augment the generated artifacts. By serving as reference points, these historical documents guide current processes and enhance artifact quality.
- *Source Code.* Source code embodies the concrete implementation and system logic, allowing tests to align directly with actual behavior. P81 [81] adopts code structure as a prioritization objective, running tests that cover critical modules earlier in regression suites. P154 [154] incorporates code-change information into test prioritization to pinpoint error-prone components. Within REDAST, source code grounds functional tests in implementation detail, enables precise fault-detection metrics via coverage analysis, and supports fine-grained traceability.

Table 21. Additional Inputs of Selected Studies (RQ2)

| Additional Input Types | Paper IDs | Num. |
|---|---|---|
| System Implementation | [69, 108, 109, 124] | 4 |
| Requirements Documents | [93, 122, 155] | 3 |
| Source Code | [81, 154] | 2 |
| Supporting Documents | [109, 124] | 2 |
| Historical Documents | [32, 122] | 2 |

Table 22. Framework Structure Results of Selected Studies (RQ2)

| Structure | Paper IDs | Num. |
|---|---|---|
| Sequential | [1, 2, 5, 6, 8, 11–14, 16, 17, 20, 22, 24, 26, 27, 29–33, 35–37, 40, 43–46, 50, 53–55, 58, 59, 61, 62, 64, 65, 70, 75–78, 96–98, 103–107, 110–112, 114, 116, 119, 120, 122–124, 127–132, 134, 137, 138, 140–144, 146, 149, 151, 154] | 80 |
| Parallel | [3, 7, 9, 10, 19, 21, 25, 34, 39, 51, 52, 57, 66, 68, 72, 74, 80, 100, 135, 145, 147, 148] | 22 |
| Conditional | [15, 18, 28, 47, 49, 71, 73, 139, 150] | 9 |
| Loop | [4, 56, 109] | 3 |

- *Supporting Documents and Requirements Documents.* Constructed usage scenarios—derived from specific segments of the requirements document—validate requirement completeness and consistency (P93 [93]). These scenarios, in conjunction with the original requirements documentation, serve as inputs for comprehensive test-artifact generation. P122 [122] uses the legacy requirements statement to generate the knowledge graph, which functions similarly to the past source code discussed above. Retrieval-Augmented Generation (RAG) [29, 100] is an advanced technique to integrate external knowledge into the generation process. P155 [155] selectively retrieves the relevant domain knowledge from requirements documents and constructs it into a prompt template for test scenario generation.

***Framework Structure.*** From requirements specification to test artifacts, various framework designs can be applied to transformation methodologies. Within the selected REDAST framework, we categorize these methodologies into four distinct structure types:

- *Sequential Structure.* In the sequential framework, each transformation step must complete before the next begins, enforcing a strict linear progression from requirements to test artifacts. P107 [107] exemplifies this approach by converting formal specifications first into conjunctive normal form, then into assignment tables, and finally into concrete test cases, ensuring each phase builds directly on the previous one. This clear ordering simplifies reasoning about data dependencies and makes it easier to trace each output artifact back to its originating requirements element. Because there are no shortcuts or parallel branches, debugging and logging become straightforward—errors at any stage can be isolated and replayed in a reproducible manner. On the downside, purely sequential designs can incur performance bottlenecks when early stages are slow or require human validation, and they may be less resilient to changing inputs mid-process.
- *Conditional Structure* introduce decision points that allow the pipeline to branch into alternative sub-flows based on runtime checks or metadata. P73 [73] demonstrates this by defining conditions like "Need more details of requirements" and "Transformation improvements available," each triggering distinct sub-workflows to handle different scenarios. By evaluating predicates such as insufficient requirement detail or requirement complexity threshold exceeded, the system can route inputs to specialized processors (e.g., a detail enrichment module or a complexity reduction filter) before resuming the main flow. This adaptability supports a wider variety of input types and quality levels, improving robustness against unexpected or incomplete specifications. However, designing and maintaining the condition logic can add complexity, and traceability across multiple branches may require careful documentation.

- *Parallel Structure* decompose the overall task into independent subtasks that execute concurrently, significantly reducing end-to-end processing time on multicore or distributed infrastructures. For example, one processor might generate use case diagrams while another simultaneously derives executable contracts. Their results are later merged to produce comprehensive test scenarios. This division of labor not only accelerates throughput but also allows heterogeneous tools or services to be integrated without linear coupling. The primary challenge lies in cooperating of different expressions and resolving conflicts when different branches produce overlapping outputs. In P10 [10], a two-way parallel design converts requirement inputs into both use case diagrams and contract-based models concurrently, then combines them to generate contractual use case scenarios efficiently.
- *Loop structure* embed iterative cycles within the pipeline, enabling continuous refinement of intermediate artifacts based on quality metrics or validation feedback. After an initial transformation pass, a validator assesses correctness or coverage, and if the results fall below a predefined threshold, the system loops back to earlier modules with adjusted parameters or augmented data. This feedback-driven process leads to progressively higher-quality outputs but may increase overall runtime and require robust convergence controls to avoid infinite loops. P109 [109] applies this pattern by iteratively tuning guarded assertions through a validation–refinement loop, ensuring that the generated tests meet strict correctness criteria before exiting the cycle.

*5.3.3 Intermediate Representation.* The intermediate representation functions as a detailed explanation of requirements or system structure, reflecting the framework structure's complexity. In our taxonomy, intermediate representation is the extension content of augmentation / transformation stage. Specifically, intermediate representation is the step-generated artifacts during the transformation process. When reviewing the selected studies, the intermediate representation necessarily exists in a complex REDAST framework to enable a stepwise transformation. Thus, we illustrate the details of the adopted intermediate representations to understand the framework's composition better.

***Number of Intermediate Representations.*** The number of intermediate representations is a core feature of REDAST's framework, reflecting its structural complexity. To improve explainability, we illustrate the number of intermediate representations alongside the framework structure in Fig. 7. Sequential structure (66 studies) is the dominantly common framework in the "single" category, followed by conditional structure (8 studies), parallel structure (4 studies), and loop structure (2 studies), (2) parallel structure is the most common framework in "multiple" category (18 studies), where sequential is also popular (14 studies). The results show different trends in “single” and “multiple” categories. From the results, we have the following conclusion: (1) In the sequential and conditional categories, a single intermediate representation is more common. Since both conditional (viewed as a one-way path) and sequential frameworks follow logical structures, simpler intermediate processes are generally preferred. This likely reflects an effort to avoid unnecessary complexity and improve traceability. (2) In contrast, the parallel category shows a preference for multiple intermediate representations. Because parallel structures must accommodate inputs from multiple branches, the higher number of intermediate representations aligns with the inherent complexity of this structure.

***Type of Intermediate Representations.*** Since intermediate representations are typically derived from requirements, we categorize them using a schema similar to that used for requirement specifications, as shown in Table 23. The results show that structural/formal (59 studies) and model-based (35 studies) representations are most commonly used.

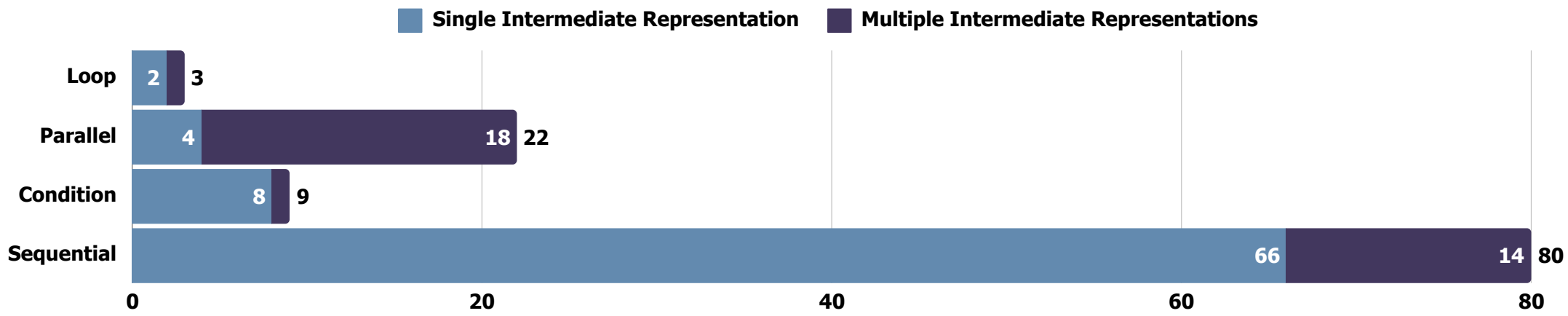

Fig. 7. Joint Distribution of Intermediate Representation and Framework Structure (RQ2)

Table 23. Type of Intermediate Representation in Selected Studies (RQ2)

| Repre. Types | Paper IDs | Num. |
|---|---|---|
| Model-based | [2, 4, 5, 10, 14, 16, 22, 25, 31, 33, 35, 36, 43, 45, 48, 49, 54, 56, 57, 59–62, 69, 70, 72–77, 80, 83–85, 87, 89, 91, 92, 95–97, 101, 105, 106, 110, 127–129, 132, 135, 140, 142–145, 149, 151, 156] | 59 |
| Graph-based | [3, 6–8, 11–13, 17–19, 23, 26–28, 42, 44, 47, 50, 53, 64, 65, 82, 86, 94, 99, 104, 111, 116, 119, 122, 131, 134, 141, 147, 148] | 35 |
| Structural/Formal | [1, 9, 15, 21, 24, 29, 30, 34, 37, 40, 41, 46, 52, 55, 58, 67, 68, 71, 78, 79, 102, 107–109, 113, 121, 123, 125, 130, 133, 139, 152, 153, 155] | 34 |
| Test-related | [20, 32, 38, 39, 51, 66, 81, 100, 103, 114, 115, 117, 124, 126, 154] | 15 |

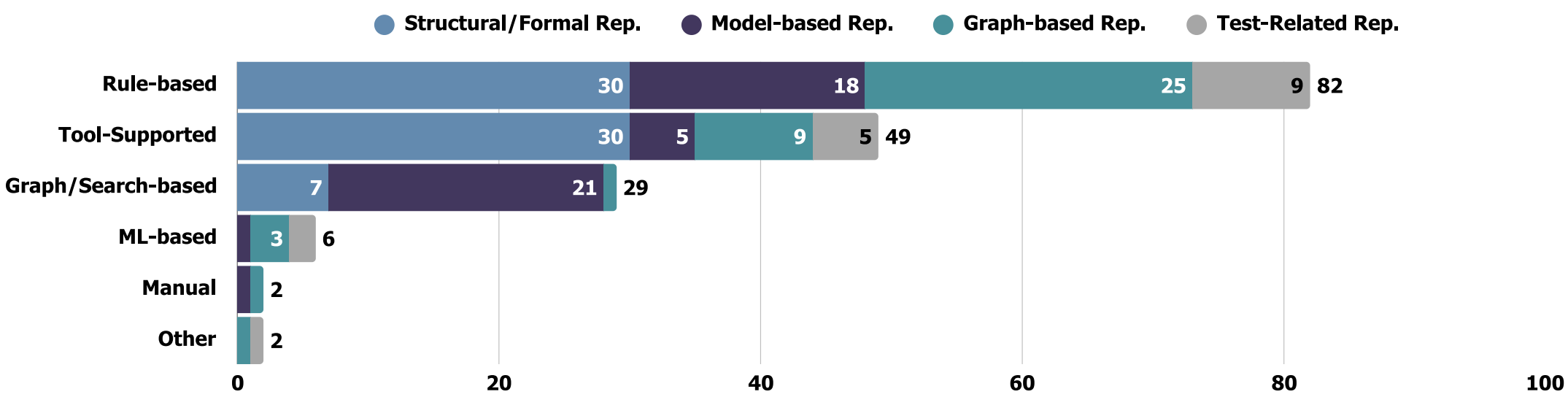

Fig. 8. Joint Distribution of Techniques in Augmentation/Transformation Stage and Intermediate Representation (RQ2)

Graph-based representations, though less prevalent (34 studies), remain widely adopted. Test-related representations appear in 15 studies. In the following, we present several examples in each category, as:

- *Model-based Representation.* P62 [62] employs a meta-model as an intermediate representation. Each use case, written in Controlled Natural Language (CNL), is decomposed into variables representing different flows. These variables are translated into meta-model elements and processed using predefined templates or rules within a Communicating Sequential Processes (CSP) framework. P70 [70] applies a semantic analysis tool to parse use cases into linguistically grounded models, identifying various types of actions.
- *Graph-based Representation* is about structuring the intermediate artifacts as nodes and edges in a graph structure. P94 [94] utilizes a knowledge graph as an intermediate representation, where nodes and edges represent requirement conditions. Concrete test cases are generated by sequentially integrating sub-graphs using graph traversal and dependency analysis. P116 [116] adopts the Trivial Graph Format (TGF) to represent decomposed elements such as actors, preconditions, and flow variants (basic and alternative). Graph traversal is then used to identify linearly independent paths for constructing test cases.

- *Structural/Formal Representation.* P155 [155] integrates LLMs for test generation, using structurally designed prompts—based on retrieved contextual information—as intermediate representations. These prompts serve as input for the LLMs, which generate test scenarios. P123 [123] conceptualizes language models as transformation matrices, where the output tokens (in textual form) act as intermediate representations. These are further processed using predefined rules to produce test artifacts.
- *Test-related Representation.* P81 [81] identifies fault-prone components in code and requirements via metric-based analysis (e.g., McCabe complexity, requirement fan-in). The analysis uncovers common fault patterns and specification weaknesses. P100 [100] identifies potential failure modes per requirement, assesses their effects, and computes associated Risk Priority Numbers (RPNs) and Requirement Priority Numbers (RePNs). These failure modes represent recurring issues from a requirements traceability perspective.

As an extension of augmentation/transformation techniques, we also analyze the joint distribution of transformation techniques (during the augmentation/transformation stage) and intermediate representations, as shown in Fig. 8.

- *Rule-based Techniques.* Structural/formal (30, 37%) and graph-based (25, 31%) dominate, with model-based close behind (18, 22%). Although test-related representations are rare overall, 9 of the 16 recorded cases (56%) arise here, confirming that hand-crafted rules remain the default when the intermediate artefact is itself a test template (e.g., Coverage Matrix in P108, or Gherkin-style stubs in P141). In practice, most rule-based pipelines still funnel requirements into behavioral meta-models—typically UML activity or state-machine fragments augmented with OCL constraints—before downstream generation.
- *Tool-Supported Techniques.* Exactly 30 cases (61%) target structural/formal artefacts such as annotated CSVs or domain-specific tabular schemas; only 9 cases (18%) handle graph-based intermediates. The drop from 25 graph uses in the rule-based family to 9 here highlights a tooling gap: robust editors or code generators for arbitrary graph transformations are still scarce. Many prototype tools, therefore, flatten graphs into adjacency lists or path tables before analysis.
- *Graph/Search-Based Techniques.* Contrary to intuition, model-based representations dominate (21, 72%), while native graph artefacts appear only once. The reason is methodological: studies often *construct* an internal search graph from a higher-level model (e.g., a SysML sequence diagram expanded into a reachability graph) rather than expecting graphs as input/output artefacts.
- *ML-Based Techniques.* Small sample, but skewed toward graph (3, 50%) and test artefacts (2, 33%). No structural/formal examples surfaced. Only a single study maps raw text to a meta-model element set. LLM pipelines, therefore, still struggle to emit well-typed model objects without an extra symbolic post-processor.
- *Manual Intervention.* Both papers rely on human curation of complex models/graphs (1 model-based, 1 graph-based) to ensure semantic fidelity before automated steps proceed.

*5.3.4 RQ2 Cross-Analysis between Requirements Types and Transformation Techniques in Different Stages.* Table 24 reveals a stark contrast between how FRs and NFRs flow through the REDAST transformation stages. In both parsing and augmentation, FRs benefit from extensive automation—tool-supported approaches alone appear in 75 cases, rule-based in 38, and even ML-based methods in 10—whereas NFRs rely far more on manual effort (7 instances) and modest rule-based (6) or tool-supported (10) support, with no ML-based pipelines at all. At the generation stage, FRs again leverage graph-based (32), rule-based (47) and tool-supported (64) techniques, while ML-based methods contribute another 7 outputs; by contrast, NFR generation sees only 3 graph-based, 2 ML-based, 8 rule-based and 10 tool-supported cases, with zero manual generation. When we isolate purely traditional NLP-driven tools in tool-supported, roughly 40 out of 135 FR workflows (30%) incorporate such techniques, compared to only 3 out of 21 NFR workflows (14%). This gap underscores that, although modern semantic methods have begun to penetrate functional testing, they are far less applied to quality attributes. Across all stages, NFRs consistently trail FRs in automated coverage, indicating that specialized pipelines or intermediate representations for NFRs remain underdeveloped and demand targeted attention.

Table 24. Joint Distributions of Functional (FR) and Non-Functional Requirements (NFR) by Transformation Technique and Stage

| Stage | Req. Type | Graph-based | ML-based | Manual | Rule-based | Tool-Supported |
|---|---|---|---|---|---|---|
| Parsing | FR | 5 | 10 | 34 | 38 | 75 |
| | NFR | 1 | 0 | 7 | 6 | 10 |
| Augmentation | FR | 5 | 10 | 34 | 38 | 75 |
| | NFR | 1 | 0 | 7 | 6 | 10 |
| Generation | FR | 32 | 7 | 1 | 47 | 64 |
| | NFR | 3 | 2 | 0 | 8 | 10 |

Table 25. Joint Distribution of Requirements Formats by Parsing Technique

| Req. Format | Graph-based | ML-based | Manual | Rule-based | Tool-Supported |
|---|---|---|---|---|---|
| **Formal** | 1 | 0 | 7 | 4 | 13 |
| **Model-Based** | 2 | 0 | 16 | 12 | 30 |
| **Other** | 0 | 0 | 3 | 2 | 2 |
| **Tabular** | 0 | 1 | 1 | 0 | 6 |
| **Textual** | 3 | 9 | 23 | 36 | 56 |

*5.3.5 RQ2 Cross-Analysis between Parsing Techniques and Requirements Formats.* The data in Table 25 highlight how parsing techniques support different requirement formats. Tool-supported parsers show the broadest applicability. From the traditional NLP-based tools or meta-model checkers, these tools can perfectly support the requirements parsing process. Rule-based approaches likewise demonstrate strong coverage of textual requirements, where well-defined pattern libraries and heuristics abound, but their effectiveness diminishes on formal and model-based artifacts, which demand more rigorous syntax and semantic handling. In contrast, ML-based parsers remain narrowly focused on NL requirements, with almost no application to formal or model-based specifications, which shows a strong supporting capability of ML techniques in processing NL requirements. Graph-based techniques, while still nascent, show promise on formal and model-based requirements: their ability to encode semantic relationships into intermediate representations could bridge the gap between pure grammar parsing and the flexibility needed for complex diagrams or logic expressions. Finally, manual parsing settles a large number in the parsing stage and is primarily employed in formal, model-based, and textual formats, which suggests that requirements parsing is still an obstacle in the REDAST process.

Table 26. Joint Distribution of Requirements Abstraction Level and Parsing/Decomposition Techniques

| Abstraction Level | Tool-Supported | Rule-based | Manual | ML-based | Graph/Search-based |
|---|---|---|---|---|---|
| **Function (3rd)** | 63 | 31 | 24 | 9 | 3 |
| **Feature (2nd)** | 28 | 14 | 18 | 2 | 2 |
| **Component (4th)** | 13 | 4 | 4 | 0 | 3 |
| **Product (1st)** | 3 | 1 | 1 | 0 | 1 |

Table 27. Joint Distribution of Abstraction Level and Augmentation/Transformation Techniques

| Abstraction Level | Tool-Supported | Rule-based | Graph/Search-based | ML-based | Manual | Other |
|---|---|---|---|---|---|---|
| **Function (3rd)** | 32 | 55 | 22 | 4 | 7 | 2 |
| **Feature (2nd)** | 21 | 32 | 4 | 3 | 3 | 0 |
| **Component (4th)** | 5 | 11 | 6 | 0 | 2 | 0 |
| **Product (1st)** | 2 | 2 | 1 | 0 | 0 | 0 |

Table 28. Joint Distribution of Abstraction Level and Generation Techniques

| Abstraction Level | Tool-Supported | Rule-based | Graph/Search-based | ML-based | Other |
|---|---|---|---|---|---|
| **Function (3rd)** | 50 | 38 | 22 | 4 | 1 |
| **Feature (2nd)** | 14 | 12 | 3 | 1 | 0 |
| **Component (4th)** | 7 | 3 | 8 | 3 | 0 |
| **Product (1st)** | 3 | 1 | 2 | 0 | 0 |

Table 29. Joint Distribution of Augmentation Techniques by Framework Structure

| Augmentation Technique | Conditional | Loop | Parallel | Sequential |
|---|---|---|---|---|
| **Graph / Search-based** | 1 | 0 | 4 | 17 |
| **Machine-Learning-based** | 1 | 0 | 0 | 2 |
| **Manual** | 0 | 0 | 1 | 1 |
| **Other** | 0 | 0 | 0 | 1 |
| **Rule-based** | 6 | 1 | 13 | 40 |
| **Tool-Supported** | 0 | 2 | 7 | 31 |

*5.3.6 RQ2 Cross-Analysis between Requirements Abstraction Levels and Parsing Techniques.* The cross-analysis in Table 26, 27, and 28 confirm that abstract requirements—especially at the Feature level—require additional contextualization or implementation before they are usable for test generation, where feature-level requirements adopted more manual parsing techniques (28.1%) than the other levels (Function-level 18.5%, Component-level 16.7%, Product-level 16.7%). Considering that the component and product levels have insufficient studies for analysis, we only consider the function- and feature-levels. As for the following stages, Function- and Component-levels proceed with lighter restructuring, where Function-level studies show near parity between transformation and generation stages and the largest tool-supported generation block, while Component-level studies rely on graph/search and tools with modest augmentation. Among technique families, rule-based and tool-supported are the most level-agnostic, appearing at all abstraction levels across all three stages.

*5.3.7 RQ2 Cross-Analysis between Augmentation Techniques and Framework Structures.* Table 29 reveals the joint distribution of the framework structure and augmentation stage. This cross-discussion aims to show the influence of augmentation techniques on the complexity of the framework structure. The augmentation techniques are mainly used in sequential structure, the simplest framework design. Parallel and conditional structures express branching test logic or alternative scenarios, and are supported by rule-based and graph-based techniques, but remain far less common.

*5.3.8 RQ2 Findings - Trend of Transformation Techniques Over the Years.* This section exhibits the results of the year-to-year trend of transformation techniques in parsing, augmentation, and generation stages.

In Fig. 9, the earliest work was almost entirely tool-supported and manual. The tool-supported techniques always take the lead from the emergence of REDAST studies. Around 2006–07, the distribution of parsing techniques became diverse. The emergence of rule-based techniques became the majority from 2008 to 2015, but they have not taken the lead in recent years. As for ML-based techniques, they appeared after 2012 and have gradually found a place in adopting parsing techniques. The graph-based techniques are not frequently adopted in the parsing stage. Surprisingly, the manual parsing is expected to disappear gradually with the development of automated techniques. However, it consistently appears even in recent years. We think the reasons could come from two parts: (1) we acknowledge that the criteria of the academic papers have increased in recent years, which requires the paper to report every step in detail. Thus, authors report their manual operation to maintain the integrity of the papers; (2) as we found, NL requirements are increasingly employed in the REDAST process, but the requirements parsing process is a challenge.

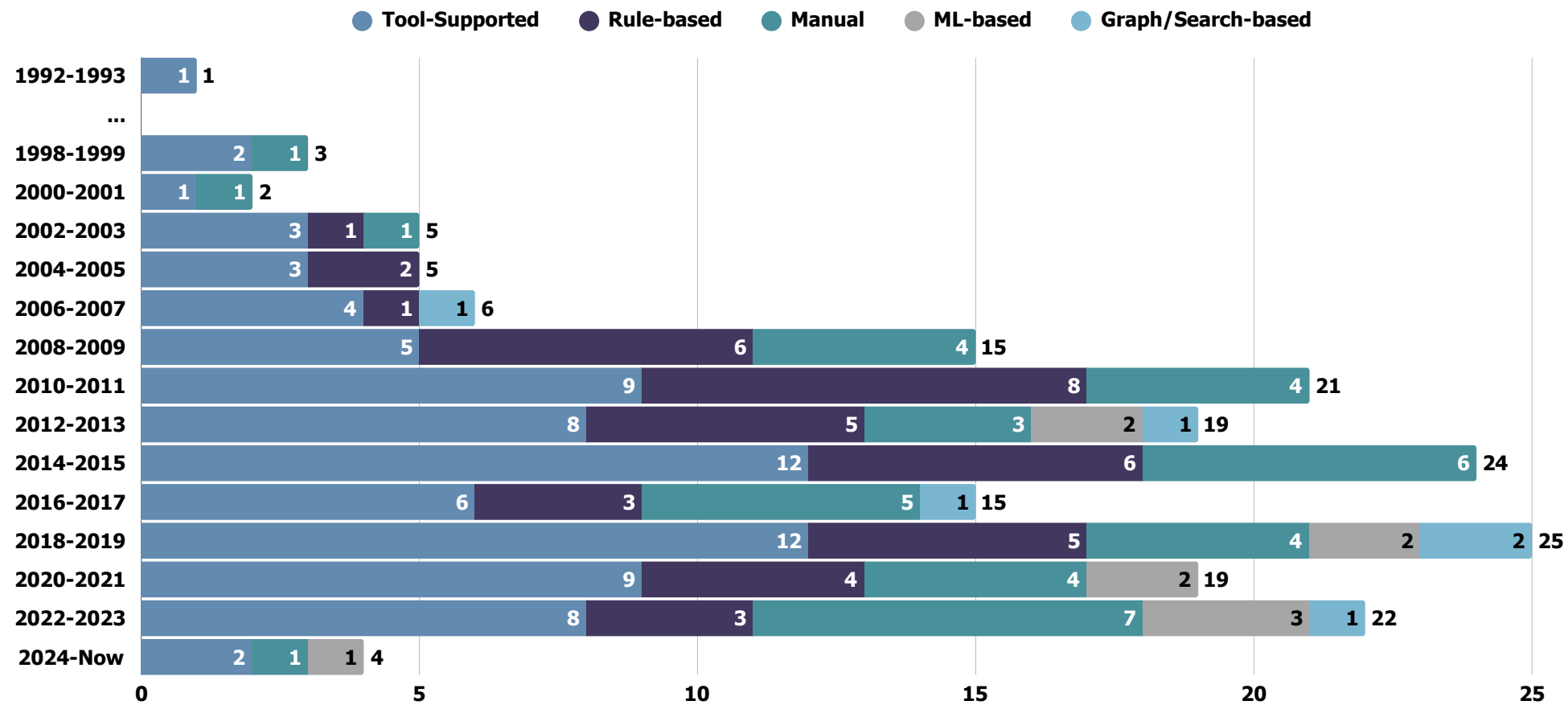

Fig. 9. Trend of Transformation Techniques in Parsing Stage by Years

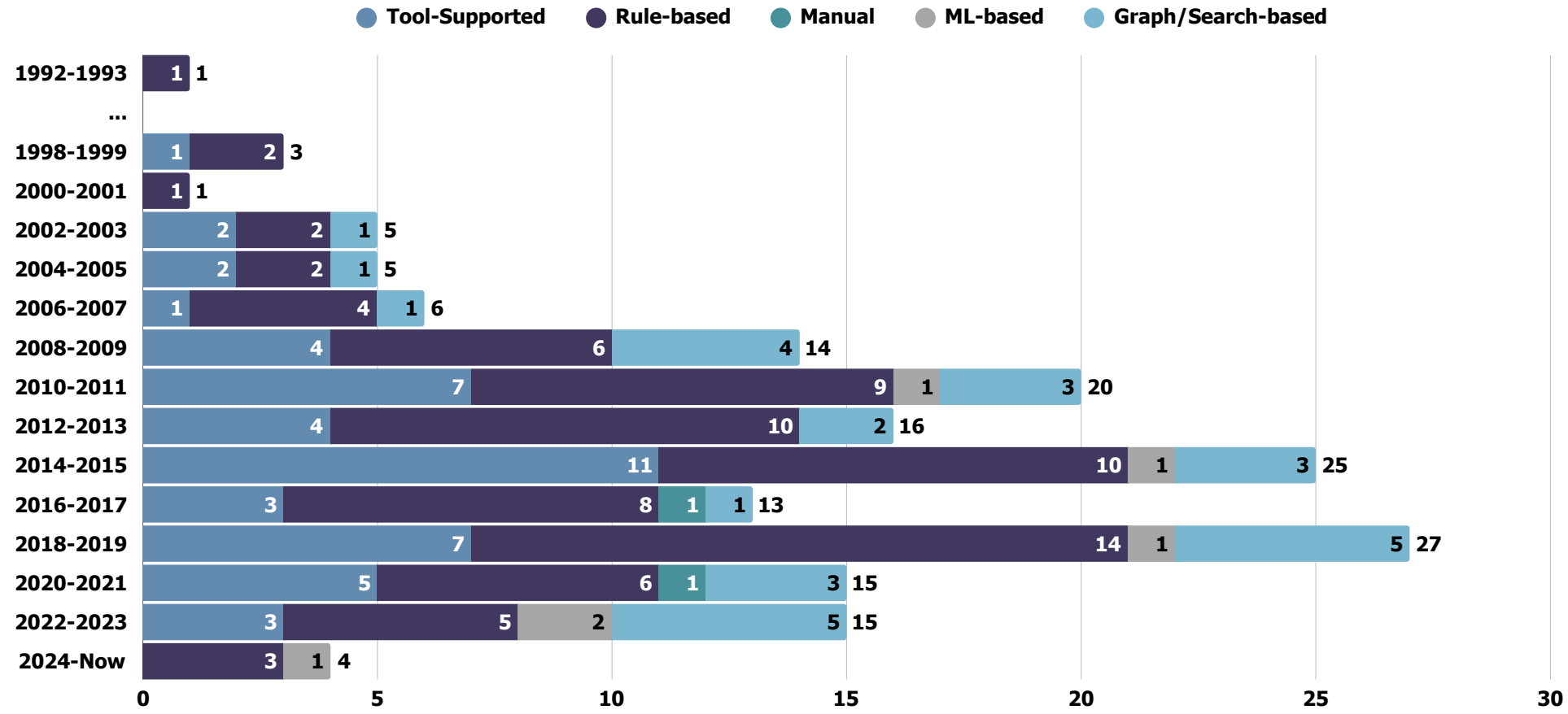

Fig. 10. Trend of Transformation Techniques in Augmentation Stage by Years

Fig. 10 shows that from the earliest REDAST work through the mid-2000s, augmentation depended largely on tool-supported frameworks and hand-coded rule engines, with manual effort filling the gaps. Around 2010, experimentation with graph- and search-based enrichment briefly accelerated, but these approaches never outpaced the more established strategies. Rule-based augmentation peaked between 2014 and 2017 as research groups codified NLP pipelines into reusable engines, then gave way to increasingly capable tool-supported platforms. Manual augmentation, once commonplace, all but vanishes after 2010, indicating that mid-level artifact enrichment has become largely automated. Machine-learning-based augmenters, however, remain almost invisible, where only a handful of studies attempt neural techniques, suggesting that current ML models struggle to capture the complex control flow and domain semantics needed for reliable test enrichment.

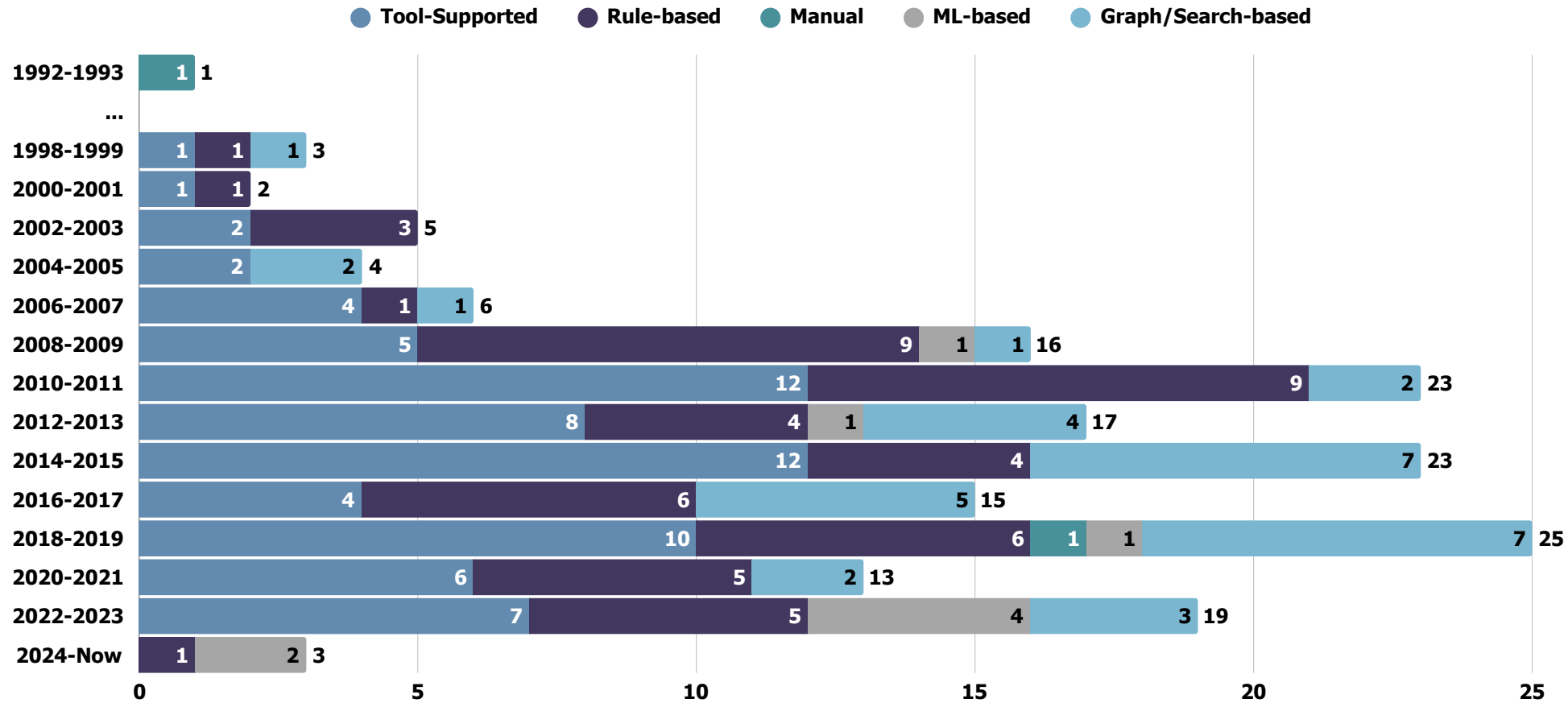

Fig. 11. Trend of Transformation Techniques in Generation Stage by Years

Fig. 11 reveals that by the late 2000s, manual generation of test scaffolds had essentially disappeared, replaced by a balance of rule-based generators and tool-supported code-emission engines. Graph- and search-based generation surged between 2008 and 2014, driven by symbolic planners and state-space explorers, but plateaued thereafter as their complexity and performance costs tempered adoption. Tool-supported generators continue to grow steadily, reflecting the rise of integrated test-script frameworks, while rule-based approaches remain a reliable fallback for custom scenarios. Machine-learning-based generation, despite its promise, is still marginal, appearing in only a few recent studies, indicating that pure neural generation has yet to meet the precision and maintainability requirements of production test suites.

**RQ2 Key Takeaways**

• Rule-based and model-based approaches dominate REDAST due to their structured, interpretable nature and low dependency on training data. While AI techniques—particularly NLP—are increasingly incorporated for automation, they primarily support rule-based and model-driven frameworks rather than serving as standalone transformation methods.

• Requirements are generally considered sufficient input for REDAST, as few studies integrate additional data sources (e.g., system implementation, source code, or historical test logs). Instead, intermediate representations serve as essential enablers for translating requirements into structured test artifacts. A significant number of studies employ multi-step transformations, making intermediate representations a key bridge between raw requirements and test automation artifacts.

Table 30. Classification of Test Artifact Types (RQ3)

| Testing Stages | Artifact | Paper IDs | Num. |
|---|---|---|---|
| Planning Stage | Test Plan | [19, 31, 46, 47, 71, 75, 87, 90, 103, 113, 117, 118, 121] | 13 |
| Designing Stage | Test Case | [1, 2, 4, 5, 9, 11, 13–16, 19, 20, 22–24, 26, 34–37, 40, 42–45, 48–50, 52, 54–56, 59–62, 68–70, 72, 74, 76–81, 83, 85, 87–89, 91, 92, 95–100, 102, 104–107, 112, 116, 117, 119, 120, 122, 123, 125, 126, 128, 130, 133–136, 146–148, 150, 152, 154, 156] | 87 |
| | Test Scenario | [6–8, 10, 12, 17, 19, 25, 28, 33, 41, 46, 51, 53, 65–67, 73, 144, 145, 155] | 21 |
| | Test Case Prioritization | [32, 38, 39, 81, 114, 115, 131, 133, 136, 137, 139, 141, 153] | 13 |
| | Test Model | [3, 18, 84, 86, 132] | 5 |
| | Test Vector | [94, 127, 129, 151] | 4 |
| Execution Stage | Test Script | [20, 21, 27, 29, 51, 57, 82, 101, 108, 110, 111, 118, 124, 138, 140, 142, 143] | 17 |
| | Test Oracle | [3, 24, 55, 63, 88, 109] | 6 |
| | Test Data / Bench | [58, 60, 64] | 3 |
| Reporting Stage | Test Analysis | [30, 149] | 2 |
| | Test Report | [93] | 1 |

• Trends show gradual automation of augmentation and generation stages, but parsing remains a bottleneck (with manual intervention) even in recent years despite advances in NLP and ML.

## 5.4 RQ3: Generated Test Artifacts in REDAST

RQ3 aims to discuss the final outcomes of the REDAST process, where we studied the test type, format, and coverage.

*5.4.1 Test Artifacts and Corresponding Stages.* This section details the different generated test artifacts among the four testing stages. We present the artifacts and their corresponding stages in Table 30. From this table, we can directly have the following conclusions: Among different stages, designing stage (120 artifacts) is the majority, followed by execution stage (25 artifacts), planning stage (13 artifacts), and reporting stage (3 artifacts), which reflects the best alignment between requirements specification and testing design. Very few artifacts target the reporting stage, because they depend on actual execution results that are outside the scope of many generation studies.

- *Planning Stage (Test Plan, 13 studies).* As the first step in the software-testing life cycle, these artefacts provide high-level, abstract objectives and strategic guidance for testing rather than detailed procedural steps. Their automatic generation, therefore, requires a thorough understanding of test goals and a clear, NL formulation of those goals. P71 [71] (1) extracts actors, processes, and objects from each NL requirement; (2) groups the decomposed requirements to reveal dependencies; and (3) applies pattern templates to each cluster to generate an ordered sequence of test steps with corresponding assertions.
- *Designing Stage (Test Case, 81 studies; Test Scenario, 19 studies; Test Case Prioritization, 13 studies; Test Model, 5 studies; Test Vector, 4 studies).* This stage accounts for the majority of generated artefacts. Test cases (including prioritised variants) represent a large majority of the outputs.
- *Execution Stage (Test Script, 16 studies; Test Oracle, 6 studies; Test Data/Bench, 3 studies).* These artefacts are the most concrete deliverables, with test scripts appearing in 64 % of the studies. Test oracles are noted as being challenging to automatically generate because they must decide correctness for complex systems [13, 28, 60]; human judgment therefore remains essential [93]. P63 [63] attempts to mitigate some of these issues by converting Restricted Use-Case Model (RUCM) specifications to OCL expressions, enabling early checks that improve correctness and completeness while reducing post-execution verification effort.

Table 31. Test Formats of REDAST Studies (RQ3)

| Test Formats | Paper IDs | Num. |
|---|---|---|
| Structural Textual Formats | [1, 10, 13, 16, 21, 23, 29, 32, 35, 41, 44, 48, 57, 59, 70, 71, 80, 83, 86, 89, 90, 93, 96–98, 113, 117, 118, 121, 144, 152, 155] | 32 |
| Other | [4, 9, 15, 21, 24, 26, 34, 46, 66, 67, 81, 100, 114–116, 119, 120, 129, 131, 133, 136, 141, 150, 151, 153, 154, 156] | 20 |
| Tabular Formats | [6, 19, 30, 31, 51, 52, 58, 68, 75, 78, 83, 102, 103, 112, 125, 128, 132, 139, 142, 147] | 20 |
| Not Specified | [4, 5, 7, 33, 38, 39, 43, 45, 61, 63, 76, 91, 99, 101, 106–108, 122, 126, 127, 148] | 20 |
| Model-Based Formats | [11, 17–20, 26, 47, 49, 53, 54, 64, 72, 77, 79, 85, 95, 104, 123] | 18 |
| Code-Based Formats | [2, 28, 37, 40, 60, 82, 92, 110, 111, 124, 140, 142–144] | 13 |
| Logic-Based Formats | [3, 14, 62, 69, 79, 80, 109, 149] | 8 |
| Standardized/Specialized Formats | [20, 27, 50, 56, 65, 74, 88] | 7 |

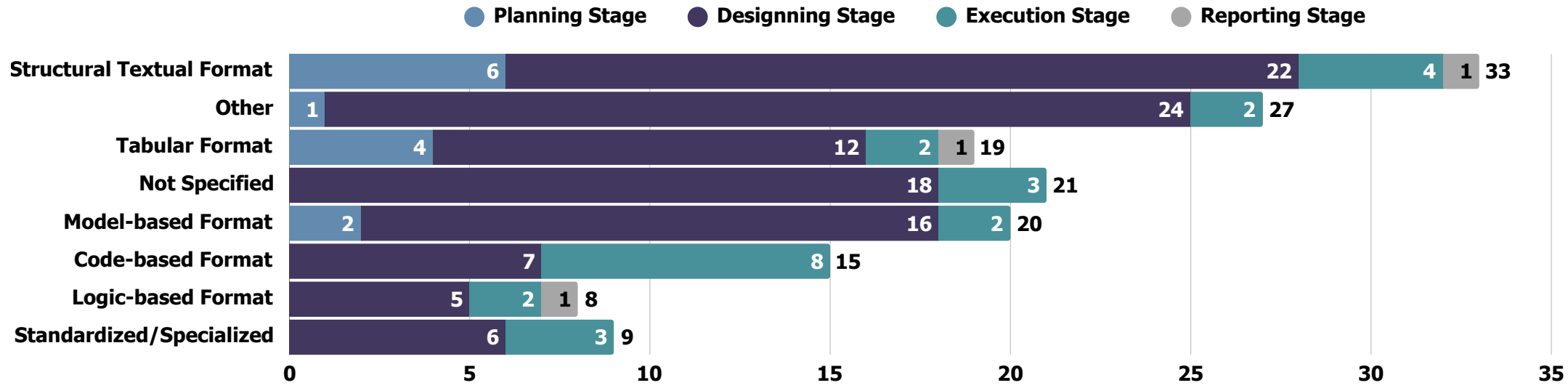

Fig. 12. Joint Distribution of Testing Formats and Stages

- *Reporting Stage (Test Analysis, 2 studies; Test Report, 1 study).* Only three studies address the reporting stage. Test analysis captures all artefacts produced for assessing testing activities, e.g., P30 [30] provides a Java prototype that analyses and reports unlinked items between test cases and requirements.

*5.4.2 Testing Formats.* This section presents the results of the testing format based on our defined categorization. The detailed results are shown in Table 12. From the results, we found that the adopted testing formats evenly distribute in these categories, where structural textual format (32 studies) is still the top choice in test specification, followed by other format (20 studies), tabular format (20 studies), not specified (20 studies), model-based format (18 studies), code-based format (13 studies), logic-based format (8 studies), standardized/specialized format (7 studies).

Besides, we also exhibit the joint distribution between testing stages and testing formats in Fig. 12, which aims to show the format preferences in different testing stages. From the results, we noticed that, (1) designing stage is the majority for most of the testing format except of code-based format; (2) planning stage primarily relies abstract testing artifacts, including structural textual, tabular, and model-based formats; (3) execution stage evenly distributed in these testing formats, where code-based testing artifacts serves the majority.

Table 32. Test Coverage of REDAST Studies (RQ3)

| Test Coverages | Paper IDs | Num. |
|---|---|---|
| Requirements Coverage | [2, 16, 22, 26, 33, 38, 39, 41, 42, 45, 49, 60, 63, 71, 72, 74, 75, 81, 83, 88, 90, 92, 94, 98, 100–102, 104, 105, 108, 109, 111, 113, 114, 116, 118–120, 122, 128, 131, 133, 136, 140, 141, 148, 150, 151, 153, 154, 156], [9, 10, 15, 19, 43, 47, 76, 82, 84] (Use Case Coverage) | 61 |
| Behavioral/Scenario Coverage | [1, 4, 6, 7, 11, 24, 32, 37, 46, 51–54, 56–58, 62, 64, 66–68, 70, 73, 77, 78, 86, 89, 93, 107, 110, 124, 125, 139, 143, 145, 146, 149, 152, 155] | 37 |
| Path Coverage | [5, 8, 12, 14, 17, 30, 35, 55, 59, 65, 85, 99, 106, 130, 138, 147] | 16 |
| Function Coverage | [13, 25, 27, 29, 61, 91, 103, 115, 117, 126, 129] | 11 |
| Not Specified | [21, 34, 48, 80, 87, 121, 132, 134, 144] | 9 |
| Decision/Branch Coverage | [50, 96, 97, 127, 135, 142] | 6 |
| Statement Coverage | [23, 28, 36, 123, 137] | 5 |
| Other | [112] (Boundary Coverage), [44] (Combinatorial Coverage), [18] (Structural Coverage) | 3 |
| Model-Based Coverage | [20, 69, 79] | 3 |
| Rule-Based Coverage | [3, 31] | 2 |
| Code Coverage | [40, 95] | 2 |

Table 33. Distribution of Testing Stages by Requirements Notation

| Testing Stage | CNLs | DSLs | NL Requirements | SCR | UML |
|---|---|---|---|---|---|
| Planning Stage | 0 | 3 | 5 | 0 | 5 |
| Designing Stage | 20 | 22 | 41 | 2 | 42 |
| Execution Stage | 5 | 10 | 6 | 0 | 5 |
| Reporting Stage | 0 | 0 | 0 | 1 | 2 |

*5.4.3 Test Coverage Methods.* Test coverage is defined as whether our test cases cover the testing objectives [108]. However, there are various methods available for this purpose. We design our categorization method for the coverage types in the selected studies and finalize our results in Table 32.

We found that requirements coverage is the most commonly used method in REDAST studies, followed by Behavioral/Scenario Coverage, Path Coverage, Functional Coverage, and so on. From the general point of view, code coverage is the most commonly adopted coverage method in automated test generation. The code coverage can simply and effectively assess the quality of generated test artifacts and measure which code are being covered in the generation [83, 84]. However, the requirements coverage is leading the trend of test coverage in REDAST, where, moreover, the behavioral/scenario, use case, statement, and decision/branch coverage are also able to be classified into requirements coverage. We believe that the requirements can serve a similar role to code. By referring to different parts of requirements, such as behavior, path, or statement, the test artifacts are also able to be assessed and measured on the requirement level.

*5.4.4 RQ3 Cross-Analysis between Requirements Notations with Testing Stages.* The cross-analysis in Table 33 shows that only UML spans every phase of test generation—from high-level planning through design and execution to final reporting—underscoring its versatility as both a human-readable and machine-actionable blueprint. CNLs and SCR, by contrast, do not appear in the planning phase, and SCR also fails to surface in execution, suggesting that neither constrained text templates nor fully formal specifications yet offer the right balance of abstraction and detail for these early and middle stages. Meanwhile, CNLs, DSLs, and NL requirements never support reporting, indicating a gap in generating the structured outputs (reports, analysis summaries) that stakeholders need to close the loop on testing.

Table 34. Joint Distribution of Requirements Abstraction Level and Test Artifacts

| Abstraction Level | Planning | Designing | | | | | Execution | | | Reporting | |
|---|---|---|---|---|---|---|---|---|---|---|---|
| | Test Plan | Test Case | Test Scenario | Test Case Pri. | Test Model | Test Vector | Test Script | Test Oracle | Test Data | Test Analysis | Test Report |
| **Function (3rd)** | 6 | 65 | 18 | 10 | 1 | 4 | 8 | 4 | 3 | 1 | 0 |
| **Feature (2nd)** | 8 | 27 | 5 | 4 | 2 | 1 | 9 | 1 | 0 | 1 | 1 |
| **Component (4th)** | 0 | 10 | 3 | 0 | 2 | 1 | 2 | 1 | 0 | 1 | 0 |
| **Product (1st)** | 0 | 4 | 0 | 0 | 0 | 0 | 0 | 1 | 0 | 0 | 0 |

Table 35. Joint Distribution of Testing Stages and Generation Techniques

| Testing Stage | Graph-based | ML-based | Other | Rule-based | Tool-Supported |
|---|---|---|---|---|---|
| Designing Stage | 34 | 9 | 1 | 43 | 56 |
| Execution Stage | 2 | 0 | 1 | 11 | 14 |
| Planning Stage | 0 | 1 | 0 | 7 | 5 |
| Reporting Stage | 0 | 0 | 0 | 0 | 2 |

Although only a handful of studies tackle planning at all, the inability of CNLs to capture high-level test objectives points to their limited expressiveness for specifying scope, risk criteria, and acceptance conditions.

*5.4.5 RQ3 Cross-Analysis between Requirements Abstraction Levels and Testing Notations.* While the test stages also reflect the abstraction levels, we hope to identify the connection between the requirements abstraction levels and the test artefacts. Feature-level and Function-level requirements produce the test artefacts in the planning stage. In the design and execution stages, the four different stages don't show a clear association. No strong relationship is visible at the stage level, where the stages look broadly similar across abstraction levels. We think these results show that the requirements, after the decomposition, augmentation, and reconstruction in the transformation process, are already converted into different representations for software testing.

*5.4.6 RQ3 Cross-Analysis between Generation Techniques and Testing Stages.* Table 35 confirms that the lion's share of generation effort is devoted to the designing stage. Tool-supported frameworks and rule-based engines together account for the vast majority of studies, leveraging their mature template libraries and pattern-matching capabilities to turn requirements into structured test cases and models. Graph-based generators—designed to traverse or synthesize state-space representations, which contribute exclusively at the design level, where their dependency graphs naturally align with scenario planning. Machine-learning approaches, though still modest in number, follow a similar trajectory. By learning to map textual requirements to mid-level artifacts, they sidestep the combinatorial complexity of lower-level code generation and the open-ended nature of high-level planning. By contrast, execution-stage generation remains underserved. Although tool-supported and rule-based techniques maintain some presence, graph and ML methods are nearly absent, suggesting that challenges such as environment configuration, API binding, and script orchestration resist purely declarative or learned mapping. Similarly, few frameworks attempt planning-stage synthesis—perhaps because turning high-level coverage goals into cohesive test strategies demands semantic abstraction and domain knowledge that most pattern-based engines do not capture. Reporting is even rarer, with only a couple of tool-supported solutions producing formatted summaries or coverage metrics.

*5.4.7 RQ3 Cross-Analysis between Requirements Input and Testing Artifacts.* Table 36 reveals important insights into the interplay between requirements formats and the generation of diverse test artifacts in REDAST literature. The cross-analysis demonstrates that textual requirements are not only the most prevalent format but also the most versatile, supporting the creation of a broad range of test artifacts including test cases, test scenarios, test plans, and prioritization artifacts. This widespread adoption is likely due to the accessibility of textual requirements in industry, their ease of use in early development phases, and the significant tool support for processing NL artifacts. Model-based requirements follow as the next most influential format, prominently contributing to the automated derivation of test cases and scenarios, which aligns with the strengths of model-driven engineering in bridging high-level requirements and executable tests. Interestingly, even though Formal, Tabular, and Other requirements formats are represented, their impact is concentrated in niche artifact types or limited to a few studies. The table also shows that test cases and scenarios are generated from a wide array of formats, reflecting the centrality of these artifacts in the software testing process and the diversity of input formats that can feed their creation. We note that the total number of artifact–format pairs in the table exceeds the number of primary studies included in the review. This is because many studies leverage multiple requirements formats. For instance, P93 [93] covers the generation of test reports, but has two different types of underlying requirements formats, i.e., both model-based (process model) and textual scenarios as input.

Table 36. Joint Distribution of Testing Artifacts and Requirements Format

| Testing Artifact | Formal | Model-Based | Other | Tabular | Textual |
|---|---|---|---|---|---|
| Test Plan | 0 | 4 | 0 | 1 | 11 |
| Test Case | 12 | 24 | 2 | 3 | 57 |
| Test Scenario | 1 | 13 | 0 | 1 | 15 |
| Test Case Prioritization | 0 | 2 | 5 | 0 | 7 |
| Test Model | 1 | 3 | 0 | 0 | 3 |
| Test Vector | 0 | 0 | 0 | 0 | 4 |
| Test Script | 3 | 9 | 0 | 1 | 9 |
| Test Oracle | 3 | 0 | 0 | 0 | 3 |
| Test Data / Bench | 1 | 0 | 0 | 0 | 3 |
| Test Analysis | 1 | 2 | 0 | 1 | 0 |
| Test Report | 0 | 1 | 0 | 0 | 1 |

In this, we also performed an analysis on the top triplets of requirements input, generation technique and the resultant test artifact. While it is difficult to present a detailed analysis here due to 80+ unique such combinations, the top 5 combinations which cover more than one-third of total combinations are all for generating test cases. Textual → Tool-Supported → Test Case (21), Model-Based → Tool-Supported → Test Case (14), Textual → Graph / Search-based → Test Case (13), Textual → Rule-based → Test Case (12) Formal → Tool-Supported → Test Case (6). The most frequent REDAST triplets reveal that test case generation from textual requirements via tool-supported techniques is the dominant pipeline. This pattern highlights the practical preference for natural language inputs and the maturity of supporting tools capable of parsing and transforming them. Graph/search-based and rule-based methods also play a notable role, especially when handling textual inputs, indicating methodological diversity in the mid-pipeline. While model-based and formal requirements appear less frequently, their consistent pairing with tool-supported approaches reflects their structured nature and alignment with formal toolchains.

*5.4.8 Findings: Trend of Test Notations, and Coverage Methods Over the Years.* The test formats by years in Fig. 13 shows an early focus on logic-based artifacts (1992–93), followed by the arrival of model-based and tabular formats in 1998–99. Code-based and structural-textual outputs then peaked around 2008–11—coinciding with the first standardized schemas—before the field shifted its emphasis to textual and tabular representations between 2012–15. From 2018 onward, structural-textual formats have dominated while code-based, logic-based, and tabular artifacts recede, signalling a lasting preference for formats that balance automated rigor with human-readable clarity.

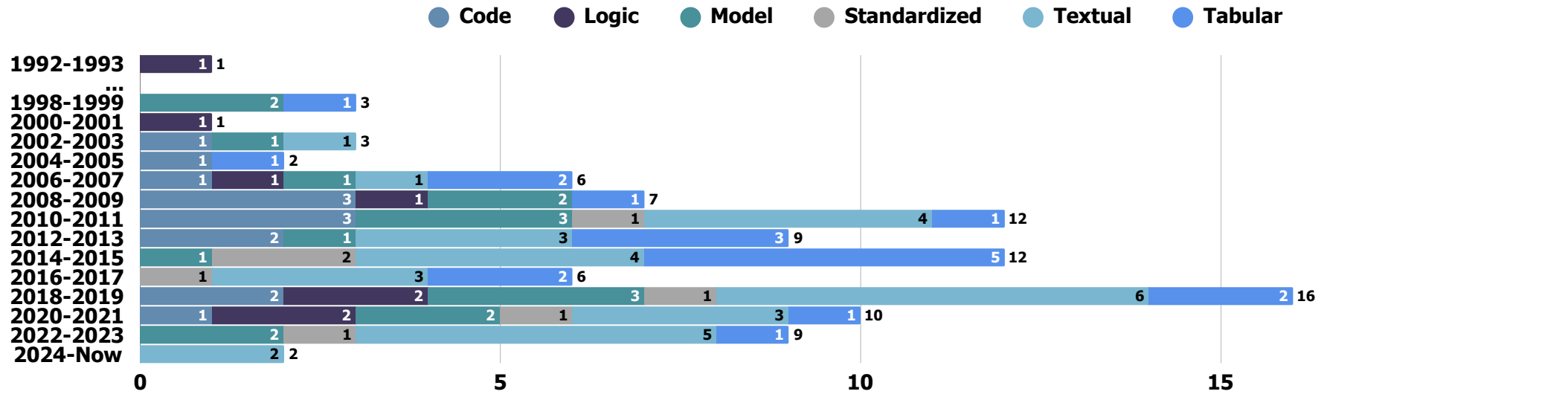

Fig. 13. Trend of Test Formats by Years

**RQ3 Key Takeaways**

• A large majority of the studies focus on concrete test cases as the outcome, followed by test scenarios. This leaves a significant gap in fully automating the testing lifecycle, especially for upfront strategy (linking requirements to high-level goals/risks) and closing the loop through automated reporting and traceability.
• Requirements coverage is the leading coverage criterion in REDAST, encompassing functional and behavioral coverage, followed by path coverage.
• Test oracles, which determine correctness and verdicts, remain very challenging to automate, as highlighted by their rare occurrence and frequent reliance on human judgment. Similarly, the automated generation of test artifacts for NFRs is underexplored.
• Textual requirements are the most versatile input, supporting the widest variety of artifact types, followed by model-based requirements which are particularly well-suited to structured test case and scenario generation.

### 5.5 RQ4: Evaluation Methods of REDAST Studies

REDAST studies aim to generate reliable and usable test artifacts, from requirements to test artifacts. However, the verification of the efficacy of the generated test artifacts is still a challenge. In this section, we plan to illustrate the related results of the selected studies to systematically discuss the evaluation methodologies for test artifacts. The results in this section consist of three parts: (1) types of evaluation methods, (2) target software platforms in the case studies, and (3) the usability of the studies.

Table 37. Benchmark Type Results of Selected Studies (RQ4)

| Benchmark Types | Paper IDs | Num. |
|---|---|---|
| Conceptual Case Demonstration | [1, 2, 4, 6–8, 10–14, 16, 17, 21, 22, 26, 27, 35, 43, 46, 48, 49, 51, 53, 62, 64, 66–69, 71, 74, 75, 77, 81, 83, 84, 86, 87, 89, 94–96, 101, 102, 107, 108, 110, 111, 115, 118, 119, 123, 127–129, 131, 136, 139, 140, 144–148, 150, 151, 154, 156] | 69 |
| Real Case Demonstration | [3, 5, 9, 12, 15, 18, 20, 23–25, 28–30, 32–34, 36–38, 42, 44, 45, 49, 50, 54, 55, 58–61, 72, 78–80, 82, 85, 88, 91, 92, 97, 98, 103–106, 109, 113, 114, 116, 118, 120–122, 124–127, 129, 130, 132, 133, 135, 137, 142, 149, 155] | 66 |
| NA | [19, 31, 39, 40, 47, 52, 57, 63, 65, 70, 73, 76, 90, 93, 99, 100, 112, 117, 134, 138, 141, 143, 153] | 23 |
| Dataset Evaluation | [18, 41, 56, 152] | 4 |

*5.5.1 Types of Evaluation Methods.* In the selected studies, we identified two primary evaluation methods: (1) case studies and (2) evaluations using given datasets. However, some studies rely on conceptually designed case studies rather than industrial demonstrations based on real-world cases. To provide a clearer classification, we restructured the categories for evaluation methods into three types: conceptual case study, industrial case study, and evaluation using given datasets. The summarized results are presented in Table 37.

- **Conceptual Case Demonstration**. These studies typically begin by designing a conceptual case and then applying REDAST methods to the designed case to demonstrate the test generation process. Commonly used conceptual cases include online systems (e.g., P11 [11], P21 [21], P119 [119]), ATMs (e.g., P12 [12], P13 [13], P35 [35]), and library systems (e.g., P6 [6]), among others. While these demonstrations effectively illustrate the methodology procedure, they often lack compulsion and persuasiveness in demonstrating the efficacy of the results. However, using conceptual cases, which are based on widely familiar scenarios, enhances the understandability of the underlying methodologies.
- **Industrial Case Demonstration**. Studies in this category demonstrate their methods by incorporating industrial cases into their experiments. By re-organizing and utilizing data extracted from industrial scenarios, REDAST methods are validated under real-world conditions, offering stronger evidence and greater persuasiveness compared to conceptual case studies. Additionally, we observed that some studies intersect across categories, combining (i) both conceptual and industrial case studies, and (ii) industrial case studies with dataset evaluations (e.g., P28 [28] on several industrial cases from public paper, P155 [155] on postal systems). These overlaps occur because industrial cases not only serve as a basis for case studies but can also be used to formulate evaluation datasets, further enhancing their utility in validating methodologies.
- **Evaluation on Given Datasets**. Evaluations conducted on public or industrial datasets provide a compelling approach to demonstrating the efficacy and usability of a method. However, transitioning from discrete cases to datasets requires significant additional effort, including data cleaning, reorganization, and formulation. This challenge arises due to two primary factors: (1) the high usability requirements of REDAST studies, which are often difficult to demonstrate effectively using certain datasets, and (2) the absence of a standardized benchmark dataset for evaluating test artifacts. As a result, most studies rely on case demonstrations rather than systematic evaluations with well-formulated datasets. Only a small number of studies (e.g., P18 [18], P39 [39], P41 [41]) employ dataset evaluations within the selected works.

The results indicate that conceptual case studies are the most commonly adopted method in the selected studies, with 69 instances, followed by industrial case studies with 66 instances, and dataset evaluations with 4 instances. Additionally, the “NA” category includes studies that do not provide any solid demonstration methods.

*5.5.2 Software Platforms in Case Demonstration.* Recognizing that case studies are the primary method of demonstration in REDAST studies, we further analyzed the details of these case studies by categorizing the software platforms used in their demonstrations (Table 38). The adopted software platforms are grouped into 17 categories, including shopping systems, resource management systems, financial systems, and more. These categories encompass the most common usage scenarios and are intended to guide the design of case studies in future research.

The results reveal that, alongside domain-specific systems such as control and automotive systems, order processing systems and ATMs are frequently selected to demonstrate test generation methodologies. These systems are often chosen because they are widely recognized and understood in the public domain, making them effective tools for illustrating methodologies in a comprehensible manner.

Table 38. Software Platforms in Case Study of Selected Studies (RQ4)

| Software Platforms | Paper IDs | Num. |
|---|---|---|
| NA | [19, 31, 39, 40, 47, 52, 57, 63, 65, 70, 73, 76, 90, 93, 99, 100, 112, 117, 134, 138, 141, 143, 153] | 23 |
| Automotive System | [5, 23, 24, 32, 45, 58, 59, 61, 79, 80, 94, 98, 103, 104, 111, 122–124, 149, 150] | 20 |
| Control System | [27, 29, 30, 42, 44, 60, 64, 78, 91, 107, 109, 116, 120, 127, 135, 137, 139, 156] | 18 |
| Safety System | [49, 50, 54, 71, 72, 78, 85, 86, 98, 104, 113, 121, 126, 127, 130, 142, 149] | 17 |
| Order System | [4, 10, 12, 16, 17, 48, 53, 55, 66–68, 110, 127–129, 147] | 16 |
| Business System | [18, 20, 23, 36, 37, 68, 77, 87–89, 119, 155] | 12 |
| Resource Management System | [5, 25, 26, 33, 34, 68, 81, 114, 116, 121, 131, 152] | 12 |
| Workforce System | [8, 22, 23, 28, 41, 46, 74, 75, 84, 92, 95, 140] | 12 |
| Healthcare System | [15, 17, 34, 38, 41, 51, 58, 82, 106, 115, 154] | 11 |
| ATM System | [12, 13, 17, 22, 35, 101, 105, 146, 148, 150] | 10 |
| Aerospace System | [28, 35, 69, 80, 96, 97, 125, 144, 151] | 9 |
| Banking System | [5, 21, 43, 68, 116] | 5 |
| Authentication System | [1, 83, 102, 108] | 4 |
| Mobile Application | [2, 34, 62, 68] | 4 |
| Library System | [6, 7, 9] | 3 |
| Booking System | [55, 56, 145] | 3 |
| Education System | [11, 118, 154] | 3 |
| Other | [132, 133, 136] | 3 |
| Database System | [3, 113] | 2 |
| Examination System | [14] | 1 |

Table 39. Usability Results of Selected Studies (RQ4)

| Usability | Paper IDs | Num. |
|---|---|---|
| The methodology explanation | Almost all of the papers enable clear methodology explanation, except [35, 36, 38, 133, 135, 151, 153] | 154 |
| Case example | [1–5, 7–9, 11–18, 21–24, 26–29, 32–44, 46, 48–50, 55, 58–62, 64, 66–69, 71, 78, 79, 82–95, 97, 98, 103–116, 118–122, 124–133, 135, 136, 138–147, 149–151, 153–156] | 120 |
| Discussion | [1, 3–13, 15, 17–29, 31–33, 35–38, 40–52, 54–56, 58–73, 75–85, 89, 91, 93, 95, 100, 109, 113, 116, 122–127, 129, 130, 132, 136, 139, 142, 147, 149–156] | 105 |
| Experiment | [2, 3, 5, 7, 9, 12–15, 17, 18, 22–24, 26–28, 30, 32, 34, 35, 37, 39–45, 52, 53, 56, 58, 59, 61, 70–72, 76, 79–82, 84, 86, 88, 92, 95, 97, 98, 103, 104, 106, 109, 113, 114, 116, 120, 122–130, 132, 133, 135–137, 149, 152, 154–156] | 77 |

*5.5.3 Examples of Evaluation or Experiments.* A critical aspect of any REDAST approach is the level of detail provided by the papers on the methodology and the evaluation of the approach. In this subsection, we classified the selected papers by separately analyzing the clarity and explanation provided in four key sections: (1) methodology explanation, (2) case examples, (3) experiments, and (4) discussions. Table 39 shows an overview of different evaluation and methodology sections covered in the selected studies. For example, the second row (Case example) lists papers that cover the description of case examples in some level of detail, which are also exemplified below for papers P159 and P24.

*Case 1: P155 - Generating Test Scenarios from NL Requirements using Retrieval-Augmented LLMs: An Industrial Study.* P155 [155] introduces an LLM-driven test scenario generation method. While LLMs offer flexible and powerful natural language generation capabilities, their limited controllability poses challenges for broader applications in REDAST studies. To address this issue and validate the methodology's reliability, P155 incorporates a case study based on an industrial usage scenario. The experiment evaluates performance using both quantitative metrics and qualitative human assessments. Quantitative metrics include standard machine translation evaluation measures such as BLEU, ROUGE, and

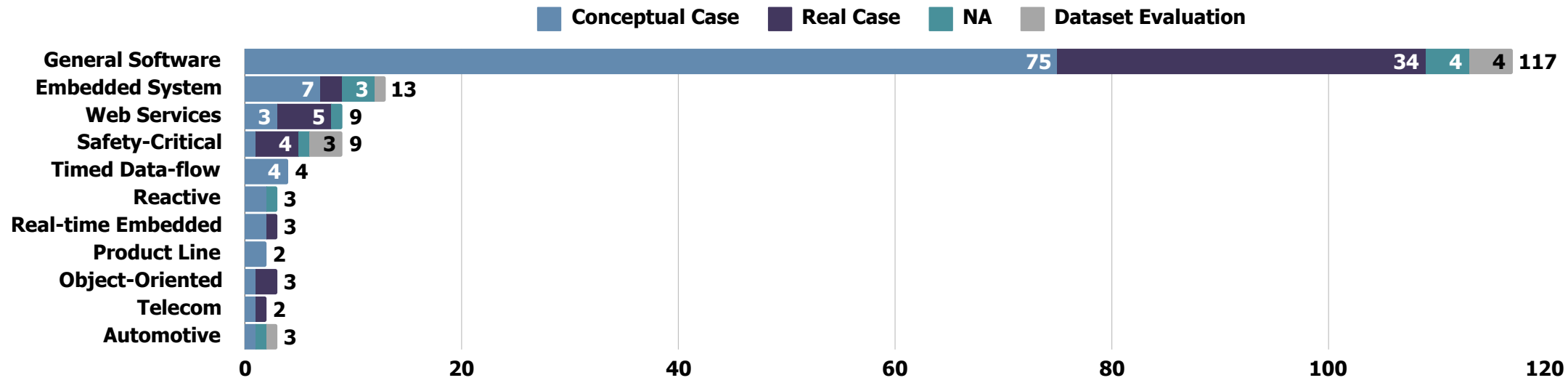

Fig. 14. Cross-Distribution of Demonstration Types and Target Software Systems

METEOR. However, the core of the evaluation is a human interview with software engineers, offering deeper insights into the practical utility of the methodology. While metrics provide a partial perspective from an NLP standpoint, the human interviews emphasize the method's effectiveness and applicability in real-world scenarios.

*Case 2: P24 - Automatic Generation of Acceptance Test Cases from Use Case Specifications: an NLP-based Approach.* P24 [24] proposes an NLP-based framework for generating executable test cases. The industrial case-based experiments address three research questions, including evaluations of correctness and manual comparisons. Notably, in the manual comparison, instead of conducting direct human interviews for the generated test artifacts, the study employs a manual comparison with test cases created by domain experts. The results of these comparisons provide a robust demonstration of the generated test cases, as their alignment with established "gold truth" test cases serves as compelling evidence of the methodology's effectiveness.

*5.5.4 RQ5 Cross-Analysis between Demonstration Types and Target Software Systems.* While researchers demonstrate the efficiency of their methods through evaluations or practical demonstrations, the target software system ultimately reflects the primary objective of REDAST studies. Our goal is to establish a connection between these demonstrations and the target software system to provide deeper insights into the preference for demonstration methods across different usage scenarios. From our findings in Fig. 14, we observed that conceptual case studies remain the most commonly adopted demonstration method. In contrast, real (industrial) case studies are generally more challenging to conduct due to the difficulty of obtaining real-world data. However, in domains such as web services, safety-critical systems, and objective-oriented systems, real case studies are predominantly chosen. These critical systems often impose stricter pass-rate criteria for test outcomes, necessitating more rigorous validation. We suggest that future researchers pay particular attention to the demonstration aspect when dealing with such systems.

**RQ4 Key Takeaways**

- Conceptual case studies are the most common evaluation method in REDAST studies. While these (example) case studies enhance understandability, they often lack strong empirical validation compared to real-world industrial cases. Industry case study evaluations are also prominent in ≈40% studies, demonstrating the real-world applicability and need for REDAST approaches.
- There are very few studies conducted on any public datasets. This showcases a gap in the standardized evaluation of REDAST approaches due to the lack of standard benchmarks.

Table 40. Limitation Results of Selected Studies (RQ5)

| Limitations | Paper IDs | Num. |
|---|---|---|
| Framework-Design Limitation | [4, 8, 11, 12, 21–23, 28, 31–33, 35, 41, 44–46, 60, 62, 64, 67, 68, 84, 86, 88, 89, 98, 102, 105, 106, 113, 114, 116, 121, 126, 140, 141, 143, 145, 146, 151, 155] | 41 |
| NA | [14–16, 18, 26, 29, 82, 87, 90, 92, 94, 95, 97, 110, 115, 118–120, 128, 129, 132–137, 153, 156] | 28 |
| Evaluation/Demonstration Limitation | [7, 19, 39, 40, 48, 74, 75, 93, 96, 98, 101, 104, 106, 117, 124, 127, 138, 141, 148, 155] | 20 |
| Scalability Limitation | [10, 30, 32, 35, 38, 47, 50, 55, 56, 66, 70, 75, 78, 80, 83, 103, 107, 121, 123, 154] | 20 |
| Input-Quality Over-Dependence | [2, 17, 34, 36, 37, 43, 44, 51, 53, 55, 65, 71, 103, 122, 139, 152] | 16 |
| Methodology Over-Complexity | [3, 54, 62, 69, 72, 73, 76, 77, 79, 81, 91, 112, 142, 144, 149] | 15 |
| Automation Gaps (Methodology) | [6, 8, 24, 27, 38, 57–59, 103, 108, 111, 112] | 12 |
| Incomplete Requirements Coverage | [13, 61, 85, 101, 104, 109, 126, 131, 150, 151] | 10 |
| Automation Gaps (Test Implementation) | [2, 5, 25, 73, 100, 117, 125, 130] | 8 |
| Automation Gaps (Requirements Specification) | [1, 9, 68, 99, 113, 146, 147] | 7 |
| Requirements Ambiguity | [42, 49, 52] | 3 |
| Implementation Limitation | [20, 63] | 2 |
| Time Cost | [20, 39] | 2 |
| Additional Specification Cost | [7] | 1 |

Table 41. Automation Level of Selected Studies (RQ5)

| Automation Levels | Paper IDs | Num. |
|---|---|---|
| Highly Automated | [2–10, 12, 16–18, 20, 22–24, 28, 29, 32, 41–44, 47–49, 51, 53, 54, 56, 57, 59, 60, 69–71, 74, 75, 77, 78, 81, 82, 86, 88–90, 94, 96–99, 101–105, 107, 108, 113, 119–122, 127–129, 131, 132, 135, 140, 141, 144, 146, 148, 149, 155] | 77 |
| Semi-Automated | [19, 26, 27, 38, 46, 58, 63, 64, 68, 83, 84, 87, 91, 92, 95, 100, 106, 109–112, 114–117, 124–126, 130, 133, 134, 136, 137, 143, 145, 147, 150, 151, 156] | 39 |
| Fully Automated | [1, 11, 13–15, 21, 25, 30, 31, 33–37, 39, 40, 45, 50, 52, 55, 61, 62, 65–67, 72, 73, 76, 79, 80, 123, 138, 139, 142, 152] | 35 |
| Low Automation | [85, 93, 118, 153, 154] | 5 |

### 5.6 RQ5: Limitation, Challenging and Future of REDAST Studies

*5.6.1 Limitations in Selected Studies.* In the selected papers, we observed that some studies explicitly discuss their limitations. This section presents the identified limitations, categorized into 14 concise types, as summarized in Table 40. Notably, 28 studies do not explicitly mention any limitations in their content. Consequently, the following discussion primarily focuses on the remaining papers that explicitly acknowledge their limitations.

Automation is a frequently mentioned limitation in the selected studies. We identified that automation limitations vary across three key areas: requirements specification, methodology conduction, and test implementation. To provide a more detailed analysis, we classified these limitations into three categories: Automation (Requirements Specification), Automation (Methodology), and Automation (Test Implementation). Given that the automation level is a critical aspect of REDAST, we present the estimated automation levels for all the selected papers in Table 41, as four levels, Fully automated - End-to-End Automation, highly automated - Automation-Dominant, semi-automated - Automation Supported, and low automated - Minimal Automation. The detailed definitions of automation are depicted in Section 4.6.

Scalability, framework design, incomplete requirements coverage, and requirements ambiguities are commonly identified limitations in the reviewed studies. Specifically, 41 studies reported incompatibilities with handling certain scenarios, 20 studies highlighted challenges in scaling to complex or larger systems, 10 studies acknowledged an inability to cover all requirements during test generation, and 3 studies discussed potential ambiguities in requirements. These limitations often stem from framework structure issues, where the methods fail to comprehensively account for diverse usage scenarios, leading to problems with generalization and applicability. For example, P38 [38] and P48 [48] discuss challenges with generalization due to difficulties in handling complex systems, while P124 [124] highlights performance gaps when dealing with systems of varying sizes. Framework design limitations also constrain methods in specific contexts, as evidenced by P146's [146] discussion of incompatibilities with non-functional requirements and P141's [141] primary focus on extra-functional properties.

Table 42. Future Direction Results of Selected Studies (RQ5)

| Future Directions | Paper IDs | Num. |
|---|---|---|
| Coverage/Requirement Extension | [5, 11, 13–18, 21–23, 28, 30, 32, 34–36, 42, 44–51, 53–55, 57–62, 65–68, 73, 75, 78, 84, 89, 92, 94, 96–104, 106, 124, 125, 127–129, 132, 140–143, 145–147, 153–155] | 72 |
| Further Validation | [5, 6, 9, 14, 15, 19–21, 26, 38, 48, 54, 59, 72, 74, 75, 84, 86, 90, 93, 97, 104–106, 109, 111, 114–116, 120–122, 133, 135, 137–141, 151, 154, 155] | 42 |
| Completeness Improvement | [24, 35, 36, 40, 45, 51, 52, 69, 71, 79, 80, 86, 89, 90, 92, 93, 108, 113, 115, 120, 121, 123, 124, 126, 129, 133, 135, 137, 144, 149, 150, 152] | 32 |
| Automation Improvement | [4–6, 10–12, 46, 53, 56, 58, 67, 72, 73, 76, 85, 92, 100, 102, 111, 134, 139, 146, 147, 149] | 24 |
| Extension of Other Techniques | [9, 19, 20, 22, 27, 30, 31, 33, 34, 37, 43, 44, 63, 70, 71, 76, 77, 85, 91, 122, 123, 142, 148, 152] | 24 |
| Extension to Other Domains or Systems | [8, 10, 24, 26, 28, 29, 33, 41, 42, 47, 56, 61, 63, 64, 66, 69, 73, 77, 79, 80, 96, 98, 103, 130] | 24 |
| Extension to Other Phases or Test Patterns | [1, 2, 7, 12, 16, 21, 25, 26, 37, 51, 52, 57, 62, 65, 74, 94, 101, 102, 107, 108, 110, 132, 148, 151] | 24 |
| Performance Improvement | [4, 27, 42, 52, 91, 102, 103, 125, 127, 128, 133, 138, 144, 150] | 14 |
| Real Tool Development | [1, 7, 13, 54, 56, 64, 78, 106, 108–110, 112, 113, 135] | 14 |
| Benchmark Construction | [3, 39, 112, 134] | 4 |
| Traceability Improvement | [31, 39, 116, 132] | 4 |
| Robustness Improvement | [18, 114, 136] | 3 |
| NA | [81–83, 87, 88, 95, 117–119, 131, 156] | 11 |

Additional limitations arise from the complexity of some framework components, including over-reliance on input quality, methodological complexity, and time inefficiencies. Over-reliance occurs when predefined rules or input formats disproportionately influence the performance of test generation, making the process vulnerable to input quality issues. For instance, P44 [44] notes that dependency relations can affect test generation accuracy, while P103 [103] highlights challenges arising from the conjunctive statement format, which complicates stable test generation. Methodological complexity and time-cost issues stem from the algorithms employed in these frameworks, which significantly increase the difficulty and runtime of the processes. For example, P79 [79] mentions that the complexity of the test generation and analysis processes impacts performance and scalability, and P20 [20] critiques the Specmate technique for its excessive complexity, which undermines runtime efficiency.

Furthermore, 20 studies identified limitations in evaluation or demonstration, emphasizing that their experiments were insufficient to validate the efficacy of methodologies in other scenarios, particularly from an industrial perspective (e.g., P7 [7], P19 [19], P74 [74]).

*5.6.2 Insight Future View from Selected Studies.* This section aims to discuss the key challenges and directions for future research identified in the selected studies. Notably, these studies share common themes regarding challenges and proposed future work, including improving or extending existing methodologies, conducting further evaluations, and addressing unresolved issues. To provide a structured analysis, the challenges and future work are categorized into several different areas. We illustrate the results in Table 42.

Extensions to other coverage criteria or requirements, test phases or patterns, and domains or systems are the three most commonly identified future directions and challenges in the reviewed studies. These directions correspond to limitations in framework design, as most frameworks are unable to address all usage scenarios, such as requirements coverage, test phases, or diverse software systems. Consequently, many studies propose expanding their scope to cover additional scenarios. For instance, P67 [67] plans to emphasize non-functional requirements in their test generation process, rather than focusing solely on functional requirements. P65 [65], having designed a method for generating test scenarios, intends to broaden their approach to encompass other phases of software development or testing. Similarly, P25 [25], which focuses on acceptance test case generation for embedded systems, plans to extend their application beyond embedded systems to enhance generalization.

Although limitations in automation were discussed in the previous section, several papers propose future plans to reduce human intervention in the test generation process. We identified 24 papers outlining plans to improve automation levels (e.g., P67 [67], P72 [72], among others).

Robustness, completeness, and traceability are widely regarded as critical factors in automated software testing research. While it is challenging to qualitatively assess these factors for a REDAST method, certain steps can be identified that may negatively impact framework robustness, completeness, or traceability. In the selected papers, some studies explicitly discuss their future directions for improving these aspects. For example, P114 [114] plans to enhance fault-handling diversity to strengthen framework robustness. P93 [93] aims to add a user input monitoring function as part of future work to improve completeness. Meanwhile, P116 [116] intends to further investigate the impact of requirement changes within the REDAST process, enabling better traceability between requirements and test artifacts.

Benchmark construction, real tool development, and further validation reflect researchers' efforts to broaden the impact of their studies. By enhancing the post-generation environment, the potential capabilities of these studies can be more thoroughly explored. For instance, P3 [3] plans to build benchmark test suites that are independent of specific model-based testing languages, which could significantly advance future research in related fields. P56 [56] aims to investigate opportunities for integrating their approach into larger industrial applications to enhance work efficiency. Additionally, recognizing that their current case study is insufficient to fully demonstrate the methodology's efficacy, P84 [84] intends to conduct more comprehensive case studies to further evaluate the performance of their method.

Some studies discuss the extension of existing techniques or performance improvements in their work, often focusing on introducing new techniques or frameworks to enhance test generation performance. For example, P19 [19] plans to expand NLP capabilities and explore more advanced automation techniques as part of their future work. Similarly, P91 [91] intends to reduce the state space of the studied model to improve the efficiency and effectiveness of their test case generation methods.

In the result, we identified that most of the studies pose an extension to other coverage criteria or other requirements (72 papers), followed by further validation (42 papers), improvement of the completeness in their future work (32 papers), and so on, which matches the results in the limitation results.

Table 43. Joint Distribution of Demonstration Types and Validation Related Future Direction or Limitation

| | Conceptual Case Study | Real Case Study | NA | Dataset Evaluation |
|---|---|---|---|---|
| **Further Validation** | 15 | 22 | 5 | 0 |
| **Limitation of Demonstration or Evaluation** | 14 | 13 | 3 | 1 |

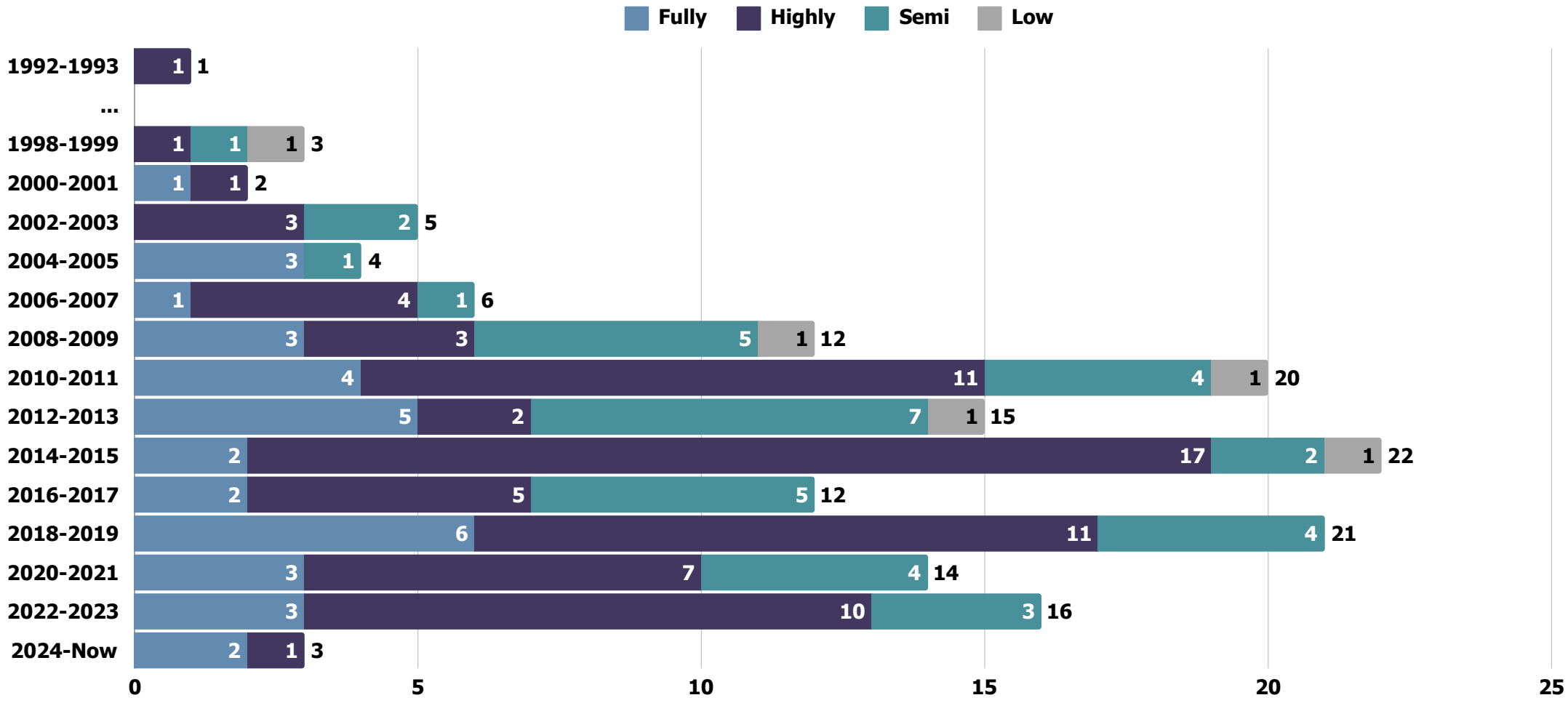

Fig. 15. Automation Level by Year

*5.6.3 RQ5 Cross-Analysis between Demonstration Types and Validation Challenges.* In analyzing the limitations and future directions of existing studies, we found that many studies identify evaluation or demonstration constraints and emphasize the need for further validation. This finding highlights the distribution of different demonstration types among studies that acknowledge validation-related limitations and future research directions. From our results in Table 43, 15 out of 69 papers that adopted conceptual case studies reported validation-related issues in their limitations or future directions, while 14 out of 69 identified similar concerns. Similarly, 22 out of 66 papers that employed real case studies reported validation-related issues in their limitations, with 13 out of 66 mentioning them in their future research directions. These findings suggest that validation challenges persist regardless of the chosen demonstration method. Additionally, we observed that papers employing real case studies more frequently reported validation-related issues. We hypothesize that this may be because conceptual case studies, being designed specifically for demonstration purposes, can more effectively represent the REDAST process in a controlled manner.

*5.6.4 Findings: Trend of Automation Level Over the Years.* As a practical and impact-driven research area, REDAST places significant emphasis on automation, particularly in the context of industrial applications. Given that technological advancements contribute to automation in REDAST, we analyze the trend of automation levels over time, as illustrated in Fig. 15. For clarity, we classify automation levels into two categories: fully and highly automated systems, which represent a high level of automation, and semi- and low-automated systems, which indicate a lower level of automation. Our findings reveal that while the proportion of low and semi-automated approaches has gradually declined, fully and highly automated methods have become increasingly dominant over the years. This trend suggests that technological advancements have progressively improved the automation level in REDAST studies. However, despite this overall improvement, the proportion of fully automated systems has not increased significantly. We attribute this to a key limiting factor: while technological advancements enhance automation capabilities, achieving full automation still requires an appropriately designed framework. This aspect is largely independent of technological progress and instead relies on methodological and architectural considerations in REDAST research.

**RQ5 Key Takeaways**

• The framework design improvement and extension is the most common limitation and future direction reported in REDAST studies. The selected papers frequently mention better coverage of the usage scenario and model configurations.
• Most REDAST studies are not fully automated, with some human invention still necessary in some key steps.
• Extension of evaluation and demonstration is always considered in the future directions. REDAST studies are strongly related to real applications, and hence, the evaluation and demonstration have a key future direction and need for evaluation in real practical settings.

## 6 Threats to Validity

***Internal Validity.*** In this study, the first author designed the SLR protocol, which was reviewed and refined collaboratively with the second, third, and fourth authors before formal implementation. The detailed topics and search strings were iteratively adjusted and executed across multiple databases to optimize the retrieval of relevant results. To accommodate the varying search policies of these databases, the search strings were customized accordingly. The selection of studies followed a multi-stage filtering process to minimize selection bias. The first round of filtering was based on titles and abstracts. The second round involved brief reading and keyword matching, while the third round consisted of a comprehensive reading of the papers. All authors validated the final selection to ensure robustness. Following study selection, a data extraction process was designed using Google Forms. All authors participated in a pilot test to refine the data extraction procedure and ensure consistency in capturing the necessary information.

***Construct Validity.*** To mitigate threats to construct validity, we conducted the search process across six widely used scientific databases, employing automated and manual search strategies. Extensive discussions among all authors were held to refine the inclusion and exclusion criteria, ensuring they effectively supported selecting the most relevant studies for this SLR. Some of the selected studies included vague descriptions of their methodologies, posing potential threats to the validity of the study. These cases were carefully reviewed and deliberated upon by the first and second authors to reach a consensus on their inclusion.

***Conclusion Validity.*** The threat to conclusion validity was minimized through a carefully planned and validated search and data extraction process. We designed the data extraction form based on the predefined RQs to ensure the extracted data aligned with our study requirements. The first author initially extracted data from a subset of selected papers using this form, after which the extracted data was reviewed and verified by the other authors. Once validated, the first author used the refined form to extract data from the remaining studies. Multiple discussions were conducted during data analysis and synthesis to determine the most effective categorization and representation of the data, ensuring robust and meaningful conclusions.

***External Validity.*** To address the threat to external validity, we employed automated and manual search strategies, adhering to widely accepted guidelines [44, 96]. Our methodology section provides a detailed explanation of the inclusion and exclusion criteria. Specifically, we focused on peer-reviewed academic studies published in English, excluding grey literature, book chapters, opinion pieces, vision papers, and comparison studies. While these criteria may exclude some potentially relevant works, they were implemented to minimize bias in the selection process. We adopted a broad inclusion approach, considering studies regardless of publication quality. Furthermore, our search encompassed publications from 1992 to the present, ensuring comprehensive coverage of advancements in REDAST.

## 7 Roadmap and Future Directions

Based on the results of our predefined RQs, we present recommendations for future research in REDAST.

***Recommendation 1: Standardizing NL Requirements: Bridging Ambiguity and Automation.*** Our analysis confirms that unconstrained NL requirements remain the most prevalent input in REDAST pipelines, yet their inherent ambiguity frequently derails fully automated parsing and planning. While introducing lightweight structural annotations—such as numbered steps, precondition–action–postcondition templates, or form-based fields—can materially reduce this ambiguity, these "structured textual" formats often prove too rigid or verbose to naturally capture high-level test objectives. UML and other model-based schemas, on the other hand, offer the richest expressiveness and enable detailed planning and end-to-end traceability, but their steep learning curve and tooling overhead limit widespread adoption, particularly in agile or cross-disciplinary teams. This persistent tension suggests that no single notation currently achieves a "win-win" balance of authoring ease, semantic precision, and tooling support. An important step towards fixing the diversity of requirements input could be developing standardized CNL templates, carefully balancing human readability with automation capabilities. Comprehensive guidelines and systematic benchmarks to support transitioning from textual to formal representations are also needed. Moreover, some studies adopt hybrid approaches combining textual flexibility with formal rigor, i.e., the RUCM [102] in P5 [5], P22 [22], P24 [24], etc., which could enhance both expressiveness and automated traceability, bridging existing adoption gaps and supporting broader applicability. For example, P22 [22] generates UML state machines and maintains bidirectional links back to each RUCM sentence, where an industrial case study produced a complete diagram in seconds and refined it to test-ready quality with much less effort.

***Recommendation 2: Extending Coverage Beyond Functional Requirements.*** Our analysis indicates that current REDAST methodologies primarily focus on FRs, while NFRs remain inadequately addressed. The systematic generation of test artifacts for NFRs presents distinct challenges due to their qualitative nature and the complexity involved in their formalization. Although the current literature broadly acknowledges the importance of NFRs, specific systematic methodologies or empirical studies targeting their automated testing are sparse. Therefore, future research should prioritize creating detailed taxonomies, standardized templates, and dedicated methodologies specifically designed to capture, formalize, and systematically transform different NFRs into executable test artifacts. Table 44 presents the comprehensive list of the studies that support software testing based on NFR. Typically, P21 [21] focuses on *Performance* perspective, where it adds "response-time" and "max-invocation-count" fields to its meta-model, then automatically converts these attributes into positive/negative event sets for state-based test-path search; P72 [72] especially addresses *Safety* NFR, where it encodes safety NFRs as OCL invariants, mutates them, and synthesises a minimal yet fault-revealing test suite that targets the mutated properties; P81 [81] discusses *Security* NFR by capturing security goals with a misuse-case template and generating attack-oriented test stubs trace-linked to each threat scenario. In conclusion, addressing this gap will ensure software systems comprehensively meet both functional and non-functional stakeholder expectations.

***Recommendation 3: Enhancing Automation Levels.*** Despite recent advances, many REDAST frameworks remain manual-intensive, particularly during the parsing stage. This limitation significantly restricts overall automation efficiency. Leveraging advanced Natural Language Processing (NLP) techniques and Large Language Models (LLMs) can substantially improve parsing automation. Some strides have been made in this context, e.g., the automated test scenario generation using retrieval-augmented generation via LLMs [155]. Researchers are thus encouraged to develop adaptive parsing frameworks capable of handling multiple requirements formats. For example, P23 [23] trains a format-agnostic classifier that detects semi-structured "if–then-else" and business-rule sentences across plain prose, use-case tables, and user stories—eliminating most hand-written parsing rules; P154 [154] clusters over its extractor, so only ambiguous clusters need labeling, greatly cutting manual effort; P123 [123] fine-tune BERT with a handful of seed examples, enabling one parser to handle raw user stories and Gherkin scenarios with high recall. Together, these studies show that adaptive, semi-supervised parsers can tackle multiple requirement formats while sharply reducing human intervention.

***Recommendation 4: Advanced Intermediate Representation Management.*** Most existing REDAST frameworks rely on single, static intermediate representations, significantly limiting flexibility and comprehensive integration capabilities, as covered in Fig. 7. This constraint hinders effective transitions between different stages of automated testing, reducing overall adaptability. Actually, software requirements are not limited to structural or unstructured expressions, but widely include dynamic or interactive expressions, such as natural language, diagrams, tables, UI prototypes, and so on [1, 66]. Establishing multi-modal intermediate representation standards can overcome this limitation, enabling more versatile and effective frameworks. Empirical research should systematically evaluate the comparative effectiveness of graph-based and structural/formal intermediate representations, including measures of coverage, performance differences, and generation complexity across different types of intermediate representations. Furthermore, several studies in the selected papers enable the auto-transformation among multiple intermediate representations. For instance, P9 [9] transforms OWL-S requirements into XML event sets, then into Petri-Net paths, allowing for the automatic transformation from formula-based to graph-based intermediate representations and producing runnable tests after two tool-driven conversions. It confirms the feasibility of different types of intermediate representations, which maintain flexibility and advantage across various formats, suggesting that automated methods for dynamically transforming between representations could provide the adaptability required to manage diverse testing contexts effectively.

***Recommendation 5: Leveraging Contextual and Additional Inputs.*** Most REDAST methodologies currently limit inputs exclusively to requirements specifications, neglecting valuable contextual information available from historical data (P32 [32], P122 [122]), existing source code (P81 [81], P154 [154]), and implementation artifacts (P69 [69], P108 [108]). Incorporating these additional inputs can significantly enhance the quality and effectiveness of generated test artifacts. Systematic integration of historical test execution logs, past bug reports, and source code information has demonstrated considerable potential in improving test generation and prioritization. With the advances in retrieval augmented LLM-based generation techniques [29, 100], which have been typically adopted in P155 [155], future approaches must prioritize the development of context-aware integration mechanisms, guided by empirical studies that clearly outline practical methods and guidelines for leveraging these additional resources effectively.

***Recommendation 6: Comprehensive Evaluation and Benchmarking.*** A prominent gap identified in this review is the lack of standardized public datasets and benchmarks, severely hindering rigorous validation and systematic comparison of REDAST methods. In RQ4, we found that dataset evaluation is rarely adopted due to the limited availability of data resources for pairing requirements with test artifacts, appearing in only 5 out of 156 papers. Developing community-driven standard benchmarks and evaluation frameworks is essential for ensuring comparability, reproducibility, and robust evaluation across studies. Promoting open-access datasets and rigorous industrial validation through comprehensive case studies and extensive field experiments can further enhance reliability and practical relevance. We observed that publicly available datasets for REDAST, even in broader RE domains, are extremely limited [42, 65]. Such standardization efforts will be critical in guiding future research toward measurable and widely accepted outcomes.

***Recommendation 8: Developing Real-World Tools and Ensuring Industry Adoption.*** Limited practical adoption of REDAST methodologies primarily results from the absence of scalable, user-friendly, and industry-grade tooling (RQ4 analysis). This lack of practical tools significantly restricts the transfer of research advancements into industry practice. Future research should prioritize developing robust, user-centric, and extensible tools aligned closely with real-world software engineering practices. Partnerships with industry stakeholders, supported by usability testing and field studies, are critical to ensuring these tools meet practical needs and facilitate widespread adoption. Enhancing tool accessibility and usability through careful design and continuous validation will bridge the existing gap between academic research and industry practice.

***Recommendation 9: Roadmap for Generation Techniques Across Testing Stages.*** Our cross-analysis in RQ3 reveals that current REDAST frameworks excel in design and execution level artifact generation but exhibit limitations in planning and reporting stages. Graph-based generation techniques have proven highly effective for creating richly structured test cases and scenarios from intermediate representations but remain confined to the designing stage. Rule-based and tool-supported approaches dominate across design, execution, and occasionally planning or reporting stages; however, their template-driven nature limits their effectiveness in scenarios requiring high-level abstraction or narrative outputs. ML methods have made initial progress in mapping textual requirements to test artifacts, yet their application remains minimal outside the designing stage. To address these shortcomings, future research should strategically align generation techniques with artifact types, optimizing the capabilities of graph-based planners for scenario and path derivation, leveraging rule-based and tool-supported modules for executable scripts and API integrations, and utilizing ML or LLM-based approaches for high-level planning and reporting artifacts. Developing "generation adapters" or orchestration layers that intelligently select appropriate generation techniques for different artifacts will foster an integrated, cohesive, end-to-end REDAST ecosystem.

***Recommendation 10: Roadmap for Applying Advanced LLM Techniques in REDAST.*** LLMs-based techniques, in recent years, have significantly promoted the research progress in various fields [40, 107]. However, in our selected REDAST studies, we cannot identify many studies that adopt LLM techniques. P155 [155] adopts retrieval-augmentation generation (RAG) by introducing GPT4 as the backbone. P23 [23], P52 [52], P94 [94], P122 [122], and P123 [123] employ BERT or transformer-based Pre-trained LMs [25, 51] with the classifier-based downstream adapters. Considering transformer-based approaches were first released in 2017, it is not a big number. By investigating the recent research in LLM-based techniques, we identify the following potential aspects about the application of LLM-based Techniques in REDAST studies,

- *Surface-Level Answers in LLMs' Generation Results.* LLMs-based techniques are known for uncertainty and uncontrollability [48, 52, 99]. While LLMs are pretrained on massive data, their extraordinary language reasoning ability, meanwhile, brings hallucinations in generative tasks, which may lead to surface-level reasoning to yield shallow answers. In the REDAST process, the test artefacts require concrete content from the targeted software, which is a great challenge for general LLMs. To address this challenge, there are some LLM techniques to improve the quality of the LLM-generated results. Fine-tuning is the general solution for this challenge [97, 104], whereby designing downstream tasks and preparing a finetuning dataset, LLMs will be instructed with specific knowledge and generate more regulated responses. However, in most of the REDAST studies, the dataset curation and computing resources greatly hinder the application of finetuning. Prompt engineering techniques, such as few-shot learning [94], in-context learning [27], and system prompt instruction [105], only require reforming from the input perspective, which can explicitly instruct and regulate LLMs with concrete examples, specific context, and persona descriptions.

- *Multi-Modal Requirements Handling Capability.* As we stated in the previous recommendation, requirements are not limited to structural or unstructured descriptions. Screenshots, voice, or figures can also be part of the requirements specifications. Multi-modal LLMs, such as GPT4o [41], Claude-4 [3], and Deepseek [50], are perfect solutions for these requirements. The multi-modal LLMs can capture not only the information within the textual prompt but also model visual or audio features for test generation.
- *Traceability and Explainability of LLM-based Techniques.* The unexplainability is one of the key disadvantages of LLMs [17, 19]. Due to the massive number of parameters, understanding the logic of LLMs is significantly more challenging compared to traditional methods. Thus, the inference of LLMs is always regarded as a black-box process, but from the input and output sides, there are some techniques that enable explainability and traceability. For the input side, Chain-of-Thought (CoT) specifically lists all the thinking routes to instruct LLMs to reason in logical steps [56, 95], which can also improve the reasoning ability and the consistency of the results. From the output perspective, some advanced LLMs enable the generation of the reasoning steps, such as Deepseek-R1 [35], OpenAI o1 [43], Qwen-3 [4], etc, where the reasoning steps enable the traceability for users to track the explicit inference process.
- *Safety and Legal Concerns in LLM-based Techniques.* Considering the software involves a lot of confidential information, safety and privacy protection should be seriously considered in the LLM's application in REDAST. The LLMs API services, due to the restriction of the API calling strategy, require that the data be distributed to the model side for processing, which violates the safety and privacy requirements. Thus, we recommend that the researchers should prioritize the small LLMs loaded locally, such as Qwen [98], Deepseek [50], Llama [33], etc. Restricting data to confidential devices is the ultimate solution to ensure compliance with safety and legal requirements.
- *LLMs-based Evaluation for REDAST Approaches.* LLMs have recently been discovered as judges for different evaluation tasks, which are named LLM-as-a-judge [34, 49]. In REDAST studies, most of the evaluations still require case studies or expert evaluations. LLM-as-a-judge is expected to replace part of the manual effort in the evaluations. However, before applying the LLM-as-a-judge to the evaluation, further demonstration of the effectiveness and the alignment with expert judgment is necessary. Besides, the ground truth as a reference for LLM-as-a-judge should be proposed

## 8 Conclusion

[3] https://www.anthropic.com/claude-4-system-card
[4] https://qwenlm.github.io/blog/qwen3/

REquirements-Driven Automated Software Testing is a crucial yet complex area of contemporary software engineering research, with significant potential to enhance the efficiency and effectiveness of software testing processes. However, as evidenced by our systematic literature review, the absence of comprehensive guidelines and clearly articulated methodologies continues to pose substantial challenges. This SLR provides an extensive analysis of technical solutions and approaches proposed for REDAST across diverse software systems.

From an initial pool of 27,333 papers, our rigorous multi-stage filtering process identified 156 relevant studies. These studies were thoroughly analyzed across five dimensions: requirements input, transformation techniques, test artifacts, evaluation methods, and reported limitations. Despite the current studies in REDAST, there are still some unresolved problems: (1) There are no golden solutions for requirements specifications in the REDAST process, where each requirements specification reflects different pros and cons due to their usage scenario. It hinders the generalization ability of parsing techniques across different domains and projects; (2) Lack of golden solutions, e.g., taxonomies, templates, and reference pipelines in handling NFR classes (e.g., performance, safety, security, usability) into executable artifacts and oracles; (3) The parsing of the requirements still strongly relies on manual resolving; (4) Single, static intermediate representations limit end-to-end flow, where the intermediate representations are rarely supportive for multi-modal data; (5) Context-aware generation remains ad hoc, with limited standard retrieval schemas, conflict resolution, and provenance/trust signals when integrating code, logs, bugs, and prior tests; (6) Benchmarks, datasets, and shared oracles are scarce, hindering reproducibility and fair comparison; (7) Industry-grade tools and reference architectures are immature, with gaps in usability, scalability, on-prem deployment, role-based review, and audit trails; (8) Stage coverage is uneven—generation excels at design/execution but lags in planning and reporting; (9) LLM fidelity and controllability remain unresolved (hallucination, shallow reasoning, weak grounding), limiting test-ready outputs; (10) Multi-modal requirements are not first-class citizens—pipelines rarely co-parse text, UI wireframes, diagrams, and tables into unified intermediate representations and oracles; (11) Traceability and explainability are brittle at scale, with inconsistent end-to-end links and minimal rationale capture; (12) Privacy, safety, and compliance pathways are unclear for organizations unable to ship data to third-party APIs, and on-prem small LLMs/policy controls are under-evaluated; (13) Evaluation with LLM-as-a-judge lacks anchors—limited evidence on agreement with experts, reliability across artifacts/stages, and bias control.

Drawing from these insights, we propose a structured research roadmap that provides recommendations with respect to the current gaps and unresolved problems. For (1) and (3), we recommend standardizing authorable yet machine-ready requirement formats and adaptive/semi-supervised parsers with drift detection to reduce manual resolving. For (2), we call for NFR taxonomies, templates, and reference pipelines that turn non-functional requirements into executable artifacts and oracles. For (4) and (10), we suggest a multi-modal, transformable intermediate representation family (text–vision co-parsers, alignment schemas, UI-aware oracles) to support end-to-end flows. For (5), we propose context adapters and provenance-aware retrieval that integrate code, logs, bugs, and prior tests with conflict resolution and trust signals. For (6) and (13), we urge community datasets and executable evaluation harnesses plus calibrated LLM-as-a-judge protocols with gold rationales and agreement reporting. For (7) and (12), we emphasize industry-grade, privacy-preserving tooling (on-prem/small LLMs, access control, redaction, audit trails) and deployment governance. For (8), we recommend integrated generation adapters that enable the transformation among different test stages. For (9) and (11), we target retrieval-grounded, schema-constrained generation with verifiers, automatic post-hoc grounding checks, and end-to-end traceability. Collectively, these recommendations—augmented by disciplined use of LLMs (RAG with provenance, multimodal understanding, tool-using orchestration, and calibrated judging)—aim to raise automation, improve artifact robustness, traceability, and completeness, strengthen benchmarking and reproducibility, and accelerate the delivery of real-world, industry-ready REDAST systems.

### Included Primary Studies

## A Appendix: NFRs Covered In Selected Studies

Table 44. NFRs in Selected Studies (RQ1)

| Paper ID | Description |
|---|---|
| P1 [1] | Introduces a conceptual metamodel for bridging software testing and requirements. Along with the metamodel for requirements pattern, they design two classes, respectively specialized for functional and NFRs. While the FRs are modeled based on Behavior-Driven Development, the NFRs model is described in a textual format. |
| P2 [2] | Introduces an additional column named System Response, which means the expected result of each step. This column uses a specially designed attribute to reflect some NFRs, such as reliability, usability and accessibility. |
| P21 [21] | Utilizes extended requirements meta-models to model not only the temporal properties (LTL), but also the non-temporal properties (i.e., response time or max-invocation-count). Besides, in the test generation process, the non-temporal requirements properties into positive and negative event sets for test-path searching. |
| P26 [26] | Encodes time constraints into SCR-style FSM, i.e., real-time rules etc. |
| P31 [31] | Additionally designs checking for NFRs and formulates the RSL-based NFRs into tests in test specification language. |
| P48 [48] | Especially designed for timed safety-critical systems, where the MC/DC covers the safety constraints lived as boolean guards. The generator applies MC/DC on every guard for all possible timed paths to generated test artifacts. |
| P57 [57] | Ddopts GRL (Goal-Requirements Language) to capture NFRs and design a testable requirements model to bind each GRL with the IUT (implementation-under-test). |
| P68 [68] | Ddopts a simple way to convert the NFRs into the other expressions, i.e., the NFRs will be restructured into pass/fail rule before the testing process. |
| P71 [71] | Designs a meta-model with an additional attribute for NFRs. Then the clustering method is applied to the dependencies and converts the NFRs into assertions for testing planning. |
| P72 [72] | The NFRs written in OCL format are mutated, where only mutants that violate an invariant but keep functional logic intact are retained and used to generate a minimal yet fault-revealing NFR test-set. |
| P76 [76] | Expands a conventional use-case/user-story meta-model for NFRs (quality attributes) such as usability, reliability, and performance. In the validation, the meta-model will undergo further analysis against the user requirements, where the "warnings" attribute reflects the missing related information and NFRs. |
| P79 [79] | Each temporal-logic property (i.e., safety, liveness, etc) is statically analyzed. A model checker is applied to synthesize test traces that hit those states first, yielding minimal test cases yet sufficiently revealing NFR violations. |
| P82 [82] | The security requirements are especially considered in this study, which is believed to be a type of NFR. This study employs use-restricted misuse-case modelling template to describe the security requirements. |
| P86 [86] | The NFRs considered in this study are the reliability-related properties. By decomposing the event logs into petri-net in the ADCV process, the results will be used to construct the software-reliability-growth model. |
| P87 [87] | Embeds performance and reliability variables, which are parts of NFRs, directly in the scenario model. Specifically, the task scenario carries the NFR-related attributes of "cycle count" and "expected response time (ERT)". |
| P91 [91] | The safety-based NFRs are formulated as property automata, wherein the corresponding system model is automatically partitioned into sub-state-spaces so that the tester can run the model checker on smaller chunks and still get counter-example traces that violate a given safety rule. |
| P100 [100] | After identifying the failure modes based on requirements, they then identify the list of effects for each failure mode, including issues, non-compliance to standards, non-functional features, performance issues, intermittent operations, and robustness issues, which are believed as NFRs. |
| P118 [118] | Adopts the similar strategy for safety requirements in P82, where this study extend the classic misuse-case idea by inserting a mitigation use case for every identified threat. |
| P141 [141] | Introduces a meta-model to model the functional and NFRs, i.e., security, dependency or conflicts. A modified PageRank method extracts all the dependencies among interdependent requirements to priority rankings based on the dependency relatedness. The priorities are then propagated to linked test cases and become the basis for selecting a minimal subset that covers the chosen extra-FRs. |
| P150 [150] | The NFRs in this paper is the high-level linear temporal logic assertions (i.e., safety or responsiveness qualities). This study introduces three coverage methods, including requirements coverage, antecedent coverage, and unique-first-cause coverage, to generate the trap properties that yield counter-example traces automatically. |
| P153 [153] | Analyzes the risk exposure under requirements-based prioritization. Each requirements are computed into a weighted priority, which describe the probability and severity for operational risks. In the test generation and execution process, the execution order aligns with the weight ranking, reflection that requirements whose failure would hurt quality attributes such as reliability or availability are exercised first. |